\documentclass[11pt]{article}
 \usepackage{benstyle}
  
\usepackage{mathtools}

\usepackage[font={small}]{caption}
\usepackage{cite}

\usepackage{url}

\newcommand{\Romnum}[1]{\uppercase\expandafter{\romannumeral#1}}

\begin{document}

\title{Compressed Sensing and Parallel Acquisition}
\author{Il Yong Chun \\ Department of Electrical Engineering \& Computer Science \\ The University of Michigan \\ USA \and Ben Adcock \\ Department of Mathematics \\ Simon Fraser University \\ Canada }
\maketitle

\begin{abstract}
Parallel acquisition systems arise in various applications to moderate problems caused by insufficient measurements in single-sensor systems.  These systems allow simultaneous data acquisition in multiple sensors, thus alleviating such problems by providing more overall measurements.  In this work we consider the combination of compressed sensing with parallel acquisition.  We establish the theoretical improvements of such systems by providing nonuniform recovery guarantees for which, subject to appropriate conditions, the number of measurements required per sensor decreases linearly with the total number of sensors.  Throughout, we consider two different sampling scenarios -- distinct (i.e.\ independent sampling in each sensor) and identical (i.e.\ dependent sampling between sensors) -- and a general mathematical framework that allows for a wide range of sensing matrices.  We also consider not just the standard sparse signal model, but also the so-called sparse in levels signal model.  As our results show, optimal recovery guarantees for both distinct and identical sampling are possible under much broader conditions on the so-called sensor profile matrices (which characterize environmental conditions between a source and the sensors) for the sparse in levels model than for the sparse model.  To verify our recovery guarantees we provide numerical results showing phase transitions for different multi-sensor environments.
\end{abstract}

\section{Introduction}\label{s:introduction}
Many problems in signal and image processing call for the recovery of a discrete signal $x \in \bbC^{N}$ from linear measurements of the form 
\be{
\label{standard_CS}
y = A x +e,
}
where $A \in \bbC^{m \times N}$ and $e \in \bbC^{m}$ is noise.  With the development of compressed sensing (CS) over the last decade, there is now a wealth of theory and practical reconstruction algorithms that deal with recovery in the highly underdetermined regime $m \ll N$, subject to appropriate constraints on the signal $x$ (e.g.\ sparsity).

The purpose of this paper is to introduce a generalization of this work to the case where the measurement model \R{standard_CS} is replaced by a so-called \textit{parallel acquisition} model.  This model takes the form
\be{
\label{parallel_CS}
y_{c} = A_{c} x + e_c,\qquad c=1,\ldots,C,
}
where $A_{c} \in \bbC^{m_c \times N}$ is the measurement matrix modelling the sensing in the $c^{\rth}$ sensor and $e_c \in \bbC^{m_c}$ is noise.  In other words, rather than a single sensor yielding measurements of the form \R{standard_CS}, we consider the scenario where $C$ sensors act in parallel and simultaneously acquire measurements of a single signal $x$.  

Due to various practical limitations, a single sensor system \R{standard_CS} often does not provide enough measurements for a source signal to be recovered.  Parallel acquisition systems \R{parallel_CS} ameliorate this problem by allowing simultaneous data acquisition in multiple sensors, thereby providing more overall measurements.  As we explain in \S \ref{sec:Appl}, parallel acquisition models arise in a variety of applications, and are known empirically to convey a number of practical benefits; for example, acquisition time, power consumption or cost reduction, or enhanced resolution.  

The main results we prove in this paper provide the first theoretical confirmation of these empirical observations in a CS setting.  Specifically, we introduce a series of recovery guarantees which (subject to appropriate conditions) take the form
\be{
\label{Intro_guarantee}
m = \sum^{C}_{c=1} m_c \gtrsim s \times \mbox{(log factors)},
}
independent of the number of sensors $C$, where $s$ is the sparsity of the vector $x$.  In other words, the average number of measurements required per sensor 
\bes{
m_{\mathrm{avg}} \gtrsim C^{-1} s \times \mbox{(log factors)},\qquad m_{\mathrm{avg}} = C^{-1} \sum^{C}_{c=1} m_c,
}
decreases linearly in $C$ as $C$ increases, thus demonstrating the benefits of multi-sensor over single-sensor architecture.  Moreover, these results are not just of theoretical interest.  The various conditions needed for recovery guarantees to hold provide some key insight into practical issues such as optimal sensor design and alignment.

\subsection{Measurement model and recovery algorithm}
In block form, the measurement model \R{parallel_CS} can be written as
\be{
\label{block_measurements}
y = A x + e,\qquad A = \left [ \begin{array}{c} A_1 \\ \vdots \\ A_C \end{array} \right ],\qquad y =  \left [ \begin{array}{c} y_1 \\ \vdots \\ y_C \end{array} \right ],\qquad e =  \left [ \begin{array}{c} e_1 \\ \vdots \\ e_C \end{array} \right ].
}
Throughout the paper, our recovery algorithm will be the usual basis pursuit
\be{
\label{recovery_alg}
\min_{z \in \bbC^N} \| z \|_{1}\ \mbox{subject to $\| A z - y \|_2 \leq \eta$},
}
where $\eta > 0$ is such that $\| e \|_2 \leq \eta$.  Within this setup we consider two distinct classes of problem:

\subsubsection{Identical sampling}
Here the matrices $A_1,\ldots,A_C$ are dependent with $m_1 = \ldots = m_C = m/C$.  Specifically, we let $A_{c} = \tilde{A} H_c$, where $\tilde{A} \in \bbC^{m/C \times R}$ is a standard compressed sensing matrix (e.g.\ a random subgaussian matrix, subsampled isometry or random convolution) and $H_{c} \in \bbC^{R \times N}$, $c=1,\ldots,C$, are fixed, deterministic matrices.  We refer to such matrices as \textit{sensor profile matrices}.

\subsubsection{Distinct sampling}  Here the matrices $A_1,\ldots,A_C$ are independent, i.e.\ they are drawn independently from possibly different distributions.  
Typically, these will be of the form $A_c = \tilde{A}_c H_c$, where each $\tilde{A}_{c} \in \bbC^{m_c \times N_c}$ is a standard CS matrix and $H_{c} \in \bbC^{N_c \times N}$ is a sensor profile matrix. 

\subsubsection{Sensor profile matrix}
The sensor profile matrices $H_c$ model environmental conditions in the sensing problem; for example, a communication channel between $x$ and the sensors, the geometric position of the sensors relative to $x$, or the effectiveness of the sensors to $x$.  As we explain in \S \ref{sec:Appl}, this is a realistic model in practice.  We note also that the single-sensor model \R{standard_CS} is a particular case of multi-sensor model \R{parallel_CS} corresponding to $C=1$ sensors.

\subsection{Contributions}\label{ss:contributions}
Despite considering two different sensing scenarios, in \S \ref{s:setup} we introduce an abstract framework that is sufficiently general to address both simultaneously.  This is an extension of the \textit{RIPless} CS theory of Cand\`es \& Plan (see \S \ref{ss:relation}).  A key advantage of this framework is that it allows for a wide range of sensing matrices $\tilde{A}_1,\ldots,\tilde{A}_C$ (for distinct sampling) or $\tilde{A}$ (for identical sampling), including subgaussian random matrices, subsampled isometries and random convolutions.

Our main result for this framework (Theorem \ref{t:abs_recov}) demonstrates that an approximately sparse vector $x$ with support set $\Delta$ can be stably and robustly recovered from a number of measurements
\be{
\label{abs_guarantee}
m \gtrsim D \cdot \Gamma(F,\Delta) \cdot L
}
which are contaminated with noise.
Here $D$ is a number dependent on the type of sampling ($D=1$ for distinct and $D=C$ for identical), $L$ is a log term, $F$ is the distribution from which the sensing matrix $A$ is drawn  and $\Gamma(F,\Delta)$ is the so-called \textit{local coherence of $F$ relative to $\Delta$} (Definition \ref{d:coh_rel}).  A feature of \R{abs_guarantee} is that {it does not assume a signal model on the vector $x$, in a similar way to \cite{Boyer&Bigot&Weiss:15arXiv} (see \S \ref{ss:relation} for a discussion).  This is crucial, since it allows us to prove results later not just about the recovery of sparse vectors but also concerning more structured signal models.  As we discuss in \S \ref{ss:sparsity_models}, certain structured signal models arise naturally in parallel acquisition systems such as \R{block_measurements}; in particular, the so-called sparse and distributed model, which is a particular case of the sparsity in levels model introduced in \cite{Adcock&etal:16FMS}.
As we shall see throughout the paper, optimal recovery guarantees for sparse and distributed vectors are possible under broader conditions on the sensor profile matrices.  Conversely, optimal recovery guarantees for sparse vectors may not be known, or may require much stricter conditions.

\subsubsection{Distinct sampling}
Our first result for distinct sampling, Corollary \ref{c:distinct_sparsity}, gives an optimal recovery guarantee for sparse vectors of the form
\bes{
m  \gtrsim s \cdot \left ( \max_{c=1,\ldots,C} \mu(F_c) \right ) \cdot L,
}
where $F_1,\ldots,F_C$ are the distributions of the matrices $A_1,\ldots,A_C$ and $\mu(F_1),\ldots,\mu(F_C)$ are their corresponding coherences (see \S \ref{ss:background}).  Thus, provided the sampling distributions are \textit{incoherent}, i.e.\ $\mu(F_c) \lesssim 1$ for $c=1,\ldots,C$, we obtain an optimal recovery guarantee in this setting.

Our next results consider the case of sensing matrices $A_{c} = \tilde{A}_c H_c$, $c=1,\ldots,C$, where $\tilde{A}_1,\ldots,\tilde{A}_C$ are standard CS sensing matrices drawn from isotropic distributions $G_1,\ldots,G_C$ and $H_1,\ldots,H_C$ are sensor profile matrices.  For diagonal sensor profiles with the sparsity model, Corollary \ref{cor:DinstictSamp_diagCh} provides a recovery guarantee of the form
\bes{
m \gtrsim s \cdot \mu_G \cdot \left ( \max_{c=1,\ldots,C} \| H_c \|^2_{\infty} \right ) \cdot L,
}
provided the sensor profile matrices satisfy $C^{-1} \sum^{C}_{c=1} H^*_c H_c = I$, where $I$ is the $N \times N$ identity matrix.  Here $\mu_G = \max_{c=1,\ldots,C} \mu(G_c)$ is the maximum coherence of distributions $G_c$.  Hence, subject to incoherent sensing, one derives an optimal recovery guarantee provided $\| H_{c} \|_{\infty} \lesssim 1$.  This result therefore sheds light on the key issue of sensor profile design: namely, one requires profiles which do not grow too large.  In Examples \ref{eg:dinstinctSamp_identDiagCh}--\ref{ex:piecewise_const} we provide a number of different constructions for the $H_c$ which satisfy this condition.  

As we explain, unfortunately there are several sensor profiles for which the condition $\| H_{c} \|_{\infty} \lesssim 1$ is not met.  However, optimal recovery guarantees are still often possible in this setting, provided $x$ is not just sparse, but also sparse and distributed.  Corollary \ref{c:distinct_diag_levels} provides a recovery guarantee for this model, and in Examples \ref{ex:nonoverlapping} and \ref{ex:piecewise_const_II} we demonstrate how this leads to greater flexibility in the sensor profile matrices.

\subsubsection{Identical sampling}
As is to be expected, our results for identical sampling are weaker than those for distinct sampling.  In \S \ref{ss:worst_case} we present a series of worst-case bounds (i.e.\ showing no improvement as $C$ increases) for this setup.  These bounds are sharp in the sense that they are achieved by certain choices of the sensor profiles $H_c$ (see Examples \ref{ex:repeated} and \ref{ex:nonoverlapping_identical}).  Fortunately, in \S \ref{ss:identical_pcwse_const} we provide a general construction of sensor profile matrices for which optimal recovery guarantees are possible within the sparse and distributed model.  These sensor profile matrices are diagonal and have piecewise constant blocks.

\subsection{Applications} \label{sec:Appl}

Parallel acquisition techniques have been applied to enhance various practical applications, through measurement acquisition time reduction (e.g.\ in parallel magnetic resonance imaging), power consumption reduction in sensors (e.g.\ in wireless sensor networks), or recovery of higher-resolution or higher-dimensional signals (e.g.\ in multi-view imaging or light field imaging), for example.

\subsubsection{Parallel magnetic resonance imaging}\label{sss:pMRI}
The most general system model in Parallel Magnetic Resonance Imaging (pMRI; MRI with multiple receive coils) can be viewed as an example of identical sampling with diagonal sensor profiles \cite{Chun&Adcock&Talavage:15TMI,Pruessmann&Weiger&Scheidegger&Scheidegger:99MRM}. 
Numerous works have sought to apply CS to the pMRI system \cite{Chun&Adcock&Talavage:14EMBS_pMRI, Chun&Adcock&Talavage:15TMI, Guerquin-Kern&Lejeune&Pruessmann&Unser:12TMI,Knoll&Clason&Bredies&Uecker&Stollberger:12MRM,She&Chen&Liang&DiBella&Ying:14MRM} in order to accelerate MR scanning by reducing the amount of $k$-space acquired.
Here, $x$ is the unknown magnetization, $y_c$ is the vector of subsampled Fourier measurements (with the same sampling trajectories across coils) for the $c^{\rth}$ receive coil, $H_c$ is the $c^{\rth}$ coil sensitivity, and $C$ is the number of receive coils.  In this case, the model \R{block_measurements}--\R{recovery_alg} is the well-known CS SENSE technique for pMRI \cite{She&Chen&Liang&DiBella&Ying:14MRM, Chun&Adcock&Talavage:15TMI}.
Some previous work \cite{Chun&Adcock&Talavage:15TMI} has derived a worst-case bound for this model (for noiseless measurements) which is similar to the results we prove in \S \ref{sec:worst_identSamp_diag} in this paper.  These bounds, however, do not demonstrate the benefits of parallel acquisition as $C$ increases.  Fortunately, a particular consequence of our main result in \S \ref{ss:identical_pcwse_const} does precisely this.  Thus, the results in this paper provide the first theoretical justification for the improvement in terms of scan time reduction offered by CS for pMRI. 

\subsubsection{Multi-view imaging}\label{sss:multiview}
Multi-view imaging arises when $C$ cameras, aligned in different positions, simultaneously image a single object.  Following the work of \cite{Park&Wakin:12EJASP, Traonmilin&etal:15JMIV}, this can be viewed in terms of \R{block_measurements}.  In far-field multi-view imaging -- with applications to satellite imaging or unmanned aerial vehicle remote sensing -- the sensor profile matrices $H_c : \bbR^{N} \rightarrow \bbR^{N_c}$ are used to represent the geometric features of the scene, e.g.\ rotations, scalings, etc.  In near-field multi-view imaging the sensor profile matrices can be represented using the plenoptic function in order to reconstruct a three-dimensional (3D) volumetric signal \cite{Park&Wakin:12EJASP,Berent&Dragotti:07SPM}.  Likewise, super-resolution imaging, wherein a detailed image is recovered from a set of low resolution images \cite{Baboulaz&Dragotti:09TIP, Duarte&Eldar:11SP, Jiang&Huang&Wilford:14APSIP-SIP}, can also be understood in this framework.

\subsubsection{Sparsity and generalized sampling theory} \label{sss:generalSamp}
The classical Shannon Sampling Theorem states that a band-limited signal $f(t)$ can be recovered from equally-spaced samples taken at the Nyquist rate \cite{Unser:PIEEE}.  A well-known extension of this is Papoulis' generalized sampling theorem \cite{Papoulis:77TCS}, which states that a band-limited signal can be recovered from samples of $C$ appropriate linear functionals $g_{c}(t)$ of $f(t)$ taken at $1/C$ of the Nyquist rate (i.e.\ $C$ times further apart).  Our identical sampling framework gives rise to a sparse, discrete version of this theorem.  Indeed, let $f \in \bbC^{N}$ be a discrete signal and consider the linear functionals
\bes{
g_{c}(t) = \cF^{-1} \left \{ H_c x  \right \},\qquad t \in \{1,\ldots,N\},
}
where $x = \cF \{ f \}$, $\cF$ denotes the discrete Fourier transform (DFT) and $H_c$ are so-called system functions \cite{Papoulis:77TCS} (which can be viewed as diagonal sensor profile matrices in our setup).  Now let $t_{1},\ldots,t_{m/C}$ be sampling points chosen randomly from the $C$-fold downsampled grid $\{ C+1, 2C+1 , \ldots, (n-1) C+1 \}$, where $N = n C$ (this downsampling corresponds to $1/C$ of the Nyquist rate in the discrete setting).  Much like Papoulis' generalized sampling, our results in \S \ref{ss:identical_pcwse_const} provide explicit conditions on the sensor profiles $H_c$ for which a discrete signal $f$ with $s$-sparse Fourier transform $x$ can be recovered uniquely from the $m$ measurements\footnote{A bandlimited signal $f$ can be considered as a signal with a clustered sparse
(a type of sparse and distributed -- see Remark \ref{r:clustered}) Fourier transform $x$.}
\bes{
g_{c}(t_i),\qquad i=1,\ldots,m,\ c=1,\ldots,C.
}
Note that in matrix-vector form, this is equivalent to the system \R{block_measurements} with $A_c = \tilde{A} H_c$ and $\tilde{A} \in \bbC^{m/C \times N}$ being of the form $\tilde{A} = P_{\Omega} \tilde{\Psi}$, where $\tilde{\Psi} \in \bbC^{N/C \times N}$ is the $C$-fold downsampled DFT matrix and $P_{\Omega}$ is the projection matrix corresponding to the indices $\Omega = \{ t_1,\ldots,t_{m/C} \}$.

\subsubsection{Other applications}
A number of other applications can also be viewed within our framework:

\vspace{1pc}\noindent \textit{(a)}\ In system identification, the problem of recovering the initial state of a high-dimensional dynamical system can be formulated in terms of \R{block_measurements}.  CS techniques have been applied to this problem to reduce the measurement burden. See observability problem  in \cite[Chpt.\ $6$]{Sanandaji:12PhD}.

\vspace{1pc}\noindent \textit{(b)}\ In wireless sensor networks (WSN), CS techniques have been applied to reduce the communication burden transmitted from wireless sensors to a fusion center \cite{Shen&etal:13SJ,Wimalajeewa&Varshney:15TSIPN,Yang&etal:13TSP}, from the perspective of multiple access channel communication architecture \cite{Haupt&etal:08SPM}.  In \cite{Choi&Park&Lee:bookCh, Oliver&Lee:11SPARS} a realistic system model is formulated for this problem, along the lines of \R{block_measurements}.

\vspace{1pc}\noindent \textit{(c)}\ In light-field imaging systems such as the Lytro \cite{Ng&atel:05Lytro} and Raytrix \cite{Raytrix:09EUPatent} plenoptic cameras provide a single-shot imaging tool for digital refocussing \cite{Ng&atel:05Lytro}, 3D volumetric imaging \cite{Nien&etal:15ICIP, Thurow&Fahringer:13PIV, Fahringer&Thurow:12AIAA}, conventional high-resolution two-dimensional (2D) imaging \cite{Bishop&Favaro:12TPAMI, Georgiev&Lumsdaine:09Adobe}, etc.  More recently, a micro lens array consisting of lenses with different focal lengths has been applied to light-field imaging to extend the plenoptic depth of field \cite{Georgiev&Lumsdaine:12SPIE, Perwass&Wietzke:12SPIE}.  This framework has been investigated in \cite{Nien:14PhD} to recover the light-field, and can be understood in terms of \R{block_measurements}.

\vspace{1pc}\noindent \textit{(d)} In synthetic aperture radar imaging, one can recover a high number of non-zeros in the signal with low-sampling-rate devices \cite{Aceska&etal:16Arxiv}. This can also be formulated in a model of the form \R{block_measurements} for certain types of partitioned sensor profile matrices.

\subsection{Relation to previous work}\label{ss:relation}

The so-called \textit{RIPless} theory developed by Cand\`es \& Plan in \cite{Candes&Plan:11IT} is well known in the CS literature (see \S \ref{ss:background} for a summary).  Our framework is a generalization of this work to multi-sensor systems.  Note that the results of \cite{Candes&Plan:11IT} become special cases of our framework corresponding to the single-sensor ($C=1$) case.  While our main result for distinct sampling with the sparsity model (Corollary \ref{c:distinct_sparsity}) is a corollary of results in \cite{Candes&Plan:11IT}, our results for identical sampling with the sparsity model (Corollaries \ref{cor:identSampl_sparsity} and \ref{c:identSampl_sparsity_circH}), and for both distinct and identical sampling with the sparse and distributed model (Corollary \ref{c:distinct_sparsity_in_levels} and Theorem \ref{t:identical_pcwse_const}), cannot be obtained in this way.
Our proofs follow a similar route to those of \cite{Candes&Plan:11IT}, albeit with some key modifications to incorporate the more complicated measurement matrices and sparsity models.
The framework introduced in this paper and its analysis are also related to several earlier works \cite{Adcock&etal:16FMS, Bigot&Boyer&Weiss:16TIT, Boyer&Bigot&Weiss:15arXiv}.  Our model is more general than that of \cite{Adcock&etal:16FMS} (which corresponds to a particular case of distinct sampling), although we use the concept of sparsity in levels introduced therein to provide recovery guarantees (see Remark \ref{r:BreakCohRelation}).  While motivated by quite different applications, the abstract model introduced in \cite{Bigot&Boyer&Weiss:16TIT,Boyer&Bigot&Weiss:15arXiv} turns out to be quite similar to ours (see Remark \ref{r:BoyerRelation}).  Our theoretical results improve on those of \cite{Bigot&Boyer&Weiss:16TIT,Boyer&Bigot&Weiss:15arXiv} in a number of ways (see Remarks \ref{r:BoyerRelation} and \ref{r:BoyerThyRelation}).
Finally, note that our results are nonuniform recovery guarantees.  For subgaussian random sensing, a series of uniform recovery guarantees -- based on the techniques of Krahmer, Rauhut \& Mendelson on suprema of chaos processes \cite{Krahmer:14CPAM} -- have recently been proved in \cite{Chun&Adcock:16arXiv-CS&PA&RIP}.

There have also been a number of other theoretical works in which different measurements are concatenated together similar to as in \R{block_measurements}.  In \cite{Eftekhari&etal:15ACHA} (see also references therein), block diagonal measurement matrices are considered, where each block consists of a subgaussian random matrix.  Such a measurement matrix can be viewed as a special case of our framework.  Unsurprisingly, the results of \cite{Eftekhari&etal:15ACHA}, being specific to subgaussian measurements, are sharper than ours (which apply to a much broader class of measurement matrices -- see \S \ref{ss:background}) would be in such an instance. See \cite{Chun&Adcock:16arXiv-CS&PA&RIP} for further details.  We note in passing some related work of Polak et al \cite{Polak&Duarte&Goeckel:15TSP}.

Finally, we remark that the model \R{block_measurements} considered in this paper is quite different to the well-known multiple measurement vector (MMV) model \cite{Baron&etal:09TIT, Cotter&etal:05TSP, Chen&Huo:06TSP, Eldar&Rauhut:10IT, Boufounos&Kutyniok&Rauhut:11IT, Duarte&Eldar:11SP} and to so-called \textit{distributed} CS \cite{Eftekhari&etal:15ACHA, Baron&etal:09TIT, Duarte&Baraniuk:12TIP}.  Rather than recovering multiple signals (possibly with a shared support), our interest lies with the recovery of a single signal $x$ from multi-sensor observations.
Note that one may be tempted to reinterpret the parallel acquisition model as an MMV problem by defining the local signals $x_{c} = H_c x$, $c=1,\ldots,C$.  Assuming these have a common sparse support, then one could apply a standard MMV solver (e.g.\ $\ell^{2,1}$-norm minimization) to recover them, followed by a least-squares fit to recover the overall signal $x$ (this is similar to the \textit{Relax.\ spJS CS SENSE} model  for parallel MRI reconstruction considered in \cite[(25)]{Chun&Adcock&Talavage:15TMI}).  However, this approach results in suboptimal recovery guarantees.  Since there are now $C$ signals $x_1,\ldots,x_C$ to recover (with generally distinct coefficients), the overall measurement condition will necessarily be of the form $m \gtrsim C \cdot s$ (plus potentially additional log factors), i.e.\ depending linearly on the number of sensors $C$ (we refer to \cite{Eldar&Rauhut:10IT, Boufounos&Kutyniok&Rauhut:11IT} for relevant theoretical results on recovery guarantees for the MMV problem).  Conversely, in this paper, by solving for the overall signal $x$ directly, we are able to obtain much stronger measurement conditions of the form \R{Intro_guarantee}, i.e.\ independent of $C$.\footnote{Ignoring sparsity, note that the system \R{block_measurements} becomes overdetermined once $m>N$ (besides in pathological cases), meaning exact recovery of any $x$.  Conversely, the MMV problem requires $C$ times as many measurements (with $m_1 \!=\! \ldots \!=\! m_C \!>\! N$) to be overdetermined. We remark in passing that our theoretical results still apply in the case $m > N$ (note that \R{recovery_alg} always has a solution since $x$ is feasible), although a least-squares fit would be a simpler approach in this case.
}

\section{Abstract framework and main result}\label{s:setup}
In this section we present our abstract framework and main result.  This framework is quite general, and will allow us to address both the distinct and identical scenarios with a wide range of different sensing matrices.

\subsection{Notation}\label{ss:notation}
Throughout, we use $\nm{\cdot}_{p}$ to denote the vector $p$-norm or its induced matrix norm (i.e., $\nm{A}_p = \sup_{\nm{x}_p = 1} \nm {A x}_p$).
We write $\ip{\cdot}{\cdot}$ for the standard inner product on $\bbC^N$.
As is conventional, we write $\nm{\cdot}_{0}$ for the $\ell^0$-norm, i.e.\ the number of nonzeros of a vector.  The canonical basis on $\bbC^N$ will be denoted by $e_1,\ldots,e_N$.  If $\Delta \subseteq \{1,\ldots,N\}$ then we use the notation $P_{\Delta}$ for both the orthogonal projection $P_{\Delta} \in \bbC^{N \times N}$ with
\bes{
(P_{\Delta}x)_{j} = \left \{ \begin{array}{cc} x_j & j \in \Delta \\ 0 & \mbox{otherwise} \end{array} \right .,\qquad x \in \bbC^N,
}
and the matrix $P_{\Delta} \in \bbC^{|\Delta| \times N}$ with
\bes{
(P_{\Delta}x)_{j} = x_j,\quad j \in \Delta,\qquad x \in \bbC^N.
}
The precise meaning will be clear from the context.  Distinct from the index $i$, we denote the imaginary unit by $\I$.  In addition, we use the notation $A \lesssim B$ or $A \gtrsim B$ to mean there exists a constant $c>0$ independent of all relevant parameters (in particular, the number of sensors $C$) such that $A \leq c B$ or $A \geq c B$ respectively.

\subsection{Background}\label{ss:background}
In order to elucidate our framework, we first recall the RIPless CS setup introduced in \cite{Candes&Plan:11IT} for the case of single-sensor measurements.  Let $\{ e_i \}^{m}_{i=1}$ be the canonical basis of $\bbC^m$ and $F$ be a distribution of vectors in $\bbC^N$.  It is assumed that $F$ is isotropic in the following sense:
\be{
\label{CPiso}
\bbE (a a^*) = I,\qquad a \sim F,
}
where $\bbE$ denotes expectation. The sensing matrix $A$ is now constructed by drawing $m$ vectors i.i.d.\ from $F$ and setting
}
\bes{
A = \frac{1}{\sqrt{m}} \sum^{m}_{i=1} e_i a^*_i.
}
Note that this setup is quite general and includes many types of measurement matrices found in CS literature.  These include subgaussian random matrices (see, for example, \cite{Candes&Plan:11IT, BaraniukSimpleRIP:08CA, Foucart&Rauhut:book}), bounded orthonormal systems \cite{Foucart&Rauhut:book, Rauhut:BookCh}, subsampled isometries \cite{Candes&Plan:11IT, Adcock&etal:16FMS, Adcock&Hansen:15FCM, Foucart&Rauhut:book}, and certain types of random convolutions \cite{Romberg:09SIAMJIS}\footnote{Note that the constructions of \cite{Rauhut:09SPARS, Rauhut&Rombert&Tropp:12ACHA}, which are based on deterministic subsampling, do not fit into this model.}, for example.

A key quantity defined in \cite{Candes&Plan:11IT} is the \textit{coherence} of $F$.  This is the smallest number such that
\be{
\label{standard_coherence}
\| a \|^2_{\infty} \leq \mu(F),\qquad a \sim F,
}
almost surely.  The main results proved in \cite{Candes&Plan:11IT} establishes that an $s$-sparse vector $x$ can be recovered from the measurements $y = A x$ using roughly $m \approx s \cdot \mu(F)$ measurements, up to log factors.  

We remark in passing that this is an example of a \textit{nonuniform} recovery guarantee: a single random draw of $A$ guarantees recovery of a fixed $s$-sparse vector $x$.  In contrast, so-called \textit{uniform} recovery guarantees ensure recovery of all $s$-sparse vectors from a single draw of $A$.  See \cite{Foucart&Rauhut:book} for a discussion.  In this paper we will only consider nonuniform recovery guarantees.

\subsection{Sensing matrices and sparsity models}\label{ss:sparsity_models}
As mentioned in \S \ref{ss:contributions}, it will be necessary in this paper to work with signal models that go beyond standard sparsity.  In this section we introduce these models and discuss why they arise naturally in parallel acquisition problems.  First we recall the definition of sparsity:

\defn{[Sparsity]
\label{d:sparsity}
A vector $z \in \bbC^N$ is $s$-sparse for some $1 \leq s \leq N$ if $\| z \|_0 \leq s$.  We write $\Sigma_{s}$ for the set of $s$-sparse vectors and, for an arbitrary $x \in \bbC^N$, write
\bes{
\sigma_{s}(x)_1 = \min \left \{ \| x - z \|_{1} : z \in \Sigma_{s} \right \},
}
for the error of the best $\ell_1$-norm approximation of $x$ by an $s$-sparse vector.
}

As discussed above, in single-sensor CS the recovery of a sparse vector $x$ from measurements $y = A x + e$ can be achieved using $m \approx s$ measurements, up to log factors, for suitable matrices $A$; for example, those arising from sampling incoherent distributions (see \S \ref{ss:background}).

In parallel acquisition with distinct sampling, it is perfectly possible to construct multi-sensor measurement matrices of the form \R{block_measurements} for which $m \approx s$ is achievable.  Indeed, we merely take each $A_{c} \in \bbC^{m_c \times N}$ to be a subgaussian random matrix.  As we shall see later, however, many other (and nontrivial) choices of the $A_c$'s will give the same optimal guarantees.

Conversely, it is also straightforward to see that in the multi-sensor setting our goal of recovering $x$ from $m \approx s$ measurements may well not be achievable for certain matrices $A_c$.  For a trivial example, suppose that each $A_{c}$ is a block matrix such that the overall matrix $A$ in \R{block_measurements} is block diagonal
\be{
\label{A_block_diag}
A = \left [ \begin{array}{ccc} \tilde{A}_1 \\ & \ddots \\ & & \tilde{A}_C \end{array} \right ],\qquad \tilde{A}_{c} \in \bbC^{m_c \times N/C}.
}
Then recovery of arbitrary $s$-sparse vectors requires $m_{c} \approx s$, since for any $c$, one may construct an $s$-sparse vector whose nonzero entries all lie in the range $\{ (c-1) N/C + 1,\ldots, c N/C \}$.  Thus the total number of measurements required in this setting is $m \approx s C$, which grows linearly with the number of sensors.

On the other hand, suppose that the vector $x$ was constrained so that not too many of its nonzero could lie in each of the subsets $\{ (c-1) N/C + 1,\ldots, c N/C \}$.  Then we can reasonably expect an optimal recovery guarantee.  This observation leads us to consider a more refined signal model than sparsity, first introduced in \cite{Adcock&etal:16FMS}, and referred to as sparsity in levels:

\defn{[Sparsity in levels]
\label{d:sparsity_lev}
Let $\cI = \{ I_1,\ldots,I_C \}$ be a partition of $\{1,\ldots,N\}$ and $\cS = (s_1,\ldots,s_C) \in \bbN^C$ where $s_c \leq | I_c|$ for $c=1,\ldots,C$.  We say that $z \in \bbC^N$ is $(\cS,\cI)$-sparse in levels if
\bes{
\left | \left \{ j : z_j \neq 0 \right \} \cap I_c \right | \leq s_c,\qquad c=1,\ldots,C.
} 
We denote the set of such vectors as $\Sigma_{\cS,\cI}$ and, for an arbitrary $x \in \bbC^N$, write
\bes{
\sigma_{\cS,\cI}(x)_1 = \min \left \{ \| x - z \|_{1} : z \in \Sigma_{\cS,\cI} \right \},
}
for the error of the best $\ell_1$-norm approximation of $x$ by an $(\cS,\cI)$-sparse vector.
}

Based on the notion of sparsity in levels, we shall also define the following:

\defn{[Sparse and distributed vectors]
\label{d:sparse_distrib}
Let $\cI = \{ I_1,\ldots,I_C \}$ be a partition of $\{1,\ldots,N\}$ and $1 \leq s \leq N$.  For $1 \leq \lambda \leq C$ We say that an $s$-sparse vector $z \in \bbC^N$ is sparse and $\lambda$-equidistributed with respect to the levels $\cI$ if $z \in \Sigma_{\cS,\cI}$ for some $\cS = (s_1,\ldots,s_C)$ satisfying
\bes{
\max_{c=1,\ldots,C} \{ s_c \} \leq \lambda s / C.
}
We denote the set of such vectors as $\Sigma_{s,\lambda,\cI}$ and, for an arbitrary $x \in \bbC^N$, write $\sigma_{s,\lambda,\cI}(x)_1$ for the $\ell_1$-norm error of the best approximation of $x$ by a vector in $\Sigma_{s,\lambda,\cI}$.
}
Note that our interest lies with the case where $\lambda$ is independent of $C$; that is, when none of the local sparsities $s_c$ greatly exceeds the average $s/C$.  
In the simple setting of \R{A_block_diag}, choosing $I_c = \{ (c-1) N/C + 1,\ldots, c N/C \}$ we see that optimal recovery is possible for sparse and distributed vectors, provided $m_{c} \approx \lambda s/C$ for each $c$, i.e.\ $m \approx \lambda s$.  Later in the paper, we will identify large classes of multi-sensor measurement matrices (not necessarily block diagonal) which can recover sparse and distributed vectors using such near-optimal numbers of measurements, but for which recovery of all sparse vectors necessarily requires a suboptimal number of measurements.

\rem{[Clustered sparse vectors]
\label{r:clustered}
It is customary to consider partitions $\cI$ where each set $I_{c}$ is of the form $\{p,p+1,\ldots,q\}$ for integers $p$ and $q$.  Yet there is no reason for this to be the case.  An interesting example is when
\bes{
I_c = (c + C \bbZ ) \cap \{1,\ldots,N \} = \{ c , c + C,\ldots,c+(n-1) C \},\qquad c=1,\ldots,C.
}
It is vectors $x$ that are clustered that turn out to be sparse and distributed with respect to this partition.  For example, suppose that 
\bes{
\Delta \subseteq \{ i,i+1,\ldots,i + \lambda s - 1\}, 
}
for some $i \in \{1,\ldots,N\}$ and $\lambda \geq 1$.  That is, the support $\Delta$ is clustered in a band of width $\lambda s$.  Then $s_c \leq \lambda s / C$ for all $c$, meaning that $x$ is sparse and $\lambda$-equidistributed with respect to this model.
}

Beside parallel acquisition, the sparsity in levels model has recently found use in a number of different applications.  These include MRI \cite{Adcock&etal:16FMS}, compressive imaging \cite{Roman&Hansen&Adcock:14arXiv,Adcock&etal:05bookCh}, radar \cite{Dorsch&Rauhut:16JFAA} and detection of clustered signals in WSNs (see Remark \ref{r:clustered}).  In general, any application where the sparse signals of interest tend to have specific distributions across their support falls within the remit of this model.  A particular case, introduced in \cite{Tsaig&Donoho:06SP} and developed further in \cite{Candes&Romberg:07IP, Adcock&etal:05bookCh, Adcock&Hansen&Roman:16SPL}, is sparse vectors of wavelet coefficients, where the levels correspond to the wavelet scales.

\subsection{Abstract framework} \label{ss:abs_framework}
We now introduce our abstract framework.

\subsubsection{General setup}\label{sss:general_setup}
For some $D \in \bbN$, let $F$ be a distribution on the space of $N \times D$ complex matrices.  We shall assume that $F$ is isotropic in the sense that
\be{
\label{F_iso}
\bbE (B B^*) = I,\qquad B \sim F.
}
Let $\{ e_i \}^{p}_{i=1}$ be the canonical basis of $\bbC^p$ and let $B_1,\ldots,B_p$ be a sequence of i.i.d.\ random matrices drawn from $F$.  Then we define the sampling matrix $A$ by
\be{
\label{A_F}
A = \frac{1}{\sqrt{p}} \sum^{p}_{i=1} e_i \otimes B^*_i = \frac{1}{\sqrt{p}} \left [ \begin{array}{c} B^*_1 \\ \vdots \\ B^*_p \end{array} \right ] \in \bbC^{p D \times N},
}
where $\otimes$ denotes the Kronecker product.  Note that the setup of \S \ref{ss:background} corresponds to the case $D=1$.    

\rem{
\label{r:BoyerRelation}
This framework is similar to that introduced by Bigot, Boyer \& Weiss \cite{Bigot&Boyer&Weiss:16TIT, Boyer&Bigot&Weiss:15arXiv}.  In \cite{Bigot&Boyer&Weiss:16TIT} the same sampling framework is considered, but only within the sparsity signal model.  Later, in \cite{Boyer&Bigot&Weiss:15arXiv} the authors consider sampling blocks of rows of an isometry, which is slightly less general than the framework considered in \cite{Bigot&Boyer&Weiss:16TIT} and this paper\footnote{Specifically, let $U$ be an $N \times N$ isometry and $\cJ_{1},\ldots,\cJ_{C}$ be a partition of $\{1,\ldots,N\}$ which describe the blocks of rows and consider the family of matrices of the form $\frac{1}{\sqrt{\pi_c}} P_{\cJ_{c}} U$ for $c=1,\ldots,C$, where $\pi_c$ is the probability of drawing the $c^{\rth}$ block.  We now let $F$ be such that $B \sim F$ if $B$ takes value $\frac{1}{\sqrt{\pi_c}} P_{\cJ_{c}} U$ with probability $\pi_c$.}.  However, in a similar manner to our main result (Theorem \ref{t:abs_recov}), \cite{Boyer&Bigot&Weiss:15arXiv}  also gives recovery guarantees that are local to the signal support (see Remark \ref{r:BoyerThyRelation}).  Note that  \cite{Bigot&Boyer&Weiss:16TIT, Boyer&Bigot&Weiss:15arXiv} are primarily motivated by the problem of practical sampling in MRI, in which isolated $k$-space measurements cannot be acquired, but blocks of measurements along smooth contours can be.  This is an important problem, albeit quite different to the parallel acquisition problem we consider in this paper.  We refer also to \cite{Boyer&Weiss&Bigot:14SJIS, Chauffert&etal:14arXiv, Chauffert&etal:14SJIS} for further details.
}

Before we present our main recovery guarantee result for measurements of the form \R{A_F}, we first explain why this model suffices for both sensing scenarios considered in this paper.

\subsubsection{Distinct sampling}\label{sss:general_setup_distinct}
Distinct sampling corresponds to a case with $D = 1$ and $p = m = \sum^{C}_{c=1} m_c$.  To see why, recall that in this case $A$, as given by \R{block_measurements}, consists of matrices $A_c \in \bbC^{m_c \times N}$ drawn from possibly distinct distributions.  To formalize this within the above framework we proceed as follows.  For $c=1,\ldots,C$ let $F_c $ be a distribution on $\bbC^{N}$.  We assume the distributions $F_1,\ldots,F_C$ are jointly isotropic in the sense that
\be{
\label{joint_iso}
 \sum^{C}_{c=1} \frac{m_c}{m} \bbE \left ( a_c a^*_c \right ) = I,\qquad a_c \sim F_c,\ c=1,\ldots,C.
}
Given $m_1,\ldots,m_C$, let $m = \sum^{C}_{c=1} m_c$ and suppose that $X$ is a random variable taking the values in $\{1,\ldots,C\}$ with $\bbP(X = c ) = m_c / m$ for $c=1,\ldots,C$.  Given $F_1,\ldots,F_C$ we define the new distribution $F$ on $\bbC^N$ so that, when conditioned on the event $\{ X=c \}$, $F = F_c$.  In other words, if $a \in \bbC^{N}$ denotes an arbitrary row of $A$, then $a$ arises from the distribution $F_c$ with probability $m_c/m$.  After permuting the rows of the matrix $A$ defined in \R{A_F}, we may write
\bes{
A = \frac{1}{\sqrt{m}} \left [ \begin{array}{c} A_1 \\ \vdots \\ A_C \end{array} \right ] \in \bbC^{m \times N},
}
where $A_{c} \in \bbC^{q_c \times N}$ contains the rows of $A$ drawn from the distribution $F_c$ and $q_c$ is the number of such rows.  Note that $q_c$ is a random variable which is equal to $m_c$ in expectation.  In other words, although the number of measurements taken in each sensor is random, it is roughly equal to $m_c$.

\rem{
\label{r:drawingmodel}
In practice, one may prefer a setup where exactly $m_c$ measurements are taken in the $c^{\rth}$ sensor.  It is straightforward to modify our proofs to use this model instead.  The recovery guarantees will be unchanged, except possibly in the log factor.  We opt for the setup above for simplicity, since it means that the both the distinct and identical sampling cases can be viewed as special cases of the framework introduced in \S \ref{sss:general_setup}.
}

\rem{
\label{r:BreakingCoherenceRelation}
This distinct sampling setup can be viewed as a generalization of that of \cite{Adcock&etal:16FMS}.  Indeed, in \cite{Adcock&etal:16FMS} a unitary matrix $U \in \bbC^{N \times N}$ is subsampled by choosing, for each $c=1,\ldots,C$, exactly $m_c$ rows uniformly at random from the range $\{ N_{c-1}+1,\ldots,N_c\}$, where $1 < N_1 < \ldots < N_c = N$ and $N_0 = 0$.  The $N_c$ are referred to as \textit{sampling} levels.  It is clear that (up to the drawing model -- see Remark \ref{r:drawingmodel}) this is a particular case of our setup in which each family $F_c$ consists of those rows of $U$ with indices in $\{ N_{c-1}+1,\ldots,N_c\}$.  Note that the framework of \cite{Adcock&etal:16FMS} is particularly relevant to compressive imaging problems, wherein the matrix $U$ corresponds to the cross-Grammian of the discrete Fourier and wavelet transforms.  This model arises in numerous applications, not only in MRI, and the results proved in \cite{Adcock&etal:16FMS} demonstrate how to optimally subsample Fourier space in the case of structured wavelet sparsity.  We refer also to \cite{Adcock&etal:05bookCh,Adcock&Hansen&Roman:16SPL,Roman&Hansen&Adcock:14arXiv} for further details.
}

\subsubsection{Identical sampling}\label{sss:general_setup_identical}
Identical sampling corresponds to an instance of the framework introduced in \S \ref{sss:general_setup} with $M = C$ and $p = m/C$.  Recall in identical sampling that $A$ is formed by concatenating matrices $A_c = \tilde{A} H_c$, where $\tilde{A} \in \bbC^{m/C \times R}$ is a random matrix, and $H_c \in \bbC^{R \times N}$ are fixed, deterministic matrices.  Following \S \ref{ss:background} let $G$ be a distribution on $\bbC^{R}$, isotropic in the sense of \R{CPiso}, so that $\tilde{A}$ is given by
\bes{
\tilde{A} = \frac{1}{\sqrt{p}} \sum^{p}_{i=1} e_i a^*_i,\qquad a_i \sim G.
}
We now define the distribution $F$ on the space of $R \times C$ matrices so that $B \sim F$ if
\bes{
B = \left [ H^*_1 a |  \cdots | H^*_C a  \right ],
}
where $a \sim G$.  After possible row permutations we see that \R{A_F} is equivalent to \R{block_measurements} with this choice of $F$.  Note that we require $F$ to be isotropic in the sense of \R{F_iso}, which in this case is equivalent to the condition
\be{
\label{Hc_iso}
\sum^{C}_{c=1} H^*_c H_c = I.
}

\rem{
Actually, to ensure that $F$ is isotropic, we do not require $G$ itself to be isotropic.  Rather we require only $\sum^{C}_{c=1} H^*_c \bbE(a a^*) H_c = I$, where $a \sim G$.  However, there is little loss in generality in assuming that $G$ is isotropic and \R{Hc_iso} holds.
}

\rem{
\label{r:sparsifying_transforms}
For the remainder of this paper we shall mainly consider the signal $x$ as sparse in the canonical basis.  However, a sparsifying transform can easily be incorporated into our abstract framework.  If $x$ are the sparse coefficients of a signal in an orthonormal sparsifying transform $\Psi \in \bbC^{N \times N}$, then this just corresponds to replacing the distribution $F$ by $\tilde{F}$, where $\tilde{B} \sim \tilde{F}$ if $\tilde{B} = \Psi^* B$.  Note that this does not affect the isotropy condition \R{F_iso}, since $\Psi$ is orthonormal.  The main difficulty comes when estimating the various coherences (defined in the next section) so as to provide concrete bounds for specific families of sensor profile matrices.  It is well-known that (standard) coherence is not invariant under orthonormal transforms, meaning that a separate estimation would be required for each choice of $\Psi$.  It is work in progress to estimate these coherences for problems of interest such as Fourier sampling with wavelet sparsity.  On the other hand, we note that in the special case of sensing subgaussian random vectors, it is possible to provide recovery guarantees for general sparsifying transforms with explicit conditions (albeit using different theoretical tools to those employed in this paper).  See \cite{Chun&Adcock:16arXiv-CS&PA&RIP}.
}

\subsection{Coherence definitions} 
Much as in the standard compressed sensing setup, we require a notion of coherence.  Due primarily to the issues raises in \S \ref{ss:sparsity_models}, in our setting we need to consider a number of more refined notions than simply the global coherence $\mu(F)$.

\defn{[Coherence relative to $\Delta$]
\label{d:coh_rel}
Let $F$ be as in \S \ref{sss:general_setup}  and $\Delta \subseteq \{ 1,\ldots,N\}$.  We define the local coherence of $F$ relative to $\Delta$ as
\bes{
\Gamma(F,\Delta) = \max \left \{ \Gamma_{1}(F,\Delta),\Gamma_{2}(F,\Delta) \right \},
}
where $\Gamma_1(F,\Delta)$ and $\Gamma_{2}(F,\Delta)$ are the smallest quantities such that 
\bes{
\| B B^* P_{\Delta} \|_{\infty} \leq \Gamma_1(F,\Delta),\qquad B \sim F,
}
and
\bes{
\sup_{\substack{z \in \bbC^N \\ \| z \|_{\infty} = 1}} \max_{i=1,\ldots,N} \bbE | e^*_i B B^* P_{\Delta} z |^2 \leq \Gamma_{2}(F,\Delta),\qquad B \sim F,
}
almost surely.  Note that $\Gamma_i(F,\Delta) \geq 1$, $i=1,2$, due to the assumption \R{F_iso} on $F$.
}

This notion of coherence is convenient in that it allows us to state our main results without defining a particular signal model, whether it be sparsity or sparsity in levels. When considering the latter, however, we will also need the following notion of a local coherence.

\defn{[Local coherence in levels]
\label{def:local_coh_lev}
Let $\cI = \{ I_1,\ldots,I_C\}$ be a partition of $\{1,\ldots,N\}$ and suppose that $F$ is a distribution on $\bbC^N$.  Then the $c^{\rth}$ local coherence of $F$ in levels is given by
\bes{
\mu_{c}(F) = \sqrt{\mu(F) \mu'_c(F)}, 
}
where $\mu(F)$ is the standard coherence of $F$ as defined in \R{standard_coherence} and $\mu'_c(F)$ is the smallest constant such that
\bes{
\|  P_{I_c} a \|^2_{\infty} \leq \mu'_{c}(F),\qquad a \sim F,
}
almost surely.
}

\subsection{Main theorem} \label{ss:main_result}

Our main result for the framework introduced in \S \ref{sss:general_setup} is as follows:

\thm{[Abstract recovery guarantee]
\label{t:abs_recov}
For $N,D,p \in \bbN$ with $N \geq 2$ and $p D \leq N$ let $F$ be a distribution on $\bbC^{N \times D}$ satisfying \R{F_iso} and suppose that $0 < \epsilon < 1$, $\eta \geq 0$ and $\Delta \subseteq \{1,\ldots,N\}$ with $s = |\Delta| \geq 2$.  Let $x \in \bbC^N$ and draw $A \in \bbC^{m \times N}$ according to \R{A_F}, where $m=pD$.  Then for any minimizer $\hat{x}$ of
\bes{
\min_{z \in \bbC^N} \| z \|_{1}\ \mbox{subject to $\| A z - y \|_2 \leq \eta$},
}
where $y = A x + e$ with $\| e \|_2 \leq \eta$, we have
\be{
\label{stab_robust_recovery}
\| x - \hat{x} \|_{2} \lesssim \| x - P_{\Delta} x \|_{1} + \sqrt{s} \eta,
}
with probability at least $1-\epsilon$, provided
\bes{
m \gtrsim D \cdot \Gamma(F,\Delta) \cdot L,
}
where
\be{
\label{L_def}
L =  \log(N/\epsilon) + \log(s) \log(s/\epsilon) .
}
}

The proof of this theorem is given in \S \ref{s:proof}.

\rem{
\label{r:logfactor}
Note that the log factor $L$ satisfies the trivial bound $L \lesssim \log(s) \cdot \log(N/\epsilon)$.   Moreover, since $\log(s) \leq \log(N)$ and
\eas{
\log(N /\epsilon) = \log(N) + \log(1/\epsilon) \leq \log(N) + \log(s/\epsilon) \lesssim \log(N) \log(s/\epsilon),
}
we also have the bound $L \lesssim \log(N) \cdot \log(s/\epsilon)$.
}

\rem{
\label{r:BoyerThyRelation}
Theorem \ref{t:abs_recov} is quite similar to the main result of \cite{Boyer&Bigot&Weiss:15arXiv}, although with several improvements.  First, the model proposed in \S \ref{sss:general_setup} is somewhat more general (see Remark \ref{r:BoyerRelation}).  Second, the log factor in \cite{Boyer&Bigot&Weiss:15arXiv} is $\log(s) \cdot \log(N/\epsilon)$, which is an upper bound for $L$ (see Remark \ref{r:logfactor}).  Third, Theorem \ref{t:abs_recov} also provides stability and robustness estimates via \R{stab_robust_recovery}, whereas only exact recovery of sparse vectors was established in \cite{Boyer&Bigot&Weiss:15arXiv}.  Note that both Theorem \ref{t:abs_recov} and the main result in \cite{Boyer&Bigot&Weiss:15arXiv} are local to the signal support $\Delta$; as discussed in \S \ref{ss:sparsity_models}, this is crucial in parallel acquisition.  Besides also \cite{Adcock:15arXiv}, which treats a different sampling model, we are aware of no other results in compressed sensing which give recovery guarantees local to the signal support in this way.
}

\rem{
\label{r:BreakCohRelation}
Theorem \ref{t:abs_recov} is a generalization of the main result proved in \cite{Adcock&etal:16FMS}  (see Remark \ref{r:BreakingCoherenceRelation}).  It also improves this result in several ways.  First, in \cite{Adcock&etal:16FMS} the corresponding log factor is $\log(s / \epsilon) \cdot \log(N)$, which is asymptotically larger than $L$ (see Remark \ref{r:logfactor}).  Second, the error bound in \cite{Adcock&etal:16FMS} is somewhat worse in the noise term than \R{stab_robust_recovery}. 
It is also informative to compare $L$ to the log factors in several earlier works.  In \cite[Thm.\ 1.3, \S \Romnum{1}-D]{Candes&Romberg&Tao:06TIT}, the log factor is given as $C_M^{-1} \cdot \log(N)$ for a failure probability of $\epsilon = N^{-M}$ for $M>0$, where $C_M \asymp (23(M+1))^{-1}$. In \cite[Thm.\ 1.2]{Candes&Plan:11IT}, the log factor is $(1+\beta) \log(N)$ for a failure probability of $1-6/N-6\exp(-\beta)$ for $\beta > 0$, which, after equating terms, results in a log factor of $\log^2(N)$ with failure probability $\epsilon = 12/N$.  Setting $\epsilon = N^{-\zeta}$ for some $\zeta > 0$ gives $L \lesssim ( \zeta + 1) \log(N)$, which is smaller than that of \cite{Candes&Plan:11IT} by a factor of $\log(N)$ and equivalent to that of \cite{Candes&Romberg&Tao:06TIT}.
}

Finally, we note that one downside of the additional flexibility that is gained by not specifying a signal model (e.g.\ sparsity or sparsity in levels).  This is an additional factor of $\sqrt{s}$ in the error bound of Theorem \ref{t:abs_recov} over corresponding uniform recovery bounds obtained via Restricted Isometry Property (RIP)-based analysis (which specifies the signal model) \cite{Foucart&Rauhut:book, Chun&Adcock:16arXiv-CS&PA&RIP}.  We refer to \S \ref{s:conclusion} for further discussion on this topic.

\section{Distinct sampling}
In this section, we focus on the case of distinct sampling.

\subsection{Main results for distinct sampling} \label{s:mainResults}

\cor{[Distinct sampling with sparsity model]
\label{c:distinct_sparsity}
Consider the distribution $F$ defined in \S \ref{sss:general_setup_distinct} and suppose that $x \in \bbC^{N}$, $0 < \epsilon < 1$ and $N \geq s \geq 2$.  Draw $A \in \bbC^{m \times N}$ according to \R{A_F} and let $y = A x + e$ with $\| e \|_2 \leq \eta$.  Then for any minimizer $\hat{x}$ of
\bes{
\min_{z \in \bbC^N} \| z \|_{1}\ \mbox{subject to $\| A z - y \|_2 \leq \eta$},
}
we have
\bes{
\| x - \hat{x} \|_{2} \lesssim \sigma_{s}(x)_1 + \sqrt{s} \eta,
}
with probability at least $1-\epsilon$, provided
\be{
\label{bound:dist_sparse}
m \gtrsim s \cdot \left ( \max_{c=1,\ldots,C} \mu(F_c) \right ) \cdot L,
}	
where $\mu$ is as in \R{standard_coherence}, $F_1,\ldots,F_C$ are as in \S \ref{sss:general_setup_distinct} and $L$ is as in \R{L_def}.
}

\prf{
We apply Theorem \ref{t:abs_recov} in the setting of \S \ref{sss:general_setup_distinct}.  It therefore suffices to show that $\Gamma(F,\Delta) \leq \max_{c=1,\ldots,C} \mu(F_c)$ for any set $\Delta \subseteq \{1,\ldots,N\}$ with $| \Delta | \leq s$.  Let $a_c \sim F_c$.  Then
\bes{
\| a_c a^*_c P_{\Delta} \|_{\infty} \leq \| a_c \|_{\infty} \| P_{\Delta} a_c \|_1 \leq s \| a_c \|^2_{\infty} = s \mu(F_c),
}
and therefore, after noting that $F = F_c$ conditioned on the event $\{ X = c \}$, we find that $\Gamma_{1}(F,\Delta) \leq s \max_{c=1,\ldots,C} \mu(F_c)$.  Now suppose that $a_c \sim F_c$.  Then
\bes{
\bbE | e^*_i a_c a^*_c P_{\Delta} z |^2 = z^* P_{\Delta}^* \bbE ( a_c a_c^* e_i e_i^* a_c  a_c^* ) P_{\Delta} z  \leq \| a_c \|^2_{\infty} ( P_{\Delta} z )^* \bbE(a_c a^*_c) P_{\Delta} z.
}
Hence if $B \sim F$ we get
\bes{
\bbE | e^*_i B B^* P_{\Delta} z |^2 = \sum^{C}_{c=1} \frac{m_c}{m} \bbE | e^*_i a_c a^*_c P_{\Delta} z |^2 \leq \max_{c} \| a_c \|^2_{\infty} (P_{\Delta} z)^* \left ( \sum^{C}_{c=1} \frac{m_c}{m} \bbE \left ( a_c a^*_c \right ) \right ) P_{\Delta} z.
}
Applying \R{joint_iso} we deduce that
\bes{
\bbE | e^*_i B B^* P_{\Delta} z |^2 \leq  \max_{c} \| a_c \|^2_{\infty} \| P_{\Delta} z \|^2_2 \leq s \max_{c} \| a_c \|^2_{\infty} \| z \|^2_{\infty},
}
and therefore $\Gamma_{2}(F,\Delta) \leq s \max_{c} \| a_c \|^2_{\infty} = s \max_{c=1,\ldots,C} \mu(F_c)$.  To complete the proof, we now let $\Delta$ be the index set of the largest $s$ entries of $x$ in absolute value, so that $\| x - P_{\Delta} x \|_{1} = \sigma_{s}(x)_1$.
}

This result is general, yet quite useful for many practical CS applications.  In essence, it shows that if $C$ different sensing mechanisms are combined together then the number of measurements required per sensor decreases linearly in $C$ as $C$ increases, provided each sensor is itself good for CS -- that is, provided each sensor has low coherence ($\mu(F_c) \approx 1$) -- and the combined sensors are jointly isotropic.

We now consider the sparsity in levels model.  For this, we first require the following definition:

\defn{[Local sparsity relative to $\cS$ and $F_1,\ldots,F_C$]
\label{def:rel_local_sparsity}
Let $F_1,\ldots,F_C$ be distributions on $\bbC^N$, $\cI = (I_1,\ldots,I_C)$ a partition of $\{1,\ldots,N\}$ and let $\cS = (s_1,\ldots,s_C) \in \bbN^C$ with $s_c \leq | I_c |$, $c=1,\ldots,C$.  We define the local sparsities $S_1,\ldots,S_C$ relative to $\cS$ and $F_1,\ldots,F_C$ as
\bes{
S_c = \sup \left \{ \bbE | a^*_c z |^2 : \ z \in \bbC^N, \| z \|_{\infty}=1,\ | \supp(z) \cap I_d | \leq s_d,\ d=1,\ldots,C \right \},\qquad c=1,\ldots,C,
}
where the expectation is taken over $a_c \sim F_c$.
}

This definition -- originally introduced in \cite{Adcock&etal:16FMS} -- is a technical construct which arises in the sparsity in levels model.  In essence it measures how localized the $c^{\rth}$ sensor can be.  As discussed in \cite{Adcock&etal:16FMS}, in the worst (i.e.\ non-localized) case it can scale with the total sparsity $s = s_1+\ldots+s_C$.  Conversely, when the sensors are completely localized, i.e.\ $\supp(a_c) = I_c$ for $a_c \sim F_c$, $S_c$ is proportional to only $s_c$.  We refer to \cite{Adcock&etal:16FMS} for a more detailed discussion.

\cor{[Distinct sampling with the sparsity in levels model] 
\label{c:distinct_sparsity_in_levels}
Let $\cI = \{ I_1,\ldots,I_C \}$ be a partition of $\{1,\ldots,N\}$ and $\cS = \{ s_1,\ldots,s_C \} \in \bbN^C$ with $ s_c \leq |I_c |$ for $c=1,\ldots,C$.  Consider the distribution $F$ defined in \S \ref{sss:general_setup_distinct} and suppose that $x \in \bbC^{N}$, $0 < \epsilon < 1$ and $N \geq s = s_1+\ldots + s_C \geq 2$.  Draw $A \in \bbC^{m \times N}$ according to \R{A_F} and let $y = A x + e$, $\| e \|_2 \leq \eta$.  Then for any minimizer $\hat{x}$ of
\bes{
\min_{z \in \bbC^N} \| z \|_{1}\ \mbox{subject to $\| A z - y \|_2 \leq \eta$},
}
we have
\bes{
\| x - \hat{x} \|_{2} \lesssim \sigma_{\cS,\cI}(x)_1 + \sqrt{s} \eta,
}
with probability at least $1-\epsilon$, provided
\be{
\label{distinct_sparsity_in_levels_bound}
m \gtrsim \max \left \{ \max_{c=1,\ldots,C} \left \{ \sum^{C}_{d=1} \mu_{d}(F_c) s_{d} \right \} , \max_{c=1,\ldots,C} \left \{ \sum^{C}_{d=1} \frac{m_{d}}{m} \mu_c(F_{d}) S_{d} \right \} \right \} \cdot L,
}
where $\mu_{1},\ldots,\mu_C$ are as in Definition \ref{def:local_coh_lev}, $F_1,\ldots,F_C$ are as in \S \ref{sss:general_setup_distinct} and $L$ is as in \R{L_def}. 
}

\prf{
Once more we seek to apply Theorem \ref{t:abs_recov}.  Let $\Delta \subseteq \{1,\ldots,N\}$ be such that $| \Delta_c | \leq s_c$ for $c=1,\ldots,C$, where $\Delta_{c} = \Delta \cap I_{c}$.  Suppose that $a_c \sim F_c$.  Then
\bes{
\| a_c a^*_c P_{\Delta} \|_{\infty} \leq \sum^{C}_{d=1} \| a_c a^*_c P_{\Delta_{d}} \|_{\infty} \leq \sum^{C}_{d=1} \| a_c \|_{\infty} \|  P_{\Delta_{d}} a_c \|_{1} \leq \sum^{C}_{d=1} \| a_c \|_{\infty} \|  P_{I_{d}} a_c \|_{\infty} s_d \leq \sum^{C}_{d=1} \mu_{d}(F_c) s_{d},
}
where the last inequality follows from Definition \ref{def:local_coh_lev}.
Therefore
\bes{
\Gamma_1(F,\Delta) \leq \max_{c=1,\ldots,C} \left \{ \sum^{C}_{d=1} \mu_{d}(F_c) s_{d} \right \}.
}
Now let $i \in I_{c}$ for some $c=1,\ldots,C$ and $a_d \sim F_d$.  Then
\bes{
\bbE | e^*_i a_{d} a^*_{d} P_{\Delta} z |^2 = z^* P_{\Delta}^*  \bbE ( a_{d}  a_{d}^* e_i e_i^* a_{d} a_{d}^* ) P_{\Delta} z  \leq \| P_{I_c} a_{d} \|^2_{\infty} ( P_{\Delta} z )^* \bbE (a_{d} a^*_{d}) P_{\Delta} z.
}
Hence if $B \sim F$ we have
\bes{
\bbE | e^*_i B B^* P_{\Delta} z |^2 = \sum^{C}_{d=1} \frac{m_{d}}{m} \bbE | e^*_i a_{d} a^*_{d} P_{\Delta} z |^2 \leq \sum^{C}_{d=1} \frac{m_{d}}{m} \|  P_{I_c} a_{d} \|^2_{\infty} \bbE | a^*_{d} P_{\Delta} z |^2 \leq \sum^{C}_{d=1} \frac{m_{d}}{m} \mu_c(F_{d}) S_{d},
}
where the last inequality follows from Definitions in \ref{def:local_coh_lev} and \ref{def:rel_local_sparsity}, and the fact that $\mu'_c(F_d) \leq \mu_c(F_d)$.  Therefore
\bes{
\Gamma_{2}(F,\Delta) \leq \max_{c=1,\ldots,C} \left \{ \sum^{C}_{d=1} \frac{m_{d}}{m} \mu_c(F_{d}) S_{d} \right \}.
}
To complete the proof, we now let $\Delta_{c}$ be the index set of the largest $s_c$ entries of $x$ restricted to $I_c$ for $c=1,\ldots,C$, so that $\| x - P_{\Delta} x \|_{1} = \sigma_{\cS,\cI}(x)_1$.
}

Importantly, the bound \R{distinct_sparsity_in_levels_bound} depends on local sparsities and coherences, rather than the global quantities $s$ and $\max_{c} \mu(F_c)$ appearing in \R{bound:dist_sparse}.  In particular, the first term expresses that the components of the sensing vectors of the $c^{\rth}$ sensor corresponding to the interval $I_d$ should be reasonably small in relation to the local sparsity $s_d$.  Whereas the second term expresses that the $c^{\rth}$ components of the sensing vectors of the $d^{\rth}$ sensor should not be too large in relation to the relative sparsity $S_d$.  Later, in Corollary \ref{c:distinct_diag_levels}, we will use these expression to obtain explicit bounds for diagonal sensor profiles $\{ H_c \}$.

\subsection{The case of diagonal sensor profile matrices} \label{sss:DistinctSamp_diagCh}

We now consider the case where
\be{
\label{distinct_diag}
A = \left [ \begin{array}{c} A_1 \\ \vdots \\ A_C \end{array} \right ] =\left [ \begin{array}{c} \tilde{A}_1 H_1 \\ \vdots \\ \tilde{A}_C H_C \end{array} \right ],
}
and the $H_c \in \bbC^{N \times N}$ are diagonal sensor profile matrices.  We assume the matrices $\tilde{A}_1,\ldots,\tilde{A}_C$ are drawn independently from (possibly different) isotropic distributions $G_c$ on $\bbC^N$, and for simplicity we shall assume that $m_c = m/C$, for $c=1,\ldots,C$ (see Remark \ref{r:mc_distinct}).  We shall assume these distributions are incoherent and use the notation
\be{
\label{mu_G}
\mu_{G} = \max_{c=1,\ldots,C} \mu(G_c).
}
To view this in the setup of \S \ref{sss:general_setup_distinct} notice that this means that the rows of $A_c$ are drawn from a distribution $F_c$ where $a_c \sim F_c$ if $a_c = H^*_c \tilde{a}_c$ for $\tilde{a}_c \sim G_c$.  Note that the joint isotropic property \R{joint_iso} is then equivalent to
\bes{
I = \sum^{C}_{c=1} \frac{m/C}{m} \bbE(a_c a^*_c) = \frac{1}{C} \sum^{C}_{c=1}  H^*_c \bbE(\tilde{a}_c \tilde{a}^*_c) H_c = \frac{1}{C} \sum^{C}_{c=1} H^*_c H_c,
}
which is referred to as the joint isometry condition for the distinct sampling scenario.

\cor{[Distinct sampling with sparsity model and diagonal sensor profiles]
\label{cor:DinstictSamp_diagCh}
Let $x \in \bbC^{N}$, $0 < \epsilon < 1$, $N \geq s \geq 2$ and suppose that $H_{c} \in \bbC^{N \times N}$, $c=1,\ldots,C$, are diagonal matrices satisfying
\be{
\label{channel_iso}
\frac{1}{C} \sum^{C}_{c=1} H^*_c H_c = I.
}
Let $G_1,\ldots,G_C$ be isotropic distributions on $\bbC^N$ and for $c=1,\ldots,C$ define $F_c$ so that $a_c \sim F_c$ if $a_c = H^*_c \tilde{a}_c$ for $\tilde{a}_c \sim G_c$.  Set $m_1 = \ldots = m_C = m/C$ and let $F$ be as in \S \ref{sss:general_setup_distinct} for this choice of $F_1,\ldots,F_C$.  Draw $A$ according to \R{A_F} and let $y = A x + e$, $\| e \|_{2} \leq \eta$.  Then for any minimizer $\hat{x}$ of
\bes{
\min_{z \in \bbC^N} \| z \|_{1}\ \mbox{subject to $\| A z - y \|_2 \leq \eta$},
}
we have
\bes{
\| x - \hat{x} \|_{2} \lesssim \sigma_{s}(x)_1 + \sqrt{s} \eta,
}
with probability at least $1-\epsilon$, provided
\be{
\label{diag_distinct_meas_cond}
m \gtrsim s \cdot \mu_{G} \cdot \left ( \max_{c=1,\ldots,C} \| H_c \|^2_{\infty} \right ) \cdot L,
}
where $\mu_{G}$ is as in \R{mu_G} and $L$ is as in \R{L_def}.
}

\prf{
We shall apply Corollary \ref{c:distinct_sparsity}.  By construction, $\mu(F_c)$ is the smallest constant such that
\bes{
\| H^*_c \tilde{a}_c \|^2_{\infty} \leq \mu(F_c),\qquad \tilde{a}_c \sim G_c.
}
Since $H_c$ is diagonal, we have $\| H^*_c \tilde{a}_c \|^2_{\infty} \leq \| H_c \|^2_{\infty} \mu(G_c)$ and the result now follows.
}

\cor{[Distinct sampling with the sparsity in levels model and diagonal sensor profiles]
\label{c:distinct_diag_levels}
Let $\cI = \{ I_1,\ldots,I_C\}$ be a partition of $\{1,\ldots,N\}$ and $\cS = \{ s_1,\ldots,s_C \} \in \bbN^C$ with $ s_c \leq |I_c |$ for $c=1,\ldots,C$.
Let $x \in \bbC^{N}$, $0 < \epsilon < 1$, $N \geq s = s_1+\ldots + s_C \geq 2$ and suppose that $H_{c} \in \bbC^{N \times N}$, $c=1,\ldots,C$, are diagonal matrices satisfying \R{channel_iso}. 
Let $A \in \bbC^{m \times N}$ be constructed as in Corollary \ref{cor:DinstictSamp_diagCh} and set $y = A x + e$, $\| e \|_{2} \leq \eta$.  Then for any minimizer $\hat{x}$ of
\bes{
\min_{z \in \bbC^N} \| z \|_{1}\ \mbox{subject to $\| A z - y \|_2 \leq \eta$},
}
we have
\bes{
\| x - \hat{x} \|_{2} \lesssim \sigma_{\cS,\cI}(x)_1 + \sqrt{s} \eta,
}
with probability at least $1-\epsilon$, provided
\be{
\label{levels_diag_cond_1}
C \gtrsim \max_{c=1,\ldots,C} \left \{  \sum^{C}_{d=1} \| H_d \|_{\infty} \| H_d P_{I_c} \|_{\infty} \right \},
}
and 
\be{
\label{levels_diag_cond_2}
m \gtrsim \mu_{G} \cdot \max_{c=1,\ldots,C} \left \{ \sum^{C}_{d=1} \| H_c \|_{\infty} \| H_c P_{I_d} \|_{\infty} s_d \right \} \cdot L.
}
where $\mu_{G}$ is as in \R{mu_G} and $L$ is as in \R{L_def}.
}

\prf{
Using Corollary \ref{c:distinct_sparsity_in_levels}, it suffices to estimate the local coherences $\mu_d(F_c)$ and the relative sparsities $S_c$.  Note that $\mu'_c(F_d) \leq \| H_d P_{I_c}  \|^2_{\infty} \mu(G_d)$ and therefore
\bes{
\mu_c(F_d) \leq \| H_d \|_{\infty} \| H_d P_{I_c} \|_{\infty} \mu(G_d).
}
Hence we obtain
\bes{
\max_{c=1,\ldots,C} \left \{ \sum^{C}_{d=1} \mu_d(F_c) s_d \right \}  \leq \mu_{G} \max_{c=1,\ldots,C} \left \{ \sum^{C}_{d=1} \| H_c \|_{\infty} \| H_c P_{I_d} \|_{\infty} s_d \right \} \leq \mu_G \sigma,
}
where $\sigma = \max_{c=1,\ldots,C} \left \{ \sum^{C}_{d=1} \| H_c \|_{\infty} \| H_c P_{I_d} \|_{\infty} s_d \right \}$.

If $a_c \sim F_c$, i.e.\ $a_c = H^*_c \tilde{a}_c$ where $\tilde{a}_c \sim G_c$, and $z \in \bbC^N$ is such that $\|z \|_{\infty} =1$ and $| \supp(z) \cap I_d | \leq s_d$, $d=1,\ldots,C$, then
\bes{
\bbE | a^*_c z |^2 = \bbE | \tilde{a}^*_c H_c z |^2 = \| H_c z \|^2_2 = \sum^{C}_{d=1} \| H_c P_{I_d} z \|^2_2 \leq \sum^{C}_{d=1} \| H_c P_{I_d} \|^2_{\infty} s_d.
}
It follows that
\bes{
S_c \leq \sum^{C}_{d=1} \| H_c P_{I_d} \|^2_{\infty} s_d  \leq \sum^{C}_{d=1} \| H_c \|_{\infty} \| H_c P_{I_d} \|_{\infty} s_d \leq \sigma
}
Therefore, we have
\bes{
\sum^{C}_{d=1} \frac{m/C}{m} \mu_c(F_d) S_d \leq \sigma \mu_{G} C^{-1} \sum^{C}_{d=1} \| H_d \|_{\infty} \| H_d P_{I_c} \|_{\infty} \lesssim \sigma \mu_G,
}
due to \R{levels_diag_cond_1}.  Therefore  \R{levels_diag_cond_2} implies \R{distinct_sparsity_in_levels_bound}, and hence the result follows from Corollary \ref{c:distinct_sparsity_in_levels}.
}

The condition \R{levels_diag_cond_1} is mainly added for convenience.  Note that it is satisfied in all examples given later.  As seen in the above proof, it could in fact be removed by replacing \R{levels_diag_cond_1}--\R{levels_diag_cond_2} with the single condition
\bes{
m  \gtrsim \mu_{G} \cdot \max_{c=1,\ldots,C} \left \{  C^{-1} \sum^{C}_{d=1} \| H_d \|_{\infty} \| H_d P_{I_c} \|_{\infty}\right \} \cdot \max_{c=1,\ldots,C} \left \{ \sum^{C}_{d=1} \| H_c \|_{\infty} \| H_c P_{I_d} \|_{\infty} s_d \right \} \cdot L.
}
We remark in passing that in recent work \cite[Cor. 3.5]{Chun&Adcock:16ITW} this bound has been improved to give an estimate that is more readily computable.  The bound therein also does not require \R{levels_diag_cond_1}.

\rem{
\label{r:mc_distinct}
For the sake of simplicity we have assumed in this section that all the $m_c$'s are equal, i.e.\ $m_c = m/C$.  However, at the expense of some more complicated estimates, one could readily allow the $m_c$'s to take distinct values.
}

\examp{[Nonoverlapping sensor profiles]
\label{ex:nonoverlapping}
Consider the case of nonoverlapping sensor profile matrices, i.e.\ $H_c \propto P_{I_c}$, where the sets $\cI = \{ I_1,\ldots,I_C\}$ give a partition of $\{1,\ldots,N\}$.  To ensure \R{channel_iso} we require the normalization $H_c = \sqrt{C} P_{I_c}$.  Hence $\| H_c \|^2_{\infty} = C$ and the measurement condition \R{diag_distinct_meas_cond} reduces to 
\bes{
m \gtrsim s \cdot C \cdot \mu_G \cdot \log(2s/\epsilon) \cdot \log(2N).
}
As we expect (recall \S \ref{ss:sparsity_models}) the recovery guarantee scales linearly with $C$ due to the properties of the sensor profile matrices.  Note that the matrix $A$ in this case is block diagonal, i.e.\ of the same form as \R{A_block_diag}.  Now consider sparsity in levels signal model based on the partition $\cI$ with local sparsities $\cS = ( s_1,\ldots,s_C)$.  Note that $\| H_c P_{I_d} \|_{\infty} = \sqrt{C} \delta_{cd}$.  Hence \R{levels_diag_cond_1} holds and therefore \R{levels_diag_cond_2} gives
\bes{
m \gtrsim \mu_{G} \cdot C \cdot \left ( \max_{c=1,\ldots,C} \{ s_c \} \right )\cdot L.
}
In particular, if $x$ is sparse and $\lambda$-equidistributed (see Definition \ref{d:sparse_distrib}), exact recovery of $x$ requires $m \gtrsim \mu_{G} \cdot \lambda \cdot s \cdot L$
measurements.  Thus we obtain an optimal recovery guarantee in the case.
}

\examp{[Almost identical sensor profiles]
\label{eg:dinstinctSamp_identDiagCh}
Suppose that $H_c = \lambda_{c} H$ for some diagonal $H \in \bbC^{N \times N}$ with $H^* H = I$ and $\lambda_{c} \in \bbC$ satisfying $\sum^{C}_{c=1} | \lambda_c |^2 = C$ (note that this example is a special case of the circulant sensor profile setup; see \S \ref{sec:distinctSamp_circCh}).  Then $\| H_c \|^2_{\infty} = | \lambda_c |^2 \| H \|^2_{\infty}$ and, since $H^* H = I$, we have $\| H_c \|^2_{\infty} = | \lambda_c |^2$. Hence the measurement condition \R{diag_distinct_meas_cond} reduces to
\bes{
m \gtrsim s \cdot \left ( \max_{c=1,\ldots,C} | \lambda_c |^2 \right ) \cdot \mu_G\cdot L.
}
This recovery guarantee behaves as expected. It is optimal (i.e.\ independent of $C$) if $| \lambda_c | \lesssim 1$ for all $c=1,\ldots,C$.  At the other extreme, if $\lambda_{c} = 0$ for $c=2,\ldots,C$ so that $\lambda_1 = \sqrt{C}$ (that is, the measurements in all the channels besides the first are zero) then the recovery guarantee grows linearly in $C$.
}

\examp{[Complex sensor profiles] \label{eg:distSamp_diagComplex}
Suppose that the $H_c$ are diagonal with unit complex entries, i.e., $H^*_c H_c = I$.  Then $\| H_c \|^2_{\infty} = 1$ and \R{diag_distinct_meas_cond} reduces to the optimal condition $m \gtrsim s \cdot \mu_G \cdot L$.  This is to be expected, since no information is lost about the support of $x$ in the operation $x \mapsto H_c x$.
}

\begin{figure}
\begin{center}
$\begin{array}{cccc}
\includegraphics[width=3.4cm]{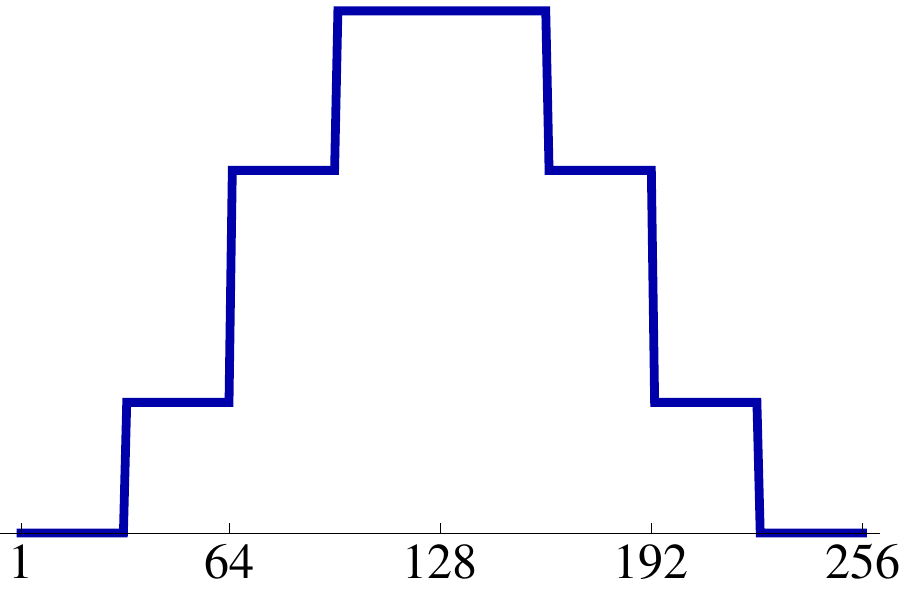} \vspace{0.75pt} &
\includegraphics[width=3.4cm]{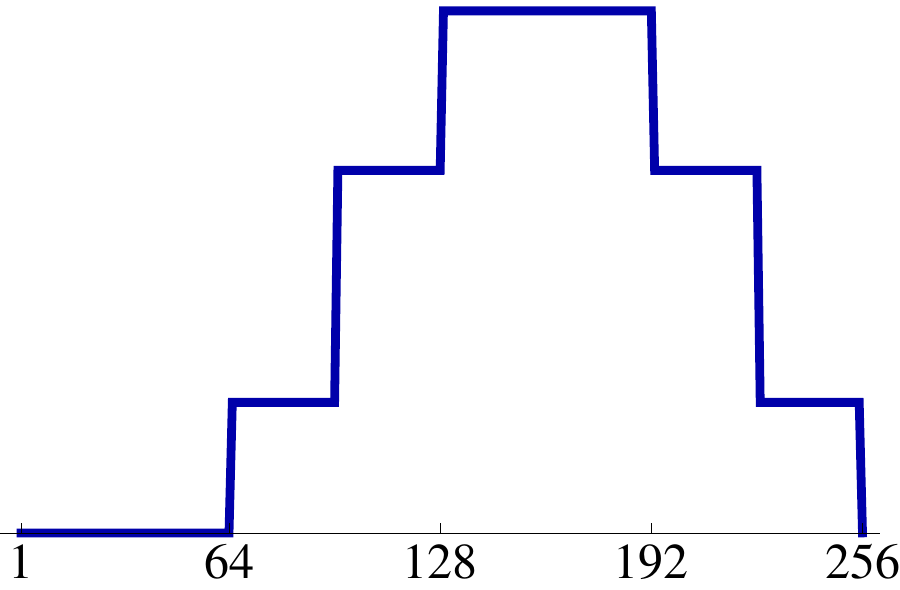} \vspace{0.75pt}&
\includegraphics[width=3.4cm]{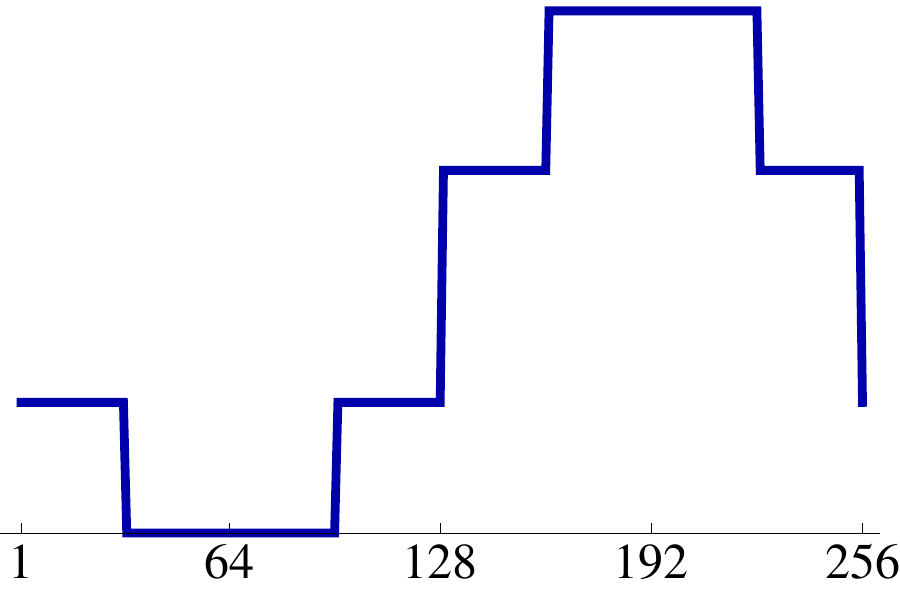} \vspace{0.75pt} &
\includegraphics[width=3.4cm]{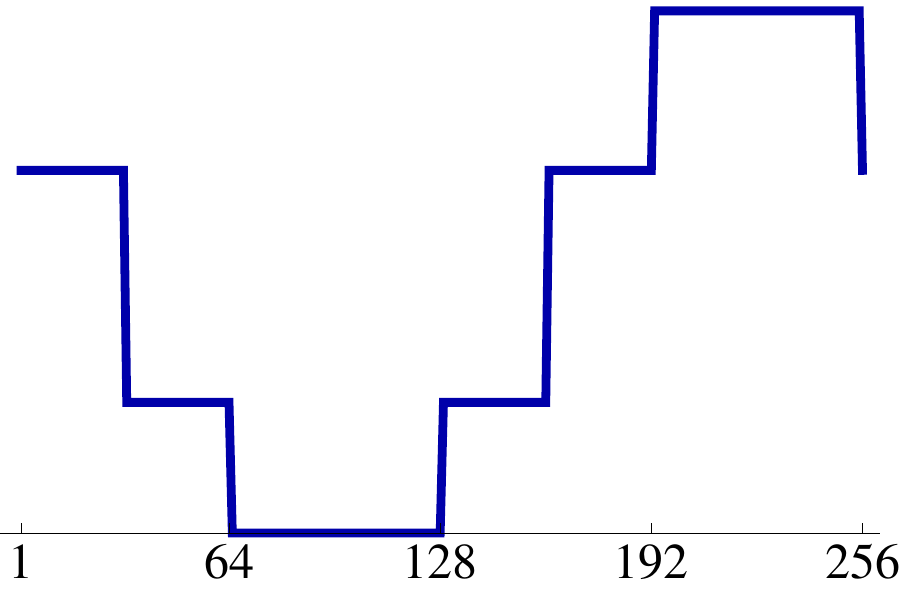} \vspace{0.75pt} \\ 
\includegraphics[width=3.4cm]{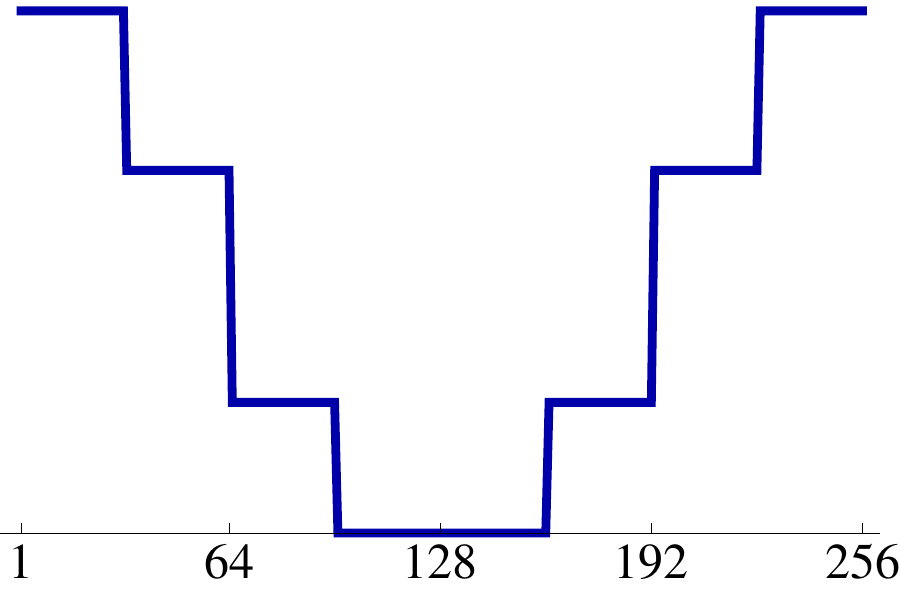} &
\includegraphics[width=3.4cm]{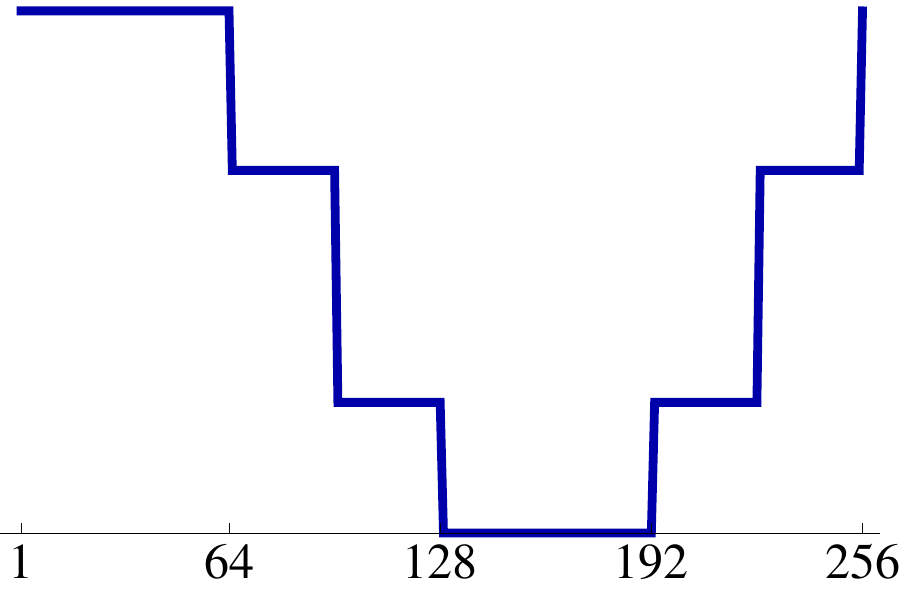} &
\includegraphics[width=3.4cm]{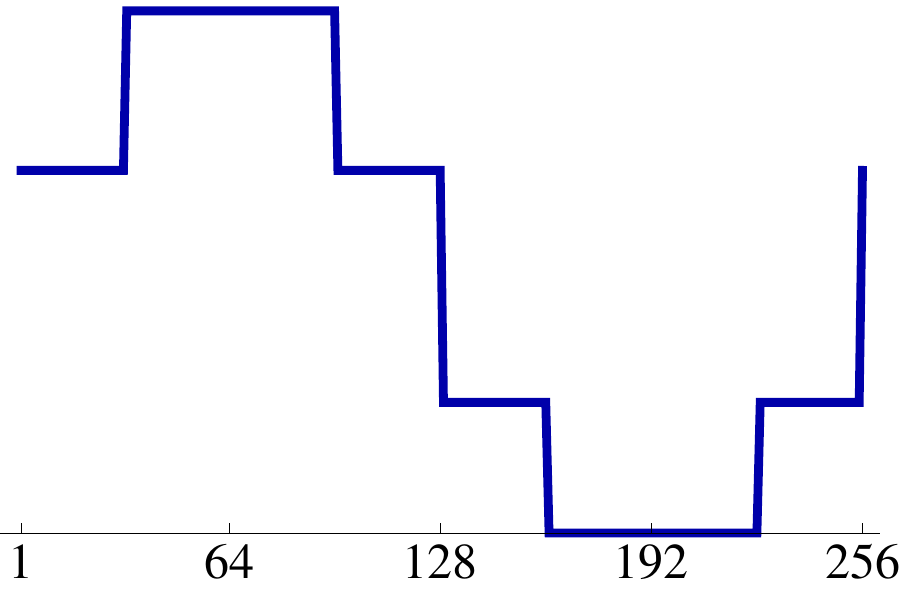} &
\includegraphics[width=3.4cm]{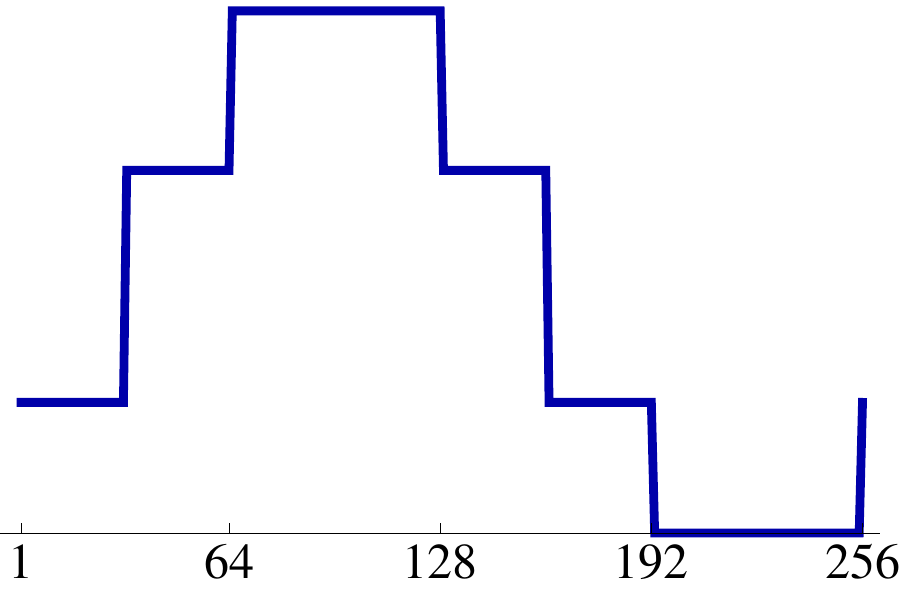} 
\end{array}$
\end{center}
\caption{An example of piecewise constant sensor profiles ($C=8$).}
\label{fig:prof:piecewise}
\end{figure}

\examp{[Piecewise constant sensor profiles I -- sparsity model]
\label{ex:piecewise_const}
Let $\cI = \{ I_1,\ldots,I_C\}$ be a partition of $\{1,\ldots,N\}$ and suppose that $V = \{ V_{c,d} \}^{C}_{c,d=1} \in \bbC^{C \times C}$ is any matrix with $\ell^2$-normalized columns, i.e.\ $\sum^{C}_{c=1} |V_{c,d} |^2 = 1$; see, for example, Fig.~\ref{fig:prof:piecewise}.  Define the sensor profile matrices 
\bes{
H_{c} = \sqrt{C} \sum^{C}_{d=1} V_{c,d} P_{I_d},
}
and observe that
\bes{
C^{-1} \sum^{C}_{c=1} H^*_c H_c = \sum^{C}_{c=1} \sum^{C}_{d=1} |V_{c,d} |^2 P_{I_d} =  \sum^{C}_{d=1} P_{I_d} = I,
}
and therefore \R{channel_iso} holds.  Note that
\bes{
\| H_c \|_{\infty} = \sqrt{C} \max_{d=1,\ldots,C} | V_{c,d} |,\qquad \| H_c P_{I_d} \|_{\infty} = \sqrt{C} | V_{c,d} |.
}
Therefore for the sparsity model, the recovery guarantee \R{diag_distinct_meas_cond} reduces to
\bes{
m \gtrsim s \cdot \mu_{G} \cdot C \cdot \mu(V) \cdot L,
}
where $\mu(V) = \max_{c,d=1,\ldots,C} | V_{c,d} |^2$ is referred to as the coherence of the matrix $V$\footnote{This is not to be confused with the other standard notion of coherence in CS literature \cite[Chpt.\ 5]{Foucart&Rauhut:book}.}.  If $V$ is incoherent, i.e.\ $\mu(V) \lesssim C^{-1}$, we obtain an optimal recovery guarantee.  We note in passing that although $V$ need not be an isometry (it only is required to have normalized columns), its coherence $\mu(V)$ still satisfies the usual bound $C^{-1} \leq \mu(V) \leq 1$.  The conclusion that $V$ should be incoherent for the sparsity model in order to get an optimal recovery guarantee is consistent with that of Example \ref{ex:nonoverlapping}.  This example is a special case of the current piecewise constant setup corresponding to the coherent matrix $V = I$.
}

\examp{[Piecewise constant sensor profiles II -- sparsity in levels model]
\label{ex:piecewise_const_II}
Consider the setup of the previous example with the sparsity in levels model.  In light of the previous comment, it may seem reasonable to expect this model to give better recover guarantees when $V$ is coherent.  Note that \R{levels_diag_cond_1} and \R{levels_diag_cond_2} in this case are equivalent to
\eas{
1 &\gtrsim \max_{c=1,\ldots,C} \left \{  \sum^{C}_{d=1} \max_{e=1,\ldots,C} | V_{d,e} | | V_{d,c} | \right \},
\\
m &\gtrsim \mu_{G} \cdot C \cdot \max_{c=1,\ldots,C} \left \{ \sum^{C}_{d=1} \max_{e=1,\ldots,C} | V_{c,e} | | V_{c,d} | s_d \right \} \cdot L.
}
These conditions do not require $V$ to be incoherent.  Clearly when $V=I$ we get the same conditions as in Example \ref{ex:nonoverlapping} for sparse and distributed vectors.  However, a more interesting instance of this setup occurs when $V$ is a circulant matrix with filter vector $w = (w_0,\ldots,w_{C-1}) \in \bbC^C$, i.e. $V_{c,d} = w_{c-d}$ taken modulo $C$.  Note that $\| w \|_2 = 1$ due to the $\ell^2$-normalization of the columns of $V$.  In this case, the above conditions are equivalent to
\eas{
1 &\gtrsim \| w \|_{\infty} \| w \|_{1},
\\
m &\gtrsim \mu_G \cdot C \cdot \| w \|_{\infty} \cdot \max_{c=1,\ldots,C} \left \{ \sum^{C}_{d=1} | w_{c-d} | s_d \right \} \cdot L.
}
In particular, if $x$ is sparse and $\lambda$-equidistributed, i.e. $s_d \leq \lambda s / C$, $d=1,\ldots,C$, then it suffices that
\eas{
1 \gtrsim \| w \|_{\infty} \| w \|_{1}, \qquad m \gtrsim \mu_{G}  \cdot \lambda \cdot s \cdot L.
}
Note that $V$ does not need to be incoherent for the condition $1 \gtrsim \| w \|_{\infty} \| w \|_{1}$ to hold.  A trivial example if the case $w = (1,0,\ldots,0)$ (which corresponds to the nonoverlapping sensor profile mentioned above).  A more interesting example is a simple banded interaction model $w = (1/\sqrt{2},1/2,0,\ldots,0,1/2)$, which corresponds to a sensor profile model where each sensor interacts with its two nearest neighbours but no others.
}

\subsection{The case of circulant sensor profile matrices}
\label{sec:distinctSamp_circCh}

We now consider the case where
\be{
\label{distinct_circ}
A = \left [ \begin{array}{c} A_1 \\ \vdots \\ A_C \end{array} \right ] =\left [ \begin{array}{c} \tilde{A}_1 H_1 \\ \vdots \\ \tilde{A}_C H_C \end{array} \right ],
}
and for each $c=1,\ldots,C$, $H_c \in \bbC^{N \times N}$ is circulant matrix with filter vector $h_{c} \in \bbC^{N}$.  Note that large classes of signal processing operators (e.g., filtering and convolution) can be represented by circulant matrices.  The matrices $\tilde{A}_1,\ldots,\tilde{A}_C$ will be as in \S \ref{sss:DistinctSamp_diagCh}.  Specifically, $\tilde{A}_1,\ldots,\tilde{A}_C$ are drawn independently from (possibly different) isotropic distributions $G_c$ on $\bbC^N$, and we assume that $m_c = m/C$, for $c=1,\ldots,C$ (see Remark \ref{r:mc_distinct}).  We shall assume these distributions are incoherent and use the notation $\mu_{G} = \max_{c=1,\ldots,C} \mu(G_c)$.  Note that the joint isotropic property \R{joint_iso} is now equivalent to $C^{-1} \sum^{C}_{c=1} H^*_c H_c = I$, which is referred to as the joint isometry condition for identical sampling scenario.

\subsubsection{Bounds based on the filter vectors $h_c$}

Our first result obtains a recovery guarantee in terms of the filter vectors $h_c$:

\cor{[Distinct sampling with sparsity model and circulant profile matrices] 
\label{cor:distinctSamp_circCh}
Let $x \in \bbC^{N}$, $0 < \epsilon < 1$, $N \geq s \geq 2$ and suppose that $H_{c} \in \bbC^{N \times N}$, $c=1,\ldots,C$, are circulant matrices satisfying
\be{
\label{eq:ChIso_DinstictSamp_circCh}
\frac{1}{C} \sum^{C}_{c=1} H^*_c H_c = I.
}
Let $A \in \bbC^{m \times N}$ be as in \R{distinct_circ}, where the matrices $\tilde{A}_1,\ldots,\tilde{A}_C \in \bbC^{m/C \times N}$ are drawn independently from isotropic distributions $G_1,\ldots,G_C$.  If $y = A x + e$ with $\| e \|_{2} \leq \eta$, then for any minimizer $\hat{x}$ of
\bes{
\min_{z \in \bbC^N} \| z \|_{1}\ \mbox{subject to $\| A z - y \|_2 \leq \eta$},
}
we have
\bes{
\| x - \hat{x} \|_{2} \lesssim \sigma_{s}(x)_1 + \sqrt{s} \eta,
}
with probability at least $1-\epsilon$, provided
\be{
\label{ref_this}
m \gtrsim s \cdot \mu_G \cdot \left( \max_{c=1,\ldots,C}  \| h_c \|_1^2 \right) \cdot L,
}
where $h_{c} \in \bbC^{N}$ is a filter vector of the circulant matrix $H_c$ for $c=1,\ldots,C$, $\mu_{G}$ is as in \R{mu_G}, and $L$ is as in \R{L_def}.
}

\prf{
We shall apply Corollary \ref{c:distinct_sparsity}.  We have $\|  H^*_c \tilde{a}_c \|_{\infty} \leq \| H^*_c \|_{\infty} \| \tilde{a}_c \|_{\infty} \leq \| h_c \|_1 \| \tilde{a}_c \|_{\infty}$.  Therefore $\mu(F_c) \leq \| h_c \|^2_1 \mu(G_c)$,
and using Corollary \ref{c:distinct_sparsity}, the result follows.
}
This theorem implies that an optimal recovery guarantee is possible, provided the filter vectors $h_c$ obey $\| h_c \|_1 \lesssim 1$ for all $c=1,\ldots,C$.  
An interesting scenario is when the filter vectors are nonnegative:

\examp{[Nonnegative filter vectors]\label{ex:nonneg_symbol}
Suppose that each filter vector $h_{c}$ has nonnegative entries (or more generally, all its entries have the same sign).  If $z = (1,\ldots,1)^{\top}$ then \R{eq:ChIso_DinstictSamp_circCh} implies that
\bes{
\| z \|^2_{2} = C^{-1} \sum^{C}_{c=1} \| H_c z \|^2_{2} = C^{-1} \sum^{C}_{c=1} \| h_c \|^2_{1} \| z \|^2_{2},
}
and therefore $\sum^{C}_{c=1} \| h_c \|^2_1 = C$.  In particular, we have the sharp bound
\bes{
\max_{c=1,\ldots,C} \| h_c \|^2_{1} \leq C,
}
and therefore a worst-case recovery guarantee of the form $m \gtrsim s \cdot C \cdot \mu_{G} \cdot L$.  To see this is sharp we may set $h_{1} = \sqrt{C} (1,0,0,\ldots,0)^{\top}$ and $h_{c} = 0$ otherwise.  Note that this means that $H_{1} = \sqrt{C} I$ and $H_{2} = \cdots = H_C = 0$.  In other words, only one sensor gives nonzero measurements.

On the other hand, it is perfectly possible to find filter vectors $h_{c}$ for which $\max_{c=1,\ldots,C} \| h_c \|^2_{1} \lesssim 1$.  A straightforward case is when $h_{c} = \lambda_{c} (1,0,\ldots,0)^{\top}$ for $\lambda_{c} \in \bbC$, in which case $\| h_c \|^2_1 = | \lambda_c |^2$.  This corresponds to $H_c = \lambda_c I$, which is exactly the model considered Example \ref{eg:dinstinctSamp_identDiagCh}.  Hence, if we pick $| \lambda_c | \approx 1$ then we obtain an optimal recovery guarantee.
Of more practical relevance, we note that a power conservation rule (e.g., $0~\textmd{dB}$ gain at $0~\textmd{Hz}$) in finite impulse response filter design corresponds exactly to the condition $\| h_c \|_1 = 1$ for $c = 1,\ldots,C$.
}

\subsubsection{Bounds based on the spectral decomposition}\label{sss:circulant_spectral}

We now give an alternative bound based on the spectral decomposition $H_c = \Phi^* \Lambda_c \Phi$, where $\Phi \in \bbC^{N \times N}$ is the unitary DFT matrix and $\Lambda_c$ is the diagonal matrix of eigenvalues of $H_c$.

\cor{[Distinct sampling with sparsity model and circulant profile matrices] 
\label{cor:distinctSamp_circCh_spect}
Let $x \in \bbC^{N}$, $0 < \epsilon < 1$, $N \geq s \geq 2$ and suppose that $H_{c} \in \bbC^{N \times N}$, $c=1,\ldots,C$, are circulant matrices satisfying
\be{
\label{eq:ChIso_DinstictSamp_circCh_Spectral}
\frac{1}{C} \sum^{C}_{c=1} \Lambda^*_c \Lambda_c = I,
}
where $\Lambda_c$ is the diagonal of eigenvalues of $H_c$ for $c=1,\ldots,C$.  Let $A \in \bbC^{m \times N}$ be as in \R{distinct_circ}, where the matrices $\tilde{A}_1,\ldots,\tilde{A}_C \in \bbC^{m/C \times N}$ are drawn independently from isotropic distributions $G_1,\ldots,G_C$.  If $y = A x + e$ with $\| e \|_{2} \leq \eta$, then for any minimizer $\hat{x}$ of
\bes{
\min_{z \in \bbC^N} \| z \|_{1}\ \mbox{subject to $\| A z - y \|_2 \leq \eta$},
}
we have
\bes{
\| x - \hat{x} \|_{2} \lesssim \sigma_{s}(x)_1 + \sqrt{s} \eta,
}
with probability at least $1-\epsilon$, provided
\be{
\label{circulant_distinct_spectral}
m \gtrsim s \cdot \left ( \max_{c}\| \Lambda_{c} \|^2_{\infty} \cdot \sigma(G_c) \right ) \cdot L,
}
where $L$ is as in \R{L_def} and $\sigma(G_c)$ is the smallest constant such that
\bes{
N^{-1} \| \Phi \tilde{a}_c \|^2_{1} \leq \sigma(G_c),
}
almost surely for $\tilde{a}_c \sim G_c$.
}

\prf{
Since $H_c = \Phi^* \Lambda_c \Phi$ and $\Phi$ is unitary, the condition $C^{-1} \sum H^*_c H_c = I$ is equivalent to \R{eq:ChIso_DinstictSamp_circCh_Spectral}.  Let $\tilde{a}_c \sim G_c$.  Then
\eas{
\| H_c \tilde{a}_c \|_{\infty} &= \max_{n=1,\ldots,N} | e^*_n \Phi^* \Lambda_{c} \Phi \tilde{a}_c | 
= \max_{n} \left | \sum^{N}_{m=1} e^*_n \Phi^* e_m (\Lambda_c)_{m,m} e^*_m \Phi \tilde{a}_c \right |
 \leq \| \Lambda_{c} \|_{\infty} \frac{1}{\sqrt{N}}  \sum_m | e^*_m \Phi \tilde{a}_c |.
}
Thus $\| H_c \tilde{a}_c \|^2_{\infty} \leq \| \Lambda_{c} \|^2_{\infty} \frac{1}{N} \| \Phi \tilde{a}_c \|^2_{1} \leq \| \Lambda_{c} \|^2_{\infty} \sigma(G_c)$.  We now apply Corollary \ref{c:distinct_sparsity}.
}

At the expense of the additional terms $\sigma(G_c)$ -- which measures the $\ell_1$-norm of the Fourier transforms of the sampling vectors $\tilde{a}_c$ -- this result gives a simpler estimate \R{circulant_distinct_spectral} for circulant sensor profile matrices in terms of their eigenvalues than \R{ref_this} which is based on their filter vectors.  As the following example shows, it is straightforward to devise nontrivial instances where \R{circulant_distinct_spectral} yields an optimal recovery guarantee:

\examp{[Fourier sensing with unit complex eigenvalues] \label{e:distF_circ_spectD}
Let the $H_c$ be any circulant matrices with unit complex eigenvalues, i.e.\ $\Lambda^*_{c} \Lambda_c = I$.  Then \R{eq:ChIso_DinstictSamp_circCh_Spectral} holds and we also have $\| \Lambda_{c} \|_{\infty} =1$.  Hence \R{circulant_distinct_spectral} yields an optimal recovery guarantee provided $\max_{c=1,\ldots,C}\sigma(G_c) \lesssim 1$.  This condition depends on the sampling distributions $G_1,\ldots,G_C$ and it is straightforward to come up with instances where it holds.  For example, suppose that each distribution $G_{c}$ samples uniformly from the columns of $\sqrt{N} \Phi^*$ (the factor $\sqrt{N}$ ensures that $G_c$ is isotropic). In other words, $\tilde{a}_c = \sqrt{N} \Phi^* e_{n}$, where $n$ is drawn uniformly at random from $\{1,\ldots,N\}$.  Then $N^{-1} \| \Phi \tilde{a}_c \|^2_{1} = \| e_n \|_{1} = 1$.  Hence $\sigma(G_c) = 1$, yielding an optimal recovery guarantee.
}

\section{Identical sampling}
As one might expect, our results for the identical sampling case are rather weaker than those for the distinct sampling case, and more sensitive to the choice of sensor profile matrices $H_c$.  We first present several worst-case guarantees which show that the required number of measurements for the sparsity model is a most linear in $C$.  Next, for the sparse and distributed model we construct a large family of nontrivial (diagonal) sensor profile matrices for which optimal recovery guarantees are possible.

\subsection{Worst-case bounds}\label{ss:worst_case}

We first provide a general worst-case bound for arbitrary (not necessarily diagonal or circulant) sensor profile matrices.  Recall that the sensor profile matrices $H_c \in \bbC^{N \times N}$ must satisfy \R{Hc_iso}, i.e.\ $\sum_{c=1}^C H_c^* H_c = I$.  If $G$ is the distribution defined in \S \ref{sss:general_setup_identical} we now also define the quantity $\mu(G,H_1,\ldots,H_C) $ to be the smallest number such that
\be{
\label{G_rel_H_c_coherence}
\max_{i,j=1,\ldots,N} \left | \sum^{C}_{c=1} e^*_i H^*_c a a^* H_c e_j \right | \leq \mu(G,H_1,\ldots,H_C),\qquad a \sim G.
}
We then have the following:

\cor{[Identical sampling with sparsity model and nondiagonal profile matrices] \label{cor:identSampl_nonDiag}
Let $x \in \bbC^{N}$, $0 < \epsilon < 1$, $N \geq s \geq 2$ and suppose that $H_{c} \in \bbC^{N \times N}$, $c=1,\ldots,C$, satisfy
\be{
\label{eq:Hc_nonDiag_iso}
\sum^{C}_{c=1} H^*_c H_c = I.
}
Let $F$ be defined as in \S \ref{sss:general_setup_identical} and draw $A$ according to \R{A_F}.  If $y = A x + e$ with $\| e \|_{2} \leq \eta$, then for any minimizer $\hat{x}$ of
\bes{
\min_{z \in \bbC^N} \| z \|_{1}\ \mbox{subject to $\| A z - y \|_2 \leq \eta$},
}
we have
\bes{
\| x - \hat{x} \|_{2} \lesssim \sigma_{s}(x)_1 + \sqrt{s} \eta,
}
with probability at least $1-\epsilon$, provided
\be{
m \gtrsim s \cdot C \cdot \mu(G,H_1,\ldots,H_C) \cdot L,
}
where $\mu(G,H_1,\ldots,H_C)$ is as in \R{G_rel_H_c_coherence} and $L$ is as in \R{L_def}.
}

\prf{
Let $B = [H_1^* a | \cdots | H_C^* a] \sim F$.  If $\Delta \subseteq \{1,\ldots,N\}$ is the set of the largest $s$ entries of $x$ in absolute value, then
\eas{
\| B B^* P_\Delta \|_{\infty} 
& = \max_{\substack{z \in \bbC^N \\ \| z \|_{\infty} = 1}} \max_{i=1,\ldots,C} \left | \sum_{j \in \Delta} z_j \sum_{c=1}^C e^*_i H_c^* a a^* H_c e_j \right |
\leq  \max_{\substack{z \in \bbC^N \\ \| z \|_{\infty} = 1}} \sum_{j \in \Delta} |z_j| \max_{i,j=1,\ldots,N} \left |  \sum_{c=1}^C e^*_i H_c^* a a^* H_c e_j \right |,
}
and therefore $\Gamma_{1}(F,\Delta) \leq s \mu(G,H_1,\ldots,H_C)$.  Also, if $z \in \bbC^N$, $\| z \|_{\infty} = 1$ then
\eas{
\bbE | e_i^* B B^* P_{\Delta} z |^2 
&= \bbE \left | \sum^{C}_{c=1} e^*_i H_c^* a a^* H_c P_{\Delta} z \right |^2
\\
& \leq \sum^{C}_{c=1} | e^*_i H^*_c a |^2 \sum^{C}_{c=1} \bbE | a^* H_c P_{\Delta} z |^2
\\
& \leq \mu(G,H_1,\ldots,H_C) \sum^{C}_{c=1} \| H_c P_\Delta z \|^2_2
\\
& = \mu(G,H_1,\ldots,H_C) \| P_\Delta z \|^2_2,
}
and therefore $\Gamma_{2}(F,\Delta) \leq s \mu(G,H_1,\ldots,H_C)$.   Applying Theorem \ref{t:abs_recov} the results follows.
}

\subsubsection{Diagonal sensor profile matrices}\label{sec:worst_identSamp_diag}
We now consider the case where the sensor profile matrices $H_c$ are diagonal:

\cor{[Identical sampling with sparsity model and diagonal profile matrices]
\label{cor:identSampl_sparsity}
Let $x \in \bbC^{N}$, $0 < \epsilon < 1$, $N \geq s \geq 2$ and suppose that $H_{c} \in \bbC^{N \times N}$, $c=1,\ldots,C$, are diagonal matrices satisfying \R{Hc_iso}, i.e.\ $\sum^{C}_{c=1} H^*_c H_c = I$.  Let $F$ be defined as in \S \ref{sss:general_setup_identical} and draw $A$ according to \R{A_F}.  If $y = A x + e$ with $\| e \|_{2} \leq \eta$, then for any minimizer $\hat{x}$ of
\bes{
\min_{z \in \bbC^N} \| z \|_{1}\ \mbox{subject to $\| A z - y \|_2 \leq \eta$},
}
we have
\bes{
\| x - \hat{x} \|_{2} \lesssim \sigma_{s}(x)_1 + \sqrt{s} \eta,
}
with probability at least $1-\epsilon$, provided
\be{
\label{bound:identSampl_sparsity}
m \gtrsim s \cdot C \cdot \mu(G) \cdot  L,
}
where $\mu(G)$ is as in \R{standard_coherence} and $L$ is as in \R{L_def}.
}

\prf{
By Corollary \ref{cor:identSampl_nonDiag} it suffices to show that $\mu(G,H_1,\ldots,H_C) \leq \mu(G)$ when the $H_c$ are diagonal.  In this case, observe that
\eas{
\left | \sum^{C}_{c=1} e^*_i H^*_c a a^* H_c e_j \right | &= |a_i | |a_j | \left | \sum^{C}_{c=1} \overline{(H_c)_{ii}} (H_c)_{jj} \right |
 \leq \| a \|^2_{\infty} \sum_{j \in \Delta} \sqrt{\sum^{C}_{c=1} | (H_c)_{i,i} |^2 } \sqrt{\sum^{C}_{c=1} | (H_c)_{j,j} |^2 } \leq \mu(G),
}
where in the last step we used \R{Hc_iso}.
}

Notice that this worst-case bound is sharp:

\examp{[Repeated sensor profiles]
\label{ex:repeated}
Suppose that $H_c = 1/\sqrt{C} H$ for some diagonal $H \in \bbC^{N \times N}$ with $H^* H = I$.  Then each sensor receives exactly the same information.  Hence there is no possibility of recovering an arbitrary $s$-sparse vector using fewer than $m \approx s$ measurements per sensor, i.e.\ $m \approx C s$ in total.
}

Similarly, for nonoverlapping sensor profiles:

\examp{[Nonoverlapping sensor profiles]
\label{ex:nonoverlapping_identical}
Consider the case of nonoverlapping sensor profile matrices, i.e.\ $H_c = P_{I_c}$, where the sets $\cI = \{ I_1,\ldots,I_C\}$ give a partition of $\{1,\ldots,N\}$.  Then the problem of recovering $x$ decouples into $C$ problems of recovering the vectors $ P_{I_c}x$.  Thus, if sparsity is the assumed model, one requires $m \approx C s$ in general, since it is possible to construct an $s$-sparse $x$ such that $\| P_{I_c} x \|_{0} = s$ for some $c$.

Note that these sensor profile matrices do yield optimal recovery guarantees for sparse and distributed vectors, as we demonstrate in \S \ref{ss:identical_pcwse_const}.
}

\subsubsection{Circulant sensor profile matrices} \label{sec:identSamp_circCh}
We now suppose the sensor profile matrices $H_1,\ldots,H_C$ are circulant with filter vectors $h_1,\ldots,h_C$.

\cor{[Identical sampling with sparsity model and circulant profile matrices]
\label{c:identSampl_sparsity_circH}
Let $x \in \bbC^{N}$, $0 < \epsilon < 1$, $N \geq s \geq 2$ and suppose that $H_{c} \in \bbC^{N \times N}$, $c=1,\ldots,C$, are circulant matrices satisfying \R{Hc_iso}, i.e.\ $\sum^{C}_{c=1} H^*_c H_c = I$.  Let $F$ be defined as in \S \ref{sss:general_setup_identical} and draw $A$ according to \R{A_F}.  If $y = A x + e$ with $\| e \|_{2} \leq \eta$, then for any minimizer $\hat{x}$ of
\bes{
\min_{z \in \bbC^N} \| z \|_{1}\ \mbox{subject to $\| A z - y \|_2 \leq \eta$},
}
we have
\bes{
\| x - \hat{x} \|_{2} \lesssim \sigma_{s}(x)_1 + \sqrt{s} \eta,
}
with probability at least $1-\epsilon$, provided
\be{
\label{eq:identSampl_circ_hc_cond}
m \gtrsim s \cdot C \cdot \mu(G) \cdot \left( \sum_{c=1}^C \| h_c \|_1^2 \right) \cdot L,
}
where $\mu(G)$ is as in \R{standard_coherence}, $L$ is as in \R{L_def} and $h_1,\ldots,h_C$ are the filter vectors of $H_1,\ldots,H_C$.
}

\prf{
We use Corollary \ref{cor:identSampl_nonDiag}.  Observe that
\eas{
\left | \sum^{C}_{c=1} e^*_i H^*_c a a^* H_c e_j \right | \leq \| a \|^2_{\infty} \sum^{N}_{k,l=1} \left | \sum^{C}_{c=1} \overline{(h_c)_{k-i}} (h_c)_{l-j} \right | \leq \| a \|^2_{\infty} \sum^{C}_{c=1} \| h_c \|^2_{1}, 
}
and therefore $\mu(G,H_1,\ldots,H_C) \leq \sum^{C}_{c=1} \| h_c \|^2_{1}$, as required.
}

Note that if the entries of the filter vectors $h_c$ are nonnegative then, in a similar manner to Example \ref{ex:nonneg_symbol}, we have $\sum_{c=1}^C \| h_c \|_1^2 = 1$.  Hence the worst-case recovery guarantee \R{eq:identSampl_circ_hc_cond} reduces to 
\bes{
m \gtrsim s  \cdot C\cdot \mu(G) \cdot L.
} 
This is clearly sharp within the setting of Corollary \ref{c:identSampl_sparsity_circH}, since one could set all the $H_{c}'s$ to be equal. 

\rem{\label{r:identSampl_circ:worst}
Unlike the case of diagonal sensor profile matrices (see \S\ref{ss:identical_pcwse_const}), it is impossible to find circulant sensor profile matrices that lead to optimal recovery guarantees in the case of identical sampling without further assumptions on not only the sensor profiles $H_c$ but also the sampling matrix $\tilde{A}$.  To see why, suppose that the matrix $\tilde{A}$ corresponds to a subsampled DFT matrix, i.e.\ $\tilde{A} = \sqrt{N} P_{\Omega} \Phi$, where $\Phi \in \bbC^{N \times N}$ is the unitary DFT matrix.  As in \S \ref{sss:circulant_spectral}, write $H_c = \Phi^* \Lambda_{c} \Phi$, where $\Lambda_{c}$ is the diagonal matrix of eigenvalues of $H_c$.  Then the measurements in the $c^{\rth}$ sensor are $y_c = \tilde{A} H_{c} x = \sqrt{N} P_{\Omega} \Lambda_{c} \Phi x = \sqrt{N} \Lambda_{c} P_{\Omega} \Phi x$.  Hence, up to multiplication by the diagonal matrix $\Lambda_c$, each sensor receives exactly the same measurements.  Note that this choice of $\tilde{A}$ is optimal for single-sensor CS, yet it is clearly suboptimal in this particular multi-sensor setting. Conversely, in \S \ref{s:NumExp} we will see numerically that sampling with a Gaussian random matrix yields optimal recovery (this empirical observation has recently been confirmed by the theoretical results of \cite{Chun&Adcock:16arXiv-CS&PA&RIP}).
}

\subsection{Bounds for piecewise constant diagonal sensor profile matrices} \label{ss:identical_pcwse_const}
We now construct a large class of sensor profile matrices for which optimal recovery guarantees are possible in the identical sampling case.  This is similar to the setup introduced earlier in Example \ref{ex:piecewise_const}.  To this end, let $\cI = \{ I_1,\ldots,I_C\}$ be a partition of $\{1,\ldots,N\}$ and suppose that $V = \{ V_{c,d} \}^{C}_{c,d=1} \in \bbC^{C \times C}$ is an isometry, i.e.\ $V^* V = I$. 
Define the sensor profile matrices
\be{
\label{Hc_pcwse_const_identical}
H_c = \sum^{C}_{d=1} V_{c,d} P_{I_d}.
}
Note that $\sum^{C}_{c=1} H^*_c H_c = \sum^{C}_{d=1} P_{I_d} \sum^{C}_{c=1} | V_{c,d} |^2 = \sum^{C}_{d=1} P_{I_d} = I$ and therefore the isometry condition \R{Hc_iso} holds in this case.

\thm{[Identical sampling with the sparsity in levels model and piecewise constant diagonal sensor profiles]
\label{t:identical_pcwse_const}
Let $\cI = \{ I_1,\ldots,I_C\}$ be a partition of $\{1,\ldots,N\}$ and $\cS  = \{s_1,\ldots,s_C \} \in \bbN^C$ with $s_c \leq | I_c |$ for $c=1,\ldots,C$.
Let diagonal matrices $H_c$ be given by \R{Hc_pcwse_const_identical} for some isometry $V \in \bbC^{C \times C}$ and suppose that $x \in \bbC^{N}$, $0 < \epsilon < 1$ and $N \geq s = s_1 + \ldots + s_C \geq 2$.
Let $F$ be defined as in \S \ref{sss:general_setup_identical} and draw $A$ according to \R{A_F}.
If $y = A x + e$ with $\| e \|_{2} \leq \eta$, then for any minimizer $\hat{x}$ of
\bes{
\min_{z \in \bbC^N} \| z \|_{1}\ \mbox{subject to $\| A z - y \|_2 \leq \eta$},
}
we have
\bes{
\| x - \hat{x} \|_{2} \lesssim \sigma_{\cS,\cI}(x)_1 + \sqrt{s} \eta,
}
with at least probability $1-\epsilon$, provided
\be{
m \gtrsim \mu(G) \cdot C \cdot \max_{c=1,\ldots,C} \{ s_c \} \cdot L,
}
where $\mu(G)$ is as in \R{standard_coherence} and $L$ is as in \R{L_def}.
}

\prf{
We shall use Theorem \ref{t:abs_recov}.  For $c=1,\ldots,C$, let $\Delta_{c}$ be the index set of the largest $s_c$ entries of $x$ in absolute value restricted to $I_c$.  Let $\Delta = \Delta_{1} \cup \cdots \cup \Delta_C$, $B \sim F$ and $z \in \bbC^N$ with $\| z \|_{\infty} = 1$.  Suppose that $i \in I_d$ for some $d=1,\ldots,C$.  Then, since $V$ is an isometry,
\eas{
| e^*_i B B^* P_{\Delta}z | = \left | \sum^{C}_{c=1} e^*_i H^*_c a a^* H_c P_{\Delta} z \right | 
 = | e^*_i a | \left | \sum^{C}_{e=1} a^* P_{I_e} z \sum^{C}_{c=1} \overline{V_{c,d}} V_{c,e} \right |
 = | e^*_i a | \left | a^* P_{I_d} z \right | \leq \| a \|^2_{\infty} s_d.
}
It now follows that $\Gamma_{1}(F,\Delta) \leq \mu(G) \max_{c=1,\ldots,C} \{ s_c \}$.  Also,
\eas{
\bbE | e^*_i B B^* P_{\Delta} z |^2 = \bbE | e^*_i a |^2 \left | a^* P_{I_d} z \right |^2 \leq \| a \|^2_{\infty} \| P_{I_d} z \|^2_2 \leq \| a \|^2_{\infty} s_d,
}
and therefore $\Gamma_{2}(F,\Delta) \leq \mu(G) \max_{c=1,\ldots,C} \{ s_c \}$ as well.  We now apply Theorem \ref{t:abs_recov}.
}

This result shows that under the sparse and $\lambda$-equidistributed model (see Definition \ref{d:sparse_distrib}) we get optimal recovery guarantees for identical sampling, i.e.\ $m \gtrsim \mu(G) \cdot \lambda \cdot s \cdot L$, provided the sensor profiles are chosen as in \R{Hc_pcwse_const_identical}.

\rem{
Within this piecewise constant model it is impossible to get optimal recovery guarantees for sparse vectors.  Indeed, suppose that $x$ is $s$-sparse with $\supp(x) \subseteq I_1$.  Then the measurements for the $c^{\rth}$ sensor are $y_c = \tilde{A} H_c x = V_{c1} \tilde{A} x$, i.e.\ each sensor obtains the same measurements up to the constant $V_{c1}$.  Therefore we require $\approx s$ measurements per sensor, and $\approx C s$ in total.  Note that this is in stark contrast to the case of distinct sampling, wherein optimal recovery guarantees for the sparsity model are possible provided $V$ is incoherent (see Example \ref{ex:piecewise_const}).
}

\examp{[Clustered sparse signal recovery with DFT-like sensor profiles] \label{eg:IdentSampl_clusteredSignal}
Consider the clustered sparse signal model (see Remark \ref{r:clustered}) with partition given by 
\be{
\label{I_c_clustered}
I_c = (c + C \bbZ) \cap \{ 1,\ldots,N\} = \{ c,c+C,\ldots,c+(n-1) C \},\qquad c=1,\ldots,C.
}
Since this is an example of the above setup, one can obtain an optimal recovery guarantee by choosing $H_c$ as in \R{Hc_pcwse_const_identical}.  Interestingly, and unlike the case where the index sets $I_c$ are blocks of integers, we can find smooth sensor profile matrices in this case.  An example of this is the following:
\bes{
(H_{c})_{jj} = \frac{1}{\sqrt{C}} \exp(2 \pi \I (c-j)/C),\quad j=1,\ldots,C,\ c=1,\ldots,C.
}
Note that is particular case of the above setup with $V \in \bbC^{C \times C}$ being the DFT matrix, i.e.\ $V_{cd} = \exp(2 \pi \I (c-d) / C ) /\sqrt{C}$.  Indeed, if $I_d$ is as in \R{I_c_clustered} and $j \in I_d$, then $(H_{c})_{jj} = \frac{1}{\sqrt{C}} \exp(2 \pi \I (c-d)/C) = V_{cd}$.  Hence $H_c$ can be written in the form \R{Hc_pcwse_const_identical}.
}

Finally, we note that in a recent work \cite[Cor. 3.6]{Chun&Adcock:16ITW} a simpler bound for identical sampling with diagonal sensor profiles has been introduced.  This gives a bound which is both computable, and can be used to avoid the linear dependence on $C$ in the measurement bound of Corollary \ref{cor:identSampl_sparsity} by using the sparsity in levels signal model.

\section{Numerical experiments} \label{s:NumExp}

\begin{figure}[!t]
\centering
\small\addtolength{\tabcolsep}{-4pt}
\begin{tabular}{cccc}
{} & {\small \hspace{.35em}$C=2$\hspace{-.35em}} & {\small $C=4$~} & {\small $C=8$~} \\
\rotatebox{90}{\parbox{10em}{\centering \small \hspace{1.5em} Magnitude ($h_c(t)$)}} &
\includegraphics[scale=0.5, trim=2.1em 2em 2.3em 1.5em, clip]{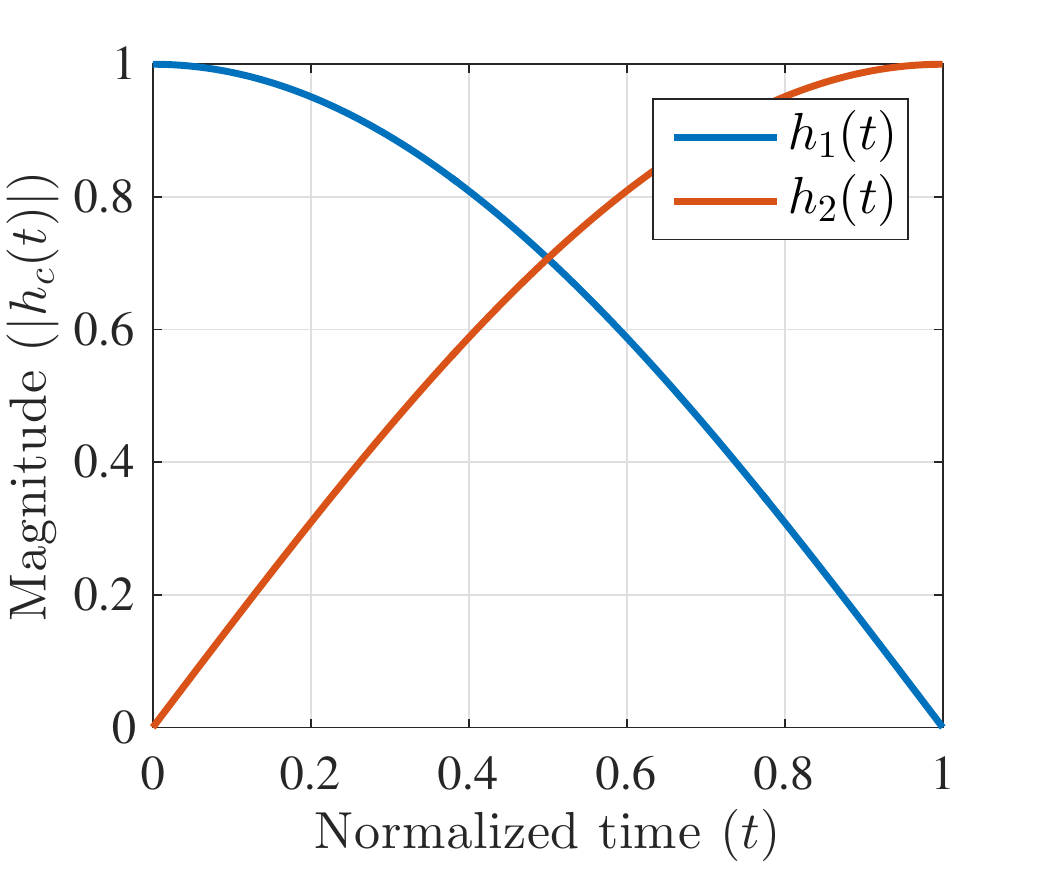}~ &
\includegraphics[scale=0.5, trim=3.6em 2em 2.3em 1.5em, clip]{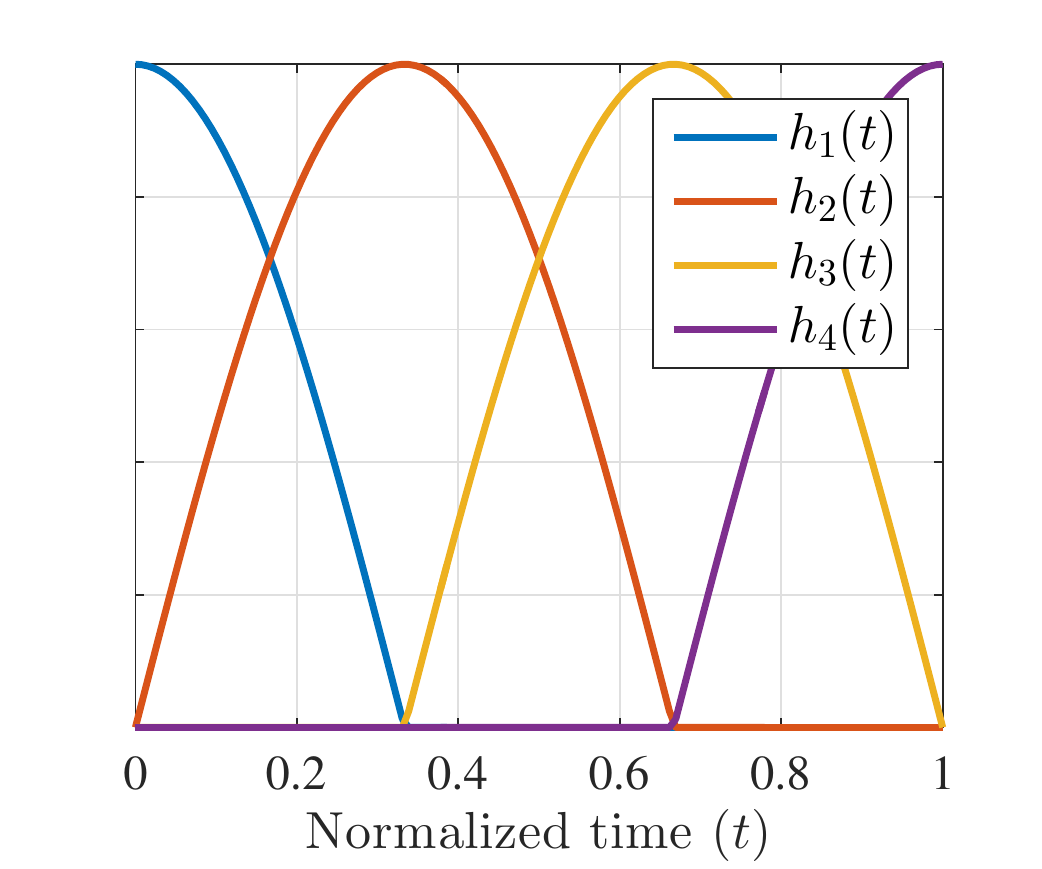}~ &
\includegraphics[scale=0.5, trim=3.6em 2em 2.3em 1.5em, clip]{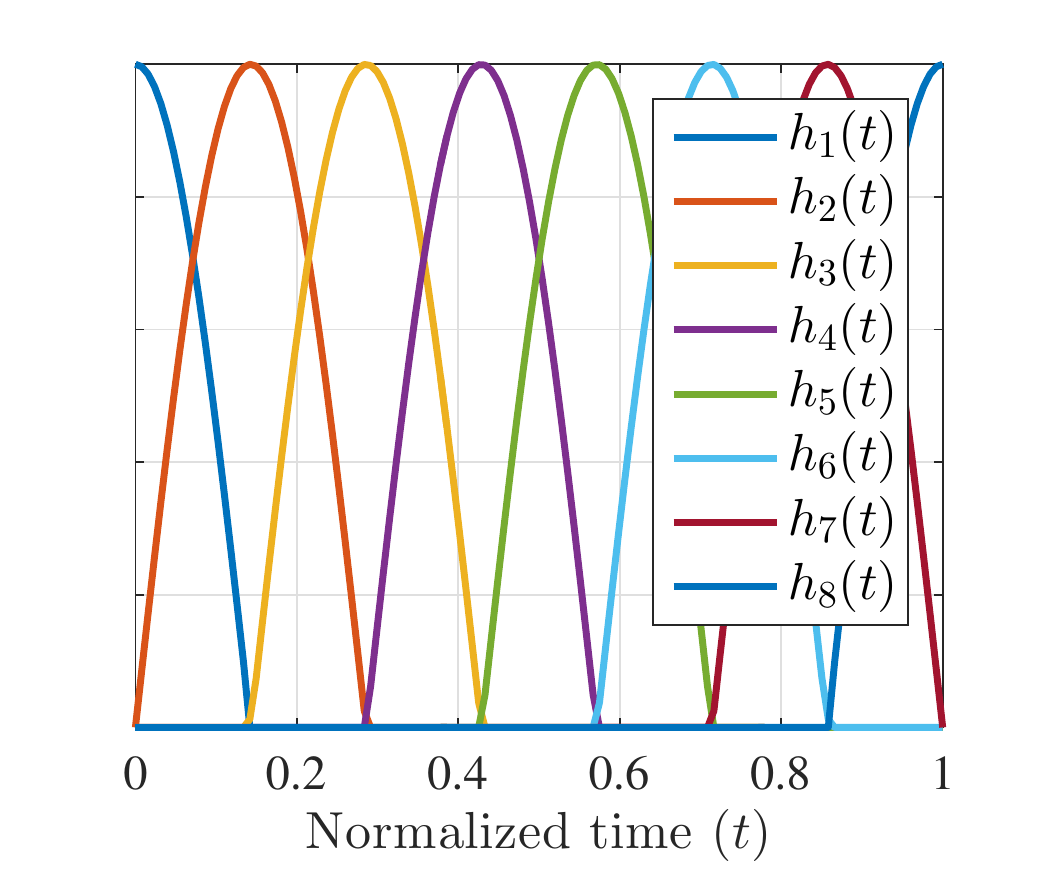} \vspace{-0.15em} \\
\multicolumn{4}{ c }{\small Normalized time ($t$)}
\end{tabular}
\caption{An example of smooth and banded sensor profiles ($C=2,4,8$).}
\label{fig:prof:band}
\end{figure}

In this section, we present empirical validation of our results using the phase transition setup (see \cite{Monajemi&etal:13PNAS} and references therein). 
The first numerical experiments consider Fourier sensing with complex diagonal sensor profile matrices.  As discussed in \S \ref{sss:pMRI}, identical sensing across sensors corresponds to a one-dimensional (1D) example of the pMRI system model with \textit{ideal} sensor profiles; that is, satisfying the joint isometry condition $\sum_c^C H_c^* H_c = I$ \cite{Chun&Adcock&Talavage:15TMI}.  The second numerical experiments consider Gaussian sensing with complex circulant sensor profile matrices.  This setup corresponds to 1D example of the multi-view imaging application (see \S \ref{sss:multiview}) with ideal sensor profiles, i.e.\ satisfying $C^{-1} \sum_c^C H_c^* H_c = I$ or $\sum_c^C H_c^* H_c = I$ for distinct or identical sampling respectively.

\subsection{Simulation setup}
The overall simulation setup is as follows. For an $s$-sparse signal $x \in \bbC^{128}$, the positions of $s$ non-zero elements are chosen uniformly at random without replacement, and the non-zero elements chosen randomly and uniformly distributed on the unit circle.  For the phase transition graph of resolution $49 \times 49$, the horizontal and vertical axes are defined by $\delta = m/CN \in (0,1)$ and $\kappa = s/N \in (0,1)$ respectively.  The empirical success fraction is calculated as $\#\{\textmd{successes}\}/\#\{\textmd{trials}\}$ with $20$ trials, where success corresponds to a relative recovery error $\| x - \hat{x} \|_{2} / \| x\|_2 < \mathrm{tol}$ for $\mathrm{tol} = 0.001$.  Throughout, we use CVX with the SDPT3 or MOSEK solver \cite{cvx, Grand&Boyd:08cvx}.

For Fourier sensing, $m/C$ rows of the DFT matrix were drawn uniformly at random without replacement and, for distinct sampling, these rows were drawn independently across sensors.  The diagonal sensor profile matrices were generated using a truncated cosine function multiplied with phase vector $\{ (c-1) 2\pi/C + 2\pi/NC,\ldots, c 2\pi/C \}$; see Fig.~\ref{fig:prof:band}.\footnote{We have chosen this banded sensor profile setup because it is more practical than the case of piecewise constant sensor profiles analyzed in the paper, particularly for pMRI with small number of receive coils (e.g., $C \leq 4$).}  For Gaussian sensing, the measurement matrices were constructed with i.i.d.\ unit Gaussian entries, and for distinct sampling, $C$ such matrices were constructed independently of each other.  The circulant sensor profile matrices were generated as $H_c = \Phi^* \Lambda_c \Phi$, where $\Phi$ is a unitary DFT matrix.  The eigenvalues $\Lambda_{c}$ were drawn independently and randomly from the unit circle, so that $\Lambda^*_c \Lambda_c = I$ \cite{Romberg:09SIAMJIS}.

\begin{figure}[!b]
\centering
\small\addtolength{\tabcolsep}{-5.25pt}
\begin{tabular}{ccccc}
{} & {\small \hspace{.7em}$C=2$\hspace{-.7em}} & {\small $C=4$} & {\small $C=8$} & {\small \hspace{-3.2em}$C=16$} \\
\rotatebox{90}{\parbox{10.5em}{\centering \small \hspace{1.5em} $\kappa = s/N$}} & 
\includegraphics[scale=0.55, trim=1.9em 4.55em 7.0em 5.0em, clip]{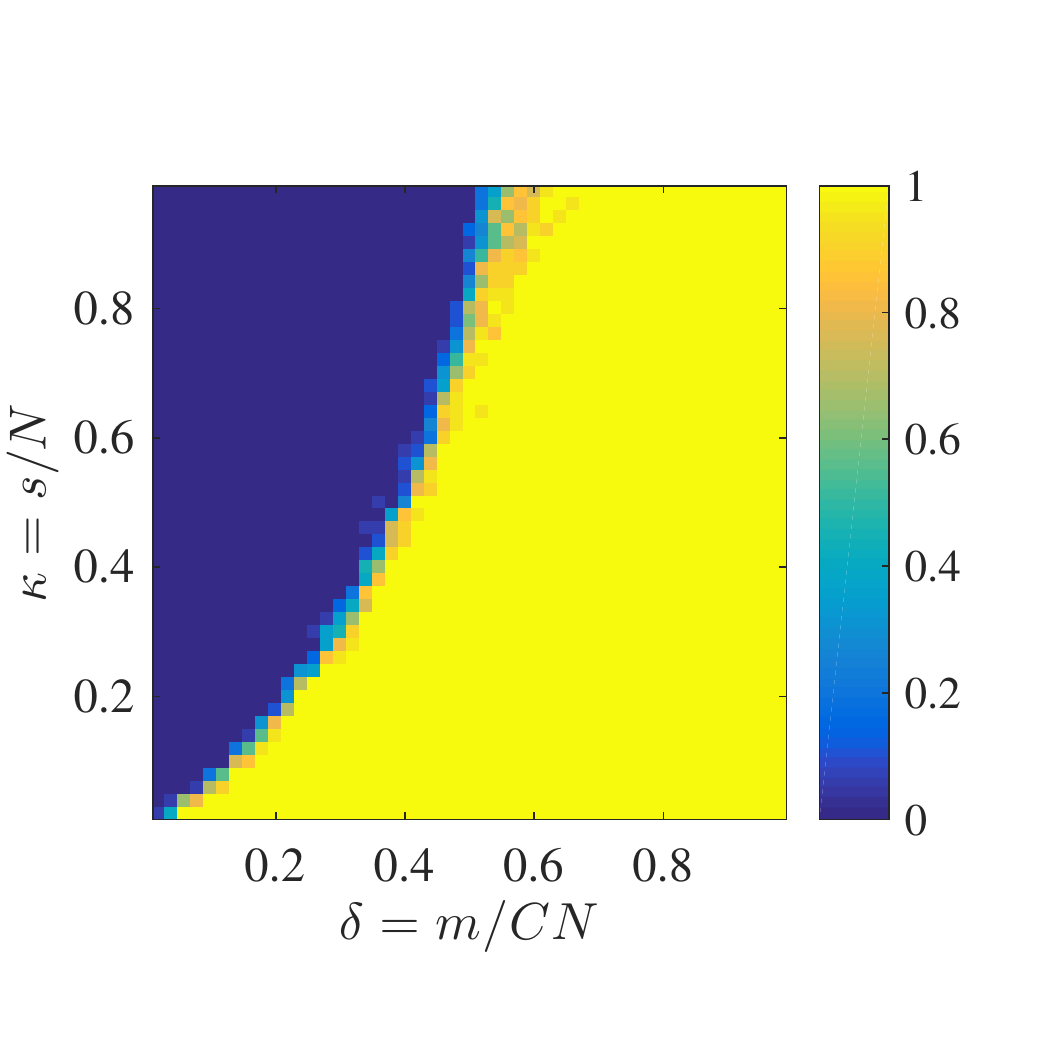} &
\includegraphics[scale=0.55, trim=4.3em 4.55em 7.0em 5.0em, clip]{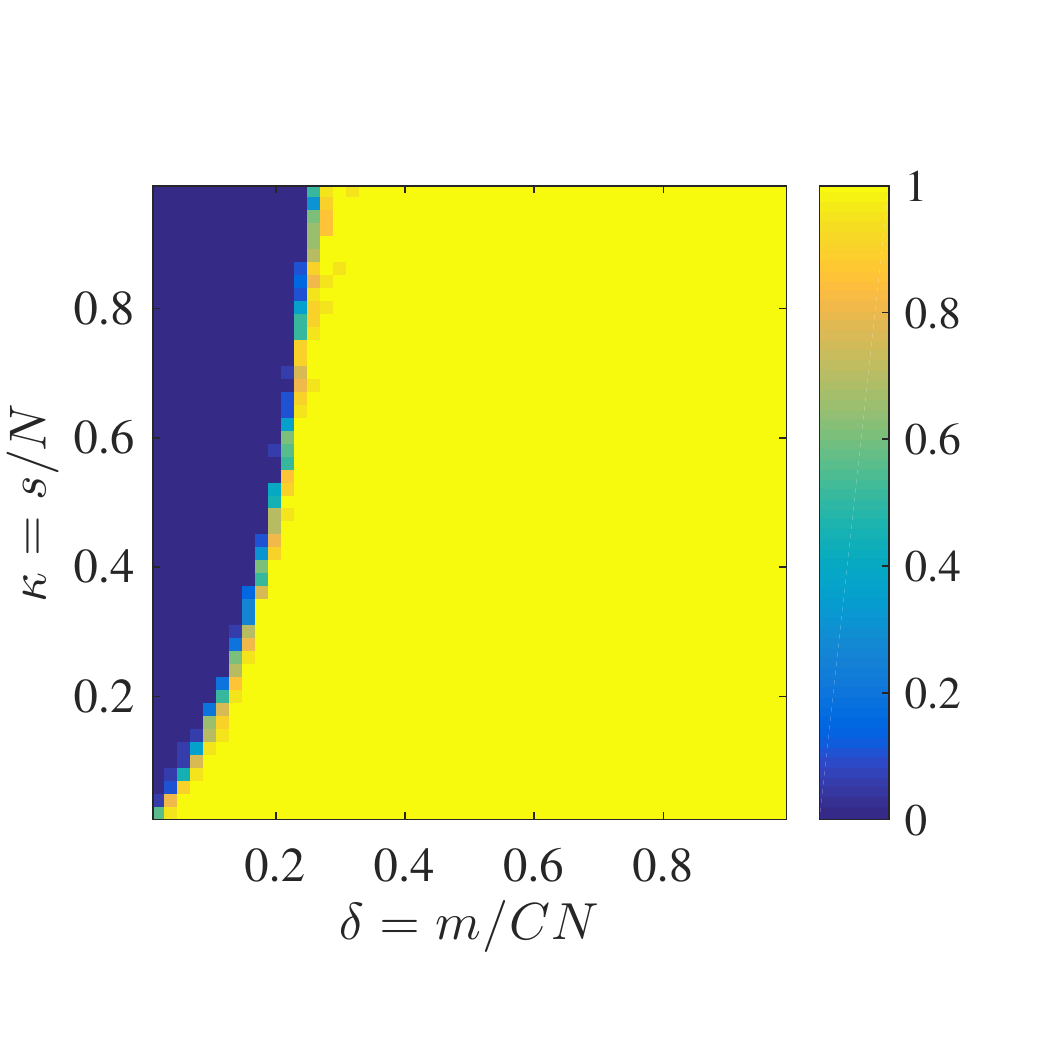} &
\includegraphics[scale=0.55, trim=4.3em 4.55em 7.0em 5.0em, clip]{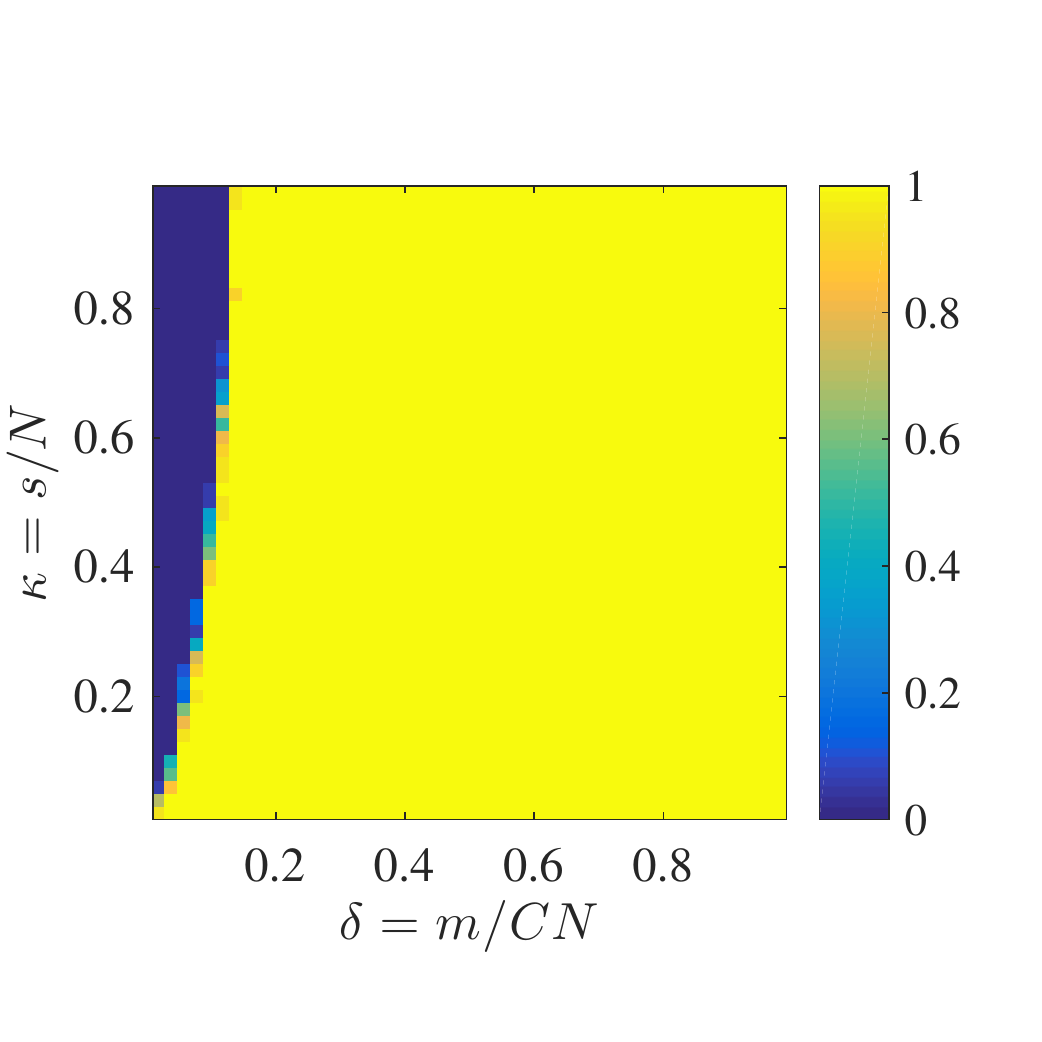} &
\includegraphics[scale=0.55, trim=4.3em 4.55em 2.25em 5.0em, clip]{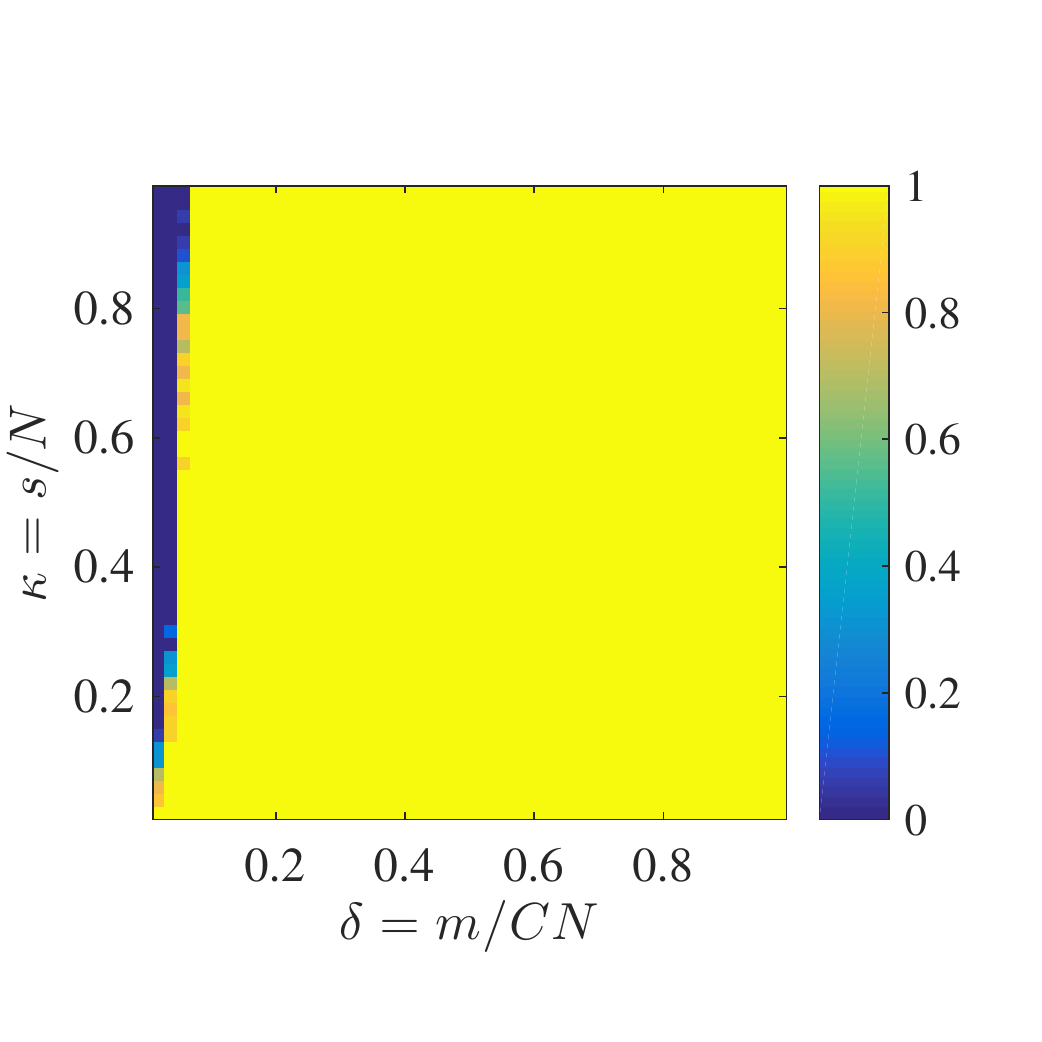} \vspace{-0.2em} \\
\multicolumn{5}{ c }{\small $\delta = m / CN$} \\
\multicolumn{5}{ c }{\small (a) Empirical phase transition for distinct sampling scenario } \\
\\
\rotatebox{90}{\parbox{10.5em}{\centering \small \hspace{1.5em} $\kappa = s/N$}} & 
\includegraphics[scale=0.55, trim=1.9em 4.55em 7.0em 5.0em, clip]{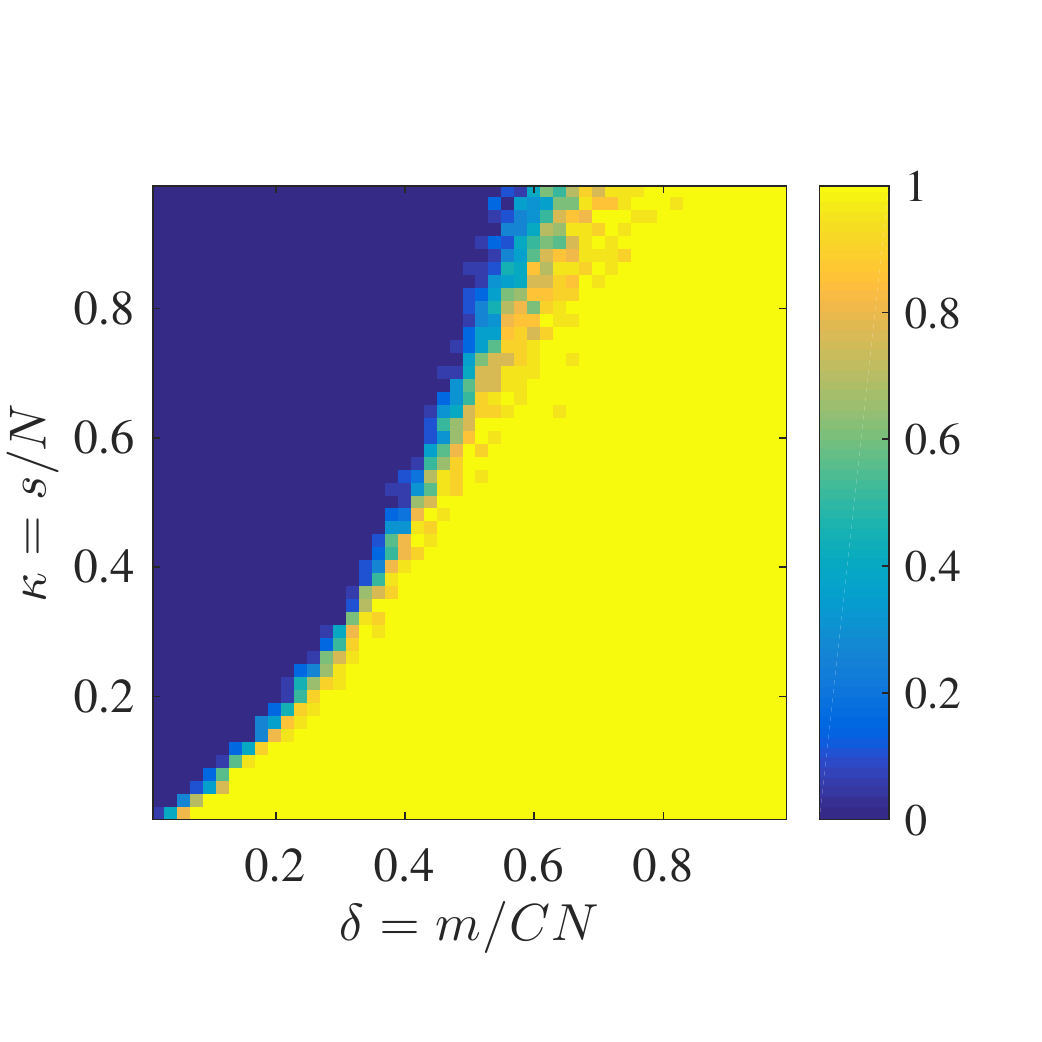} &
\includegraphics[scale=0.55, trim=4.3em 4.55em 7.0em 5.0em, clip]{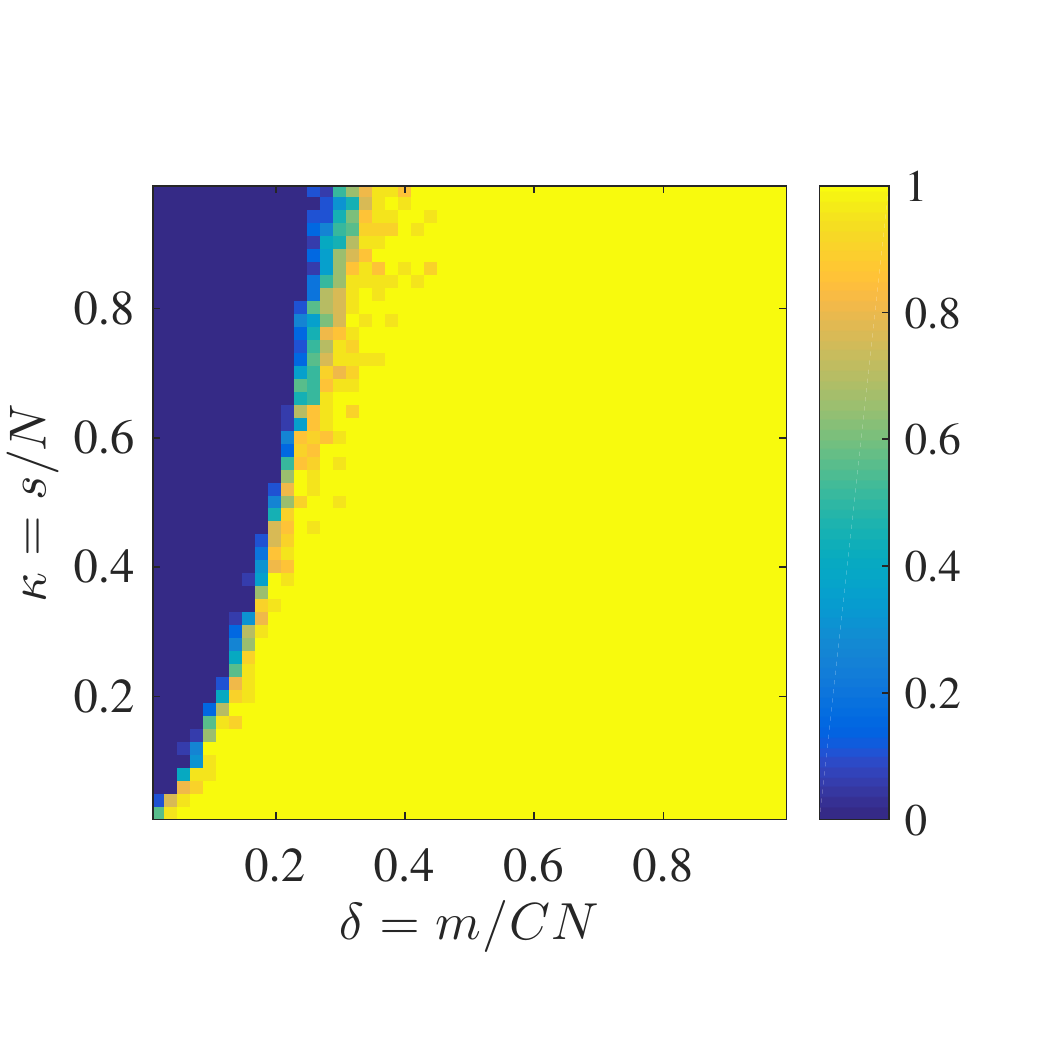} &
\includegraphics[scale=0.55, trim=4.3em 4.55em 7.0em 5.0em, clip]{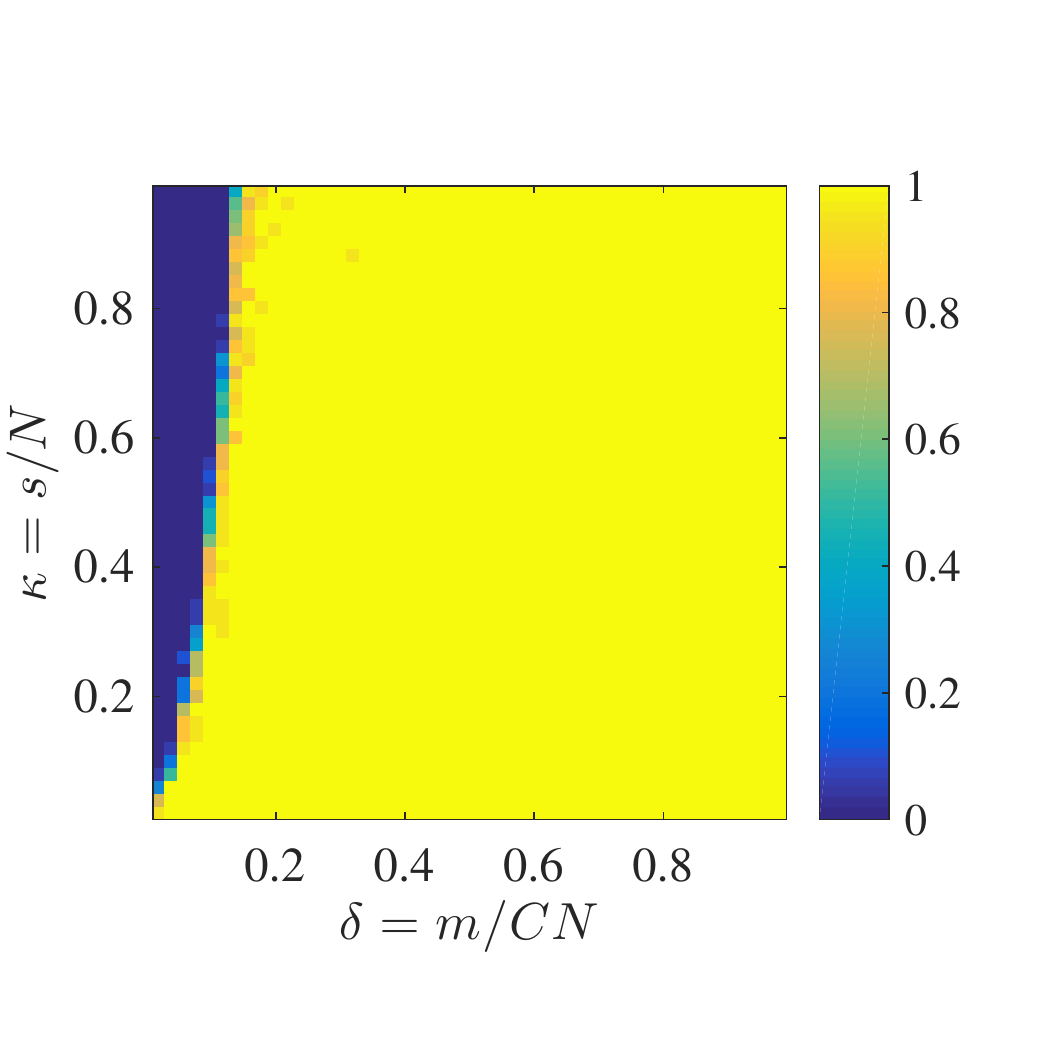} &
\includegraphics[scale=0.55, trim=4.3em 4.55em 2.25em 5.0em, clip]{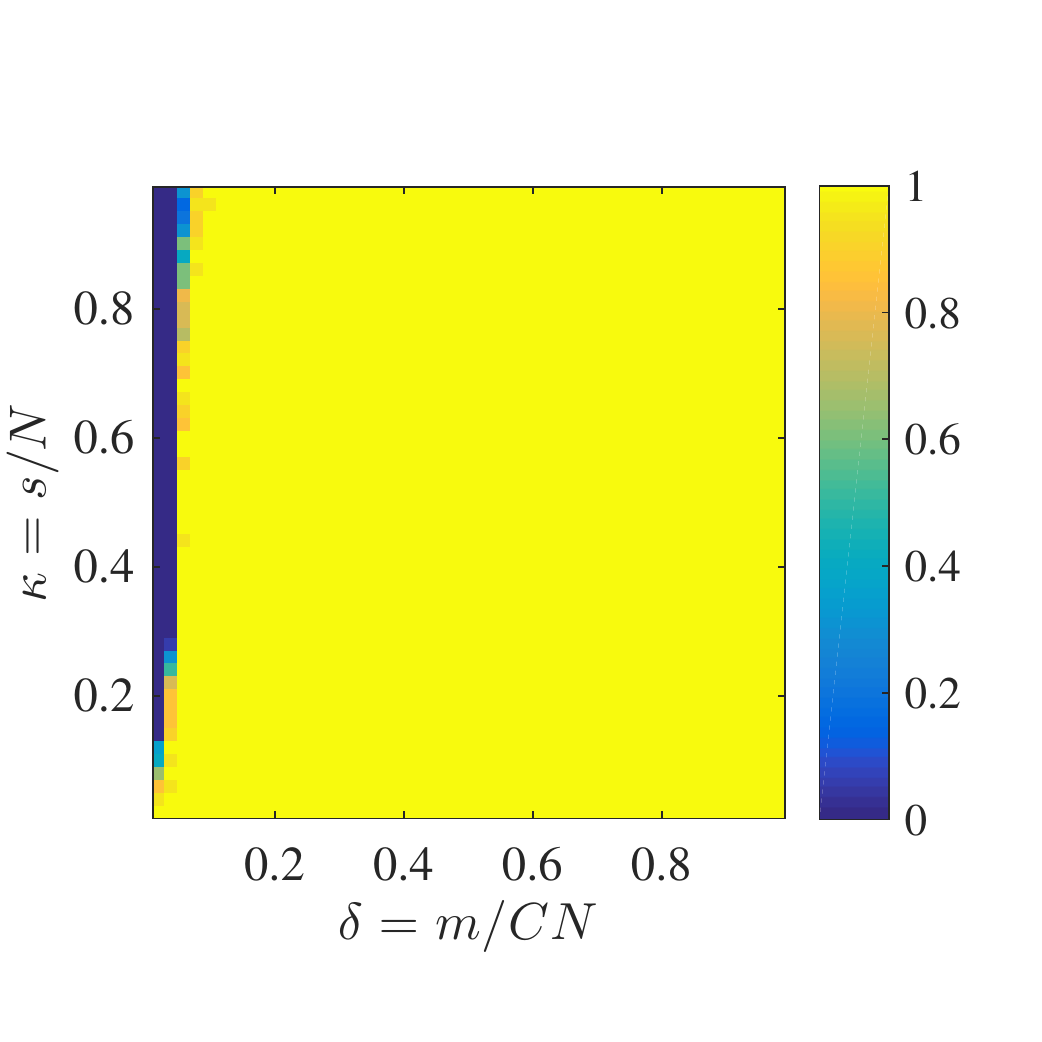} \vspace{-0.2em} \\
\multicolumn{5}{ c }{\small $\delta = m / CN$} \\
\multicolumn{5}{ c }{\small{(b) Empirical phase transition for identical sampling scenarios}} \\
\end{tabular}

\caption{Empirical phase transitions for random Fourier sensing with diagonal sensor profile matrices and $C=2,4,8,16$ sensors. 
For both sampling scenarios, the empirical probability of successful recovery increases as $C$ increase. The results in (a) are in agreement with our theoretical results. The results in (b) are in agreement with the theoretical results in \cite{Chun&Adcock:16ITW, Chun&Li&Adcock:16MMSPARSE}.}
\label{fig:PT_F_DiagH}
\end{figure}

\begin{figure}[!t]
\centering
\small\addtolength{\tabcolsep}{-5.1pt}
\begin{tabular}{ccccc}
{} & {\small \hspace{.7em}$C=2$\hspace{-.7em}} & {\small $C=4$} & {\small $C=8$} & {\small \hspace{-3.2em}$C=16$} \\
\rotatebox{90}{\parbox{10.5em}{\centering \small \hspace{1.5em} $\kappa = s/N$}} & 
\includegraphics[scale=0.55, trim=1.9em 4.55em 7.0em 5.0em, clip]{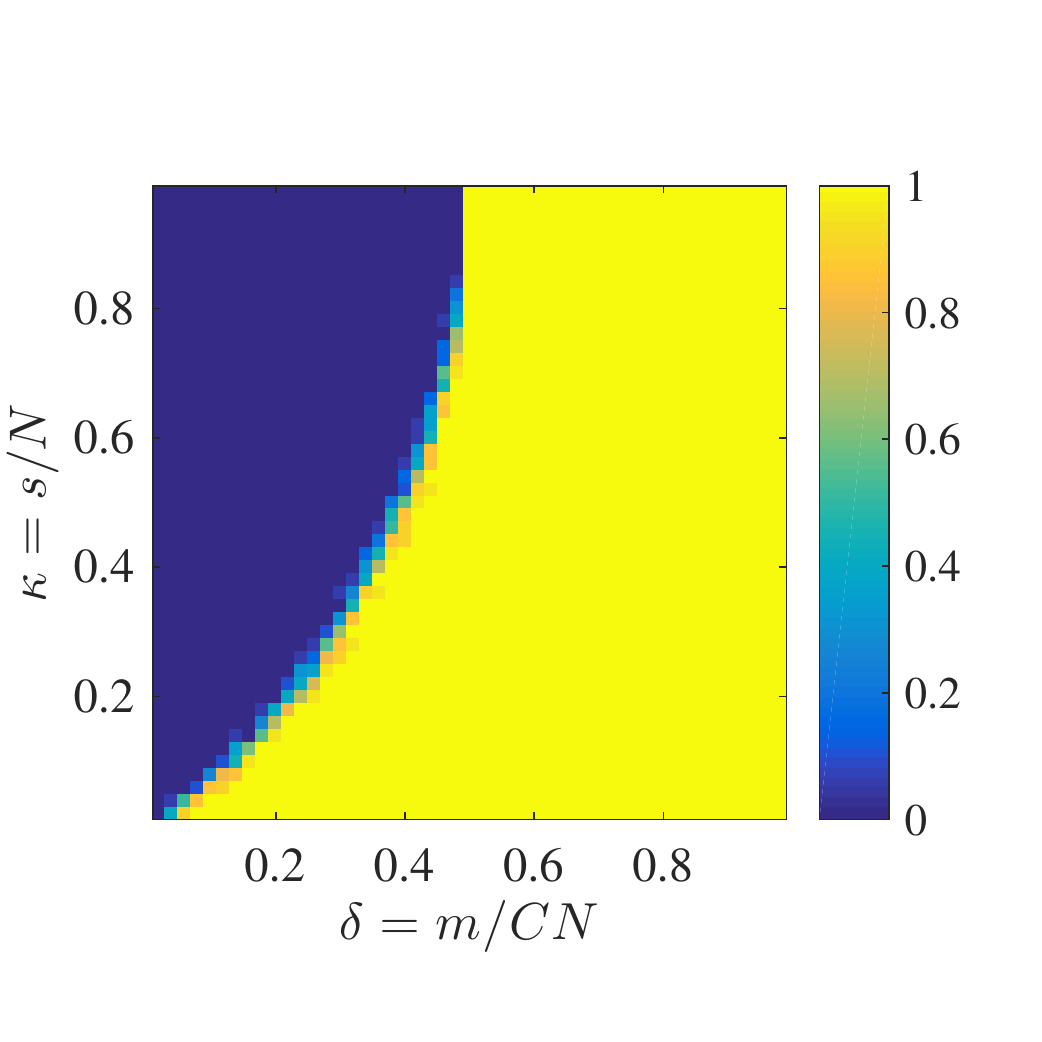} &
\includegraphics[scale=0.55, trim=4.3em 4.55em 7.0em 5.0em, clip]{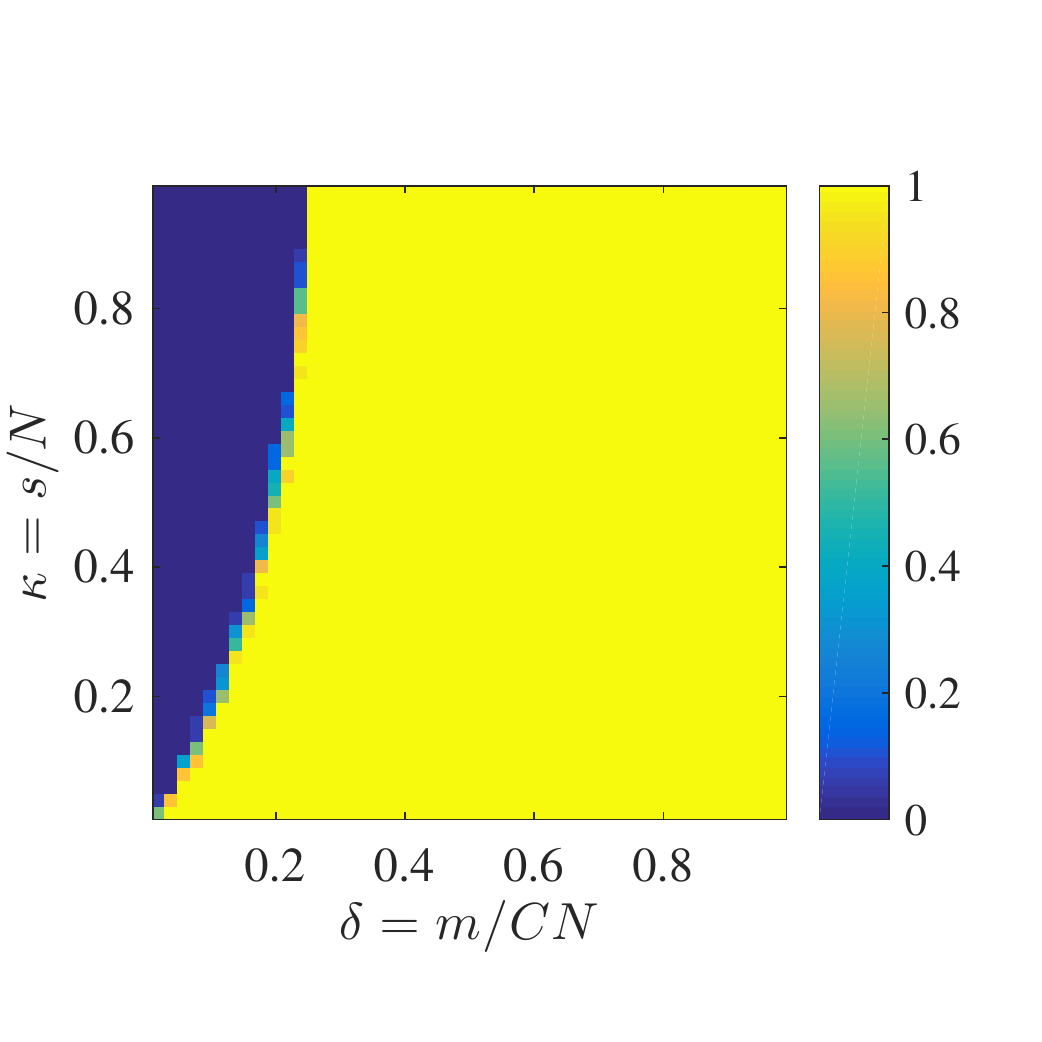} &
\includegraphics[scale=0.55, trim=4.3em 4.55em 7.0em 5.0em, clip]{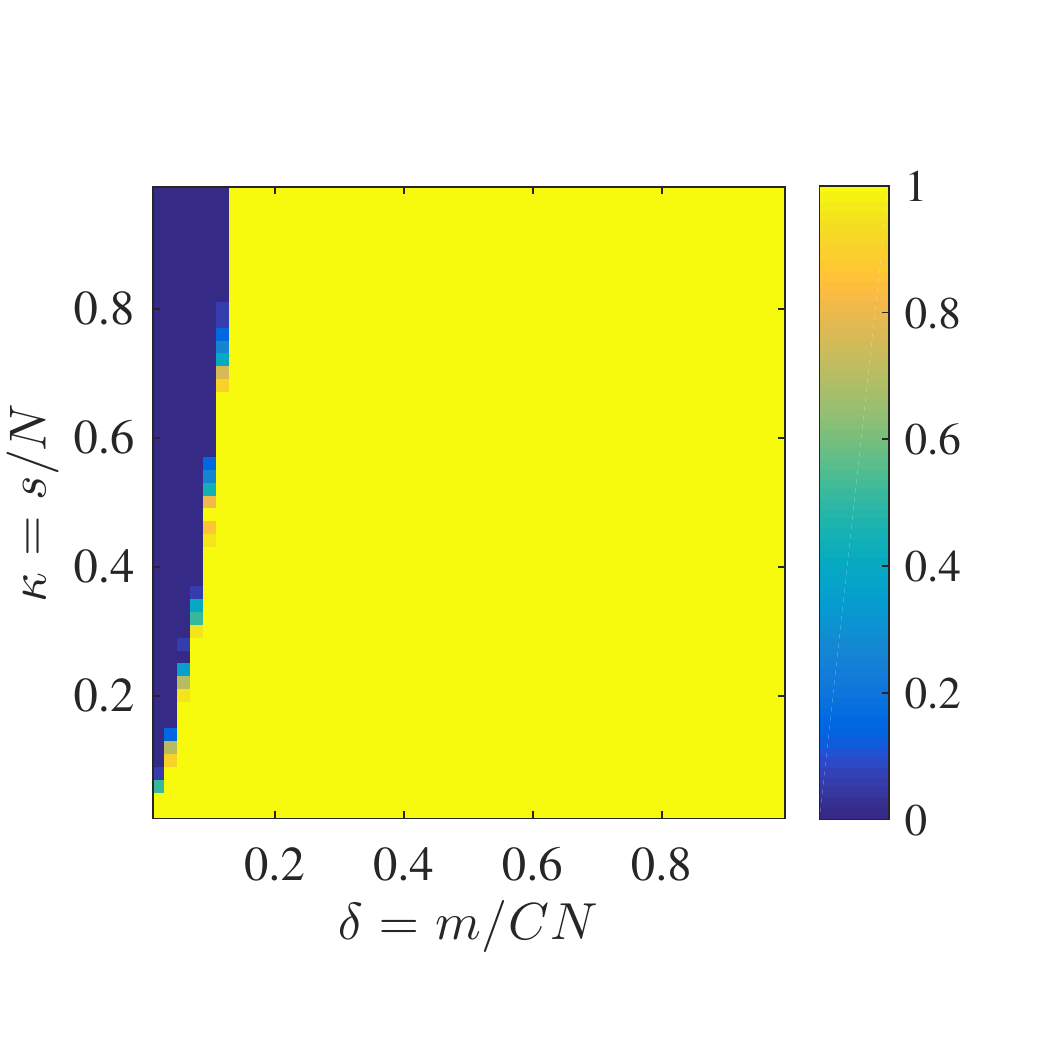} &
\includegraphics[scale=0.55, trim=4.3em 4.55em 2.25em 5.0em, clip]{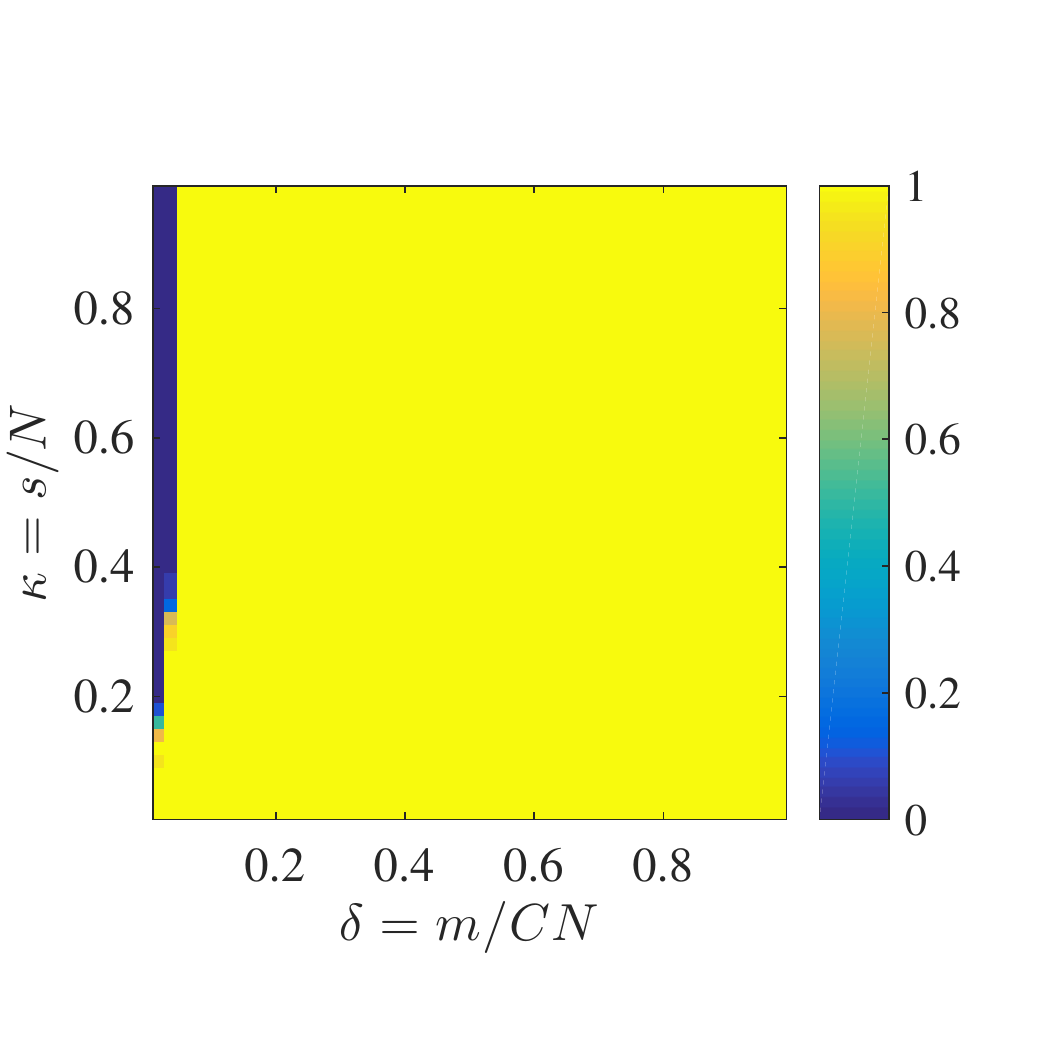} \vspace{-0.2em} \\
\multicolumn{5}{ c }{\small $\delta = m / CN$} \\
{} & {\hspace{0.75em}\small ($\mathrm{AvgP}_{\delta<0.50} = 27.88\%$) \hspace{-0.75em}} & {\small ($\mathrm{AvgP}_{\delta<0.25} = 29.92\%$)} & 
{\small ($\mathrm{AvgP}_{\delta<0.25} = 32.96\%$)} & {\small \hspace{-3.2em}($\mathrm{AvgP}_{\delta<0.12} = 49.12\%$)} \\
\multicolumn{5}{ c }{\small{(a) Empirical phase transition for distinct sampling scenario}} \\
\\
\rotatebox{90}{\parbox{10.5em}{\centering \small \hspace{1.5em} $\kappa = s/N$}} & 
\includegraphics[scale=0.55, trim=1.9em 4.55em 7.0em 5.0em, clip]{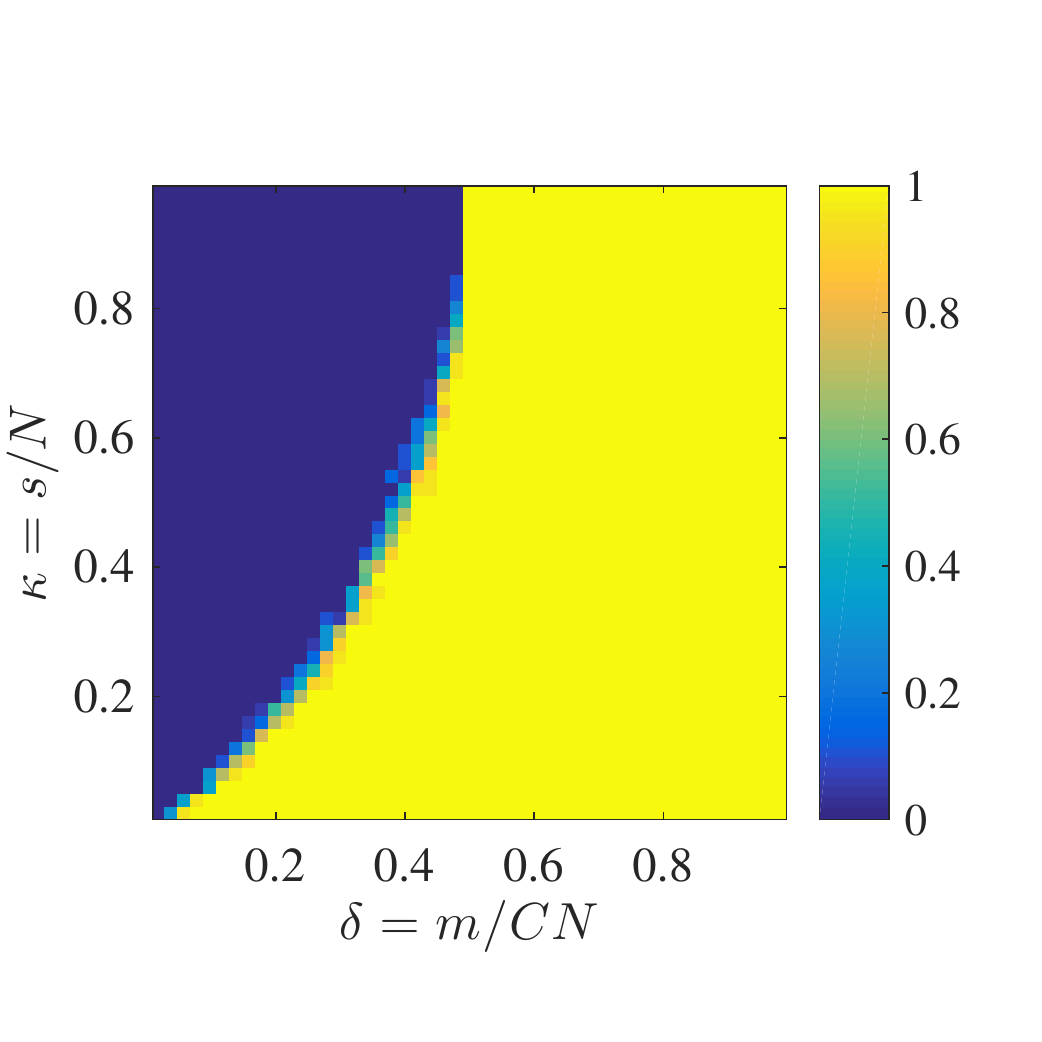} &
\includegraphics[scale=0.55, trim=4.3em 4.55em 7.0em 5.0em, clip]{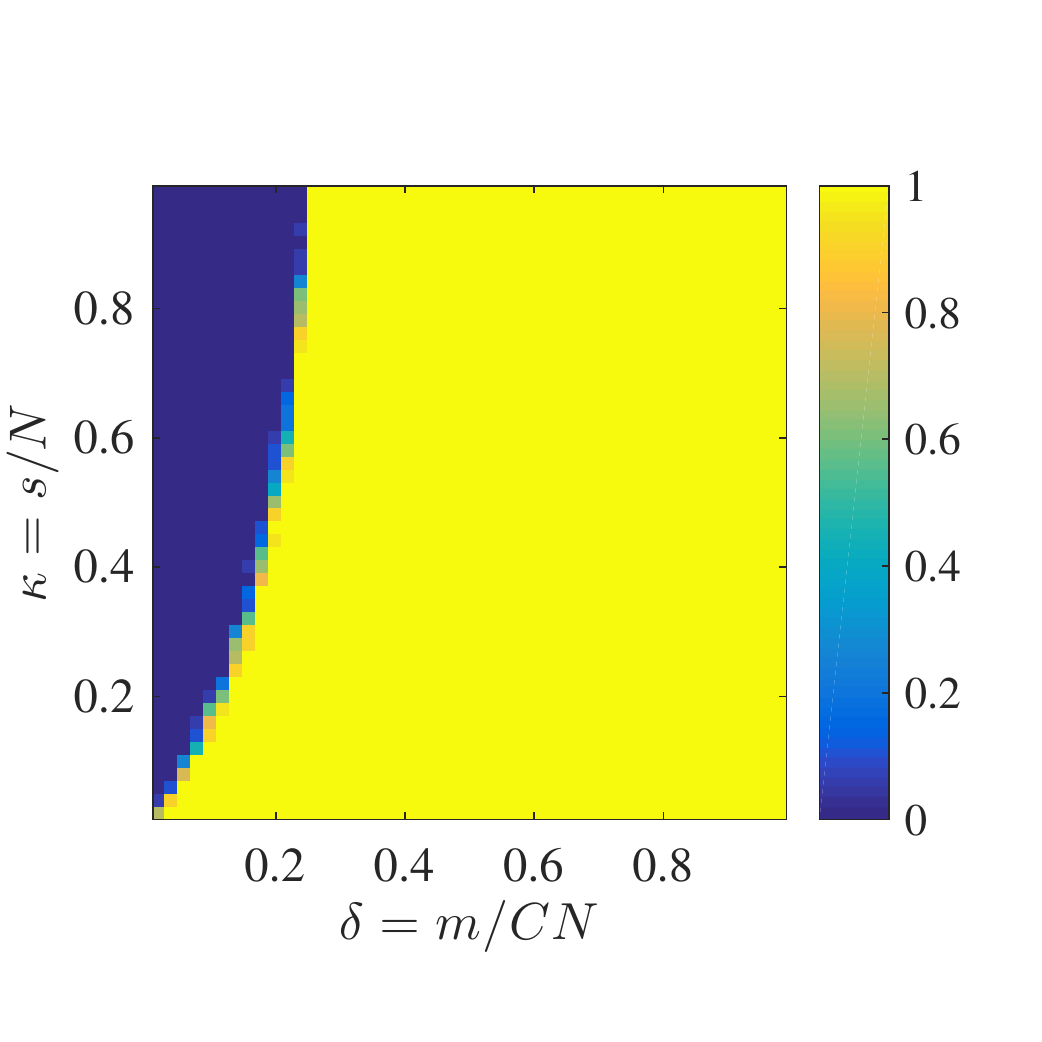} &
\includegraphics[scale=0.55, trim=4.3em 4.55em 7.0em 5.0em, clip]{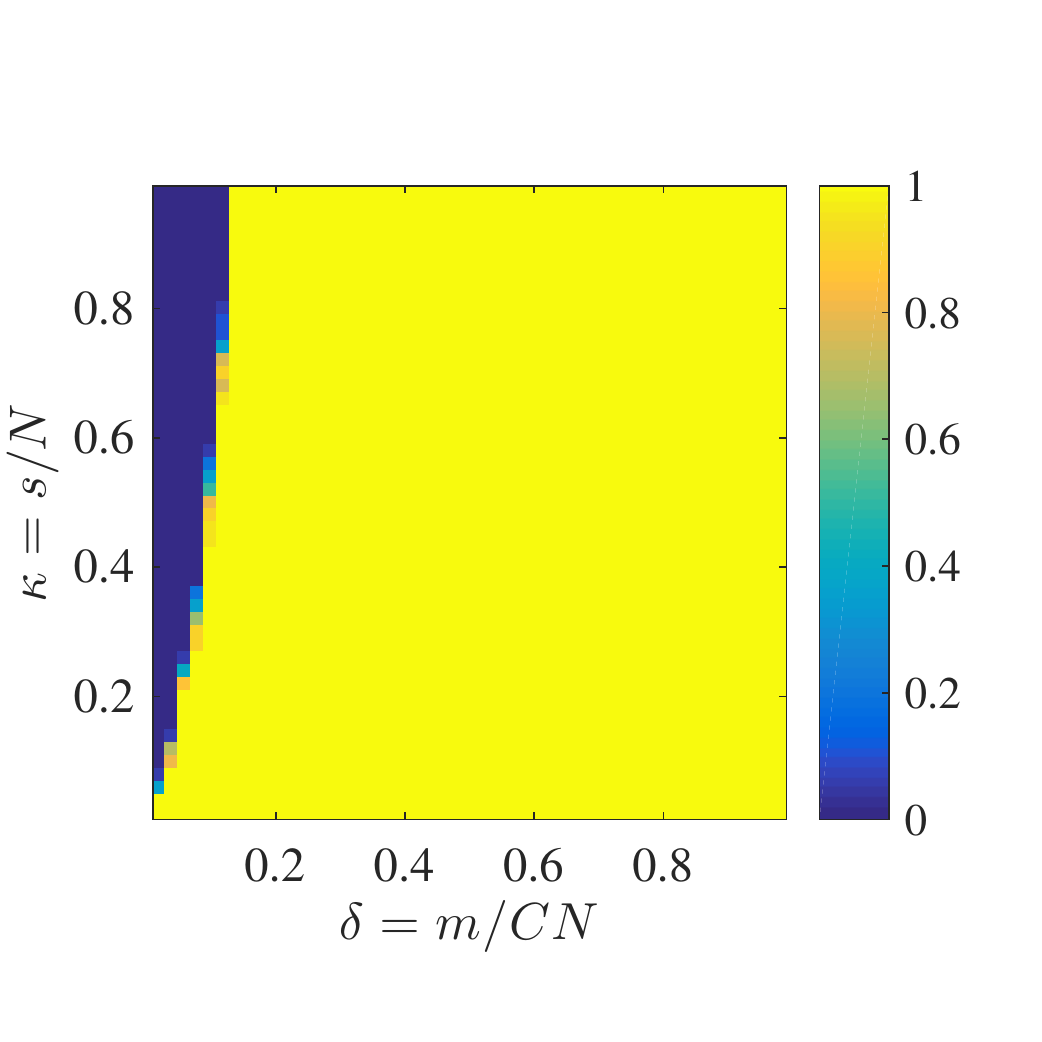} &
\includegraphics[scale=0.55, trim=4.3em 4.55em 2.25em 5.0em, clip]{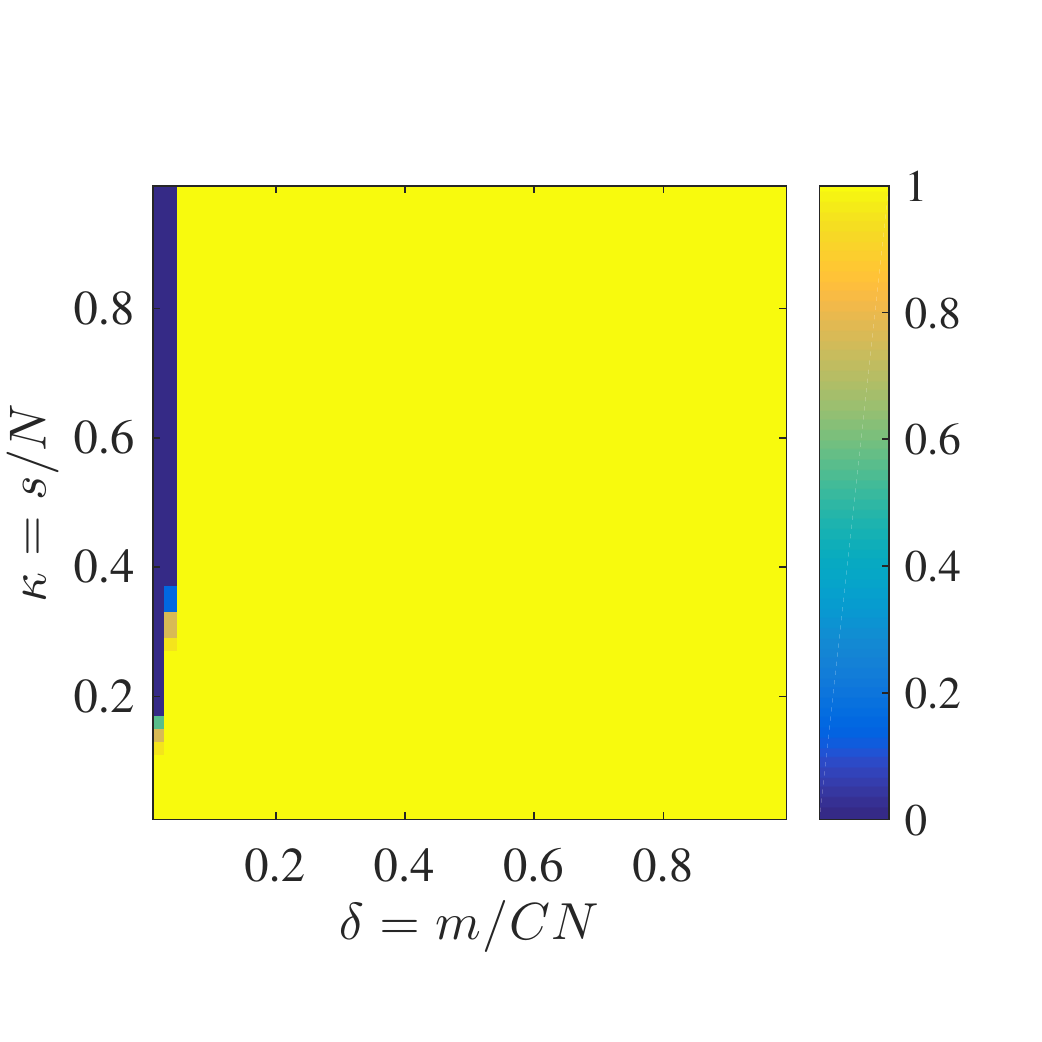} \vspace{-0.2em} \\
\multicolumn{5}{ c }{\small $\delta = m / CN$} \\
{} & {\hspace{0.75em}\small ($\mathrm{AvgP}_{\delta<0.50} = 27.86\%$)\hspace{-0.75em}} & {\small ($\mathrm{AvgP}_{\delta<0.25} = 29.74\%$)} & 
{\small ($\mathrm{AvgP}_{\delta<0.25} = 32.74\%$)} & {\small \hspace{-3.2em}($\mathrm{AvgP}_{\delta<0.12} = 48.98\%$)}\\
\multicolumn{5}{ c }{\small{(b) Empirical phase transition for identical sampling scenario}}
\end{tabular}

\caption{Empirical phase transitions for Gaussian sensing with circulant sensor profile matrices and $C=2,4,8,16$ sensors. For both sampling scenarios, the empirical probability of successful recovery increases as $C$ increases.  The results in (a) are in agreement with our theoretical results. The results in (b) are in agreement with the theoretical results in \cite{Chun&Adcock:16arXiv-CS&PA&RIP}.  The notation $\mathrm{AvgP}_{\delta<c}$ denotes the averaged probability of successful recovery for $\delta$ below $c \in (0,1)$.
}
\label{fig:PT_G_CircH}
\end{figure}

\subsection{Results and discussion}

\subsubsection{Fourier sensing with diagonal sensor profile matrices} \label{s:result_F_diagH}

Fig.\ \ref{fig:PT_F_DiagH} gives phase transitions for Fourier sensing with diagonal sensor profile matrices.  For both the distinct and identical sampling scenarios, the empirical probability of successful recovery increases as the number of sensors $C$ increases.  Moreover the rate of increase is roughly linear in $C$.  In the case of distinct sampling, this confirms the result proved in Corollary \ref{cor:DinstictSamp_diagCh}.  Interestingly, even though the sensor profile matrices are not piecewise constant (as is required in Theorem \ref{t:identical_pcwse_const}), the phase transition curves for identical sampling show a similar increase.  The transition line is somewhat more blurred, which may be as a result of the sensor profiles being not piecewise constant or the fact that randomly drawn sparse signals are only sparse and distributed in a probabilistic sense.  These results suggest that optimal recovery (i.e.\ linear decrease with $C$) is possible for identical sampling under broader conditions than those proved in this paper.

\subsubsection{Gaussian sensing with circulant sensor profile matrices} \label{s:result_G_circH}

For both the distinct and identical sampling scenarios with Gaussian sensing and circulant sensor profile matrices, the empirical probability of successful recovery increases as $C$ increases.  For distinct sampling, this confirms the result proved in Corollary \ref{cor:distinctSamp_circCh_spect}, i.e.\ the number of measurements decreases linearly in $C$ as $C$ increases.  As is to be expected, the transition curves are slightly better than for Fourier sensing (see Fig.\ \ref{fig:PT_F_DiagH}).

Interestingly, the phase transition curves for identical sampling are very similar to those for distinct sampling.  In particular, there is none of the blurring witnessed in Fourier sensing (Fig.\ \ref{fig:PT_F_DiagH}).  To highlight this, in Fig.\ \ref{fig:PT_G_CircH} we also display the $\mathrm{AvgP}$ values for each phase transition.  Note that our theoretical result for identical sampling with circulant sensor profiles (Corollary \ref{c:identSampl_sparsity_circH}) does not explain this result.  Future work we will seek to theoretically understand the significantly better empirical recovery performance observed here.

\section{Proof of Theorem \ref{t:abs_recov}}\label{s:proof}
The proof of Theorem \ref{t:abs_recov} follows similar lines to existing nonuniform recovery results (see \cite{Adcock&Hansen:15FCM,Candes&Plan:11IT}).  We first show that recovery is guaranteed by the existence of a so-called dual certificate (Lemma \ref{l:dual_certificate}), and then use a variant of the golfing scheme of Gross \cite{GrossGolfing:11TIT} to find a suitable dual certificate.

\subsection{Dual certificate}

For a vector $x \in \bbC^{N}$, we now write $\sgn(x) \in \bbC^N$ for its complex sign.  The following is a well-known result (see, for example, \cite[Thm.\ 4.33]{Foucart&Rauhut:book}):

\lem{
\label{l:dual_certificate}
Let $A \in \bbC^{m \times N}$, where $m \leq N$, and $\Delta \subseteq \{1,\ldots,N\}$.  Suppose that
\bes{
(i): \| P_{\Delta} A^*  A P_{\Delta} - P_{\Delta} \|_2 \leq \alpha,\qquad (ii):  \max_{i \notin \Delta} \left \{ \|  P_{\Delta} A^*A e_i \|_2 \right \} \leq \beta,
}
and that there exists a vector $\rho = A^*  \xi \in \bbC^N$ for some $\xi \in \bbC^m$ such that
\bes{
(iii): \| P_{\Delta} \rho - \sgn(P_{\Delta} x) \|_2 \leq \gamma,\qquad (iv): \| P^{\perp}_{\Delta} \rho \|_{\infty} \leq \theta,\qquad (v): \| \xi \| \leq \sigma \sqrt{| \Delta |},
}
for constants $0 \leq \alpha < 1$ and $\beta , \gamma, \theta, \sigma \geq 0$ satisfying $\theta + \beta \gamma / (1-\alpha) < 1$.  Let $x \in \bbC^N$, $y = A x + e$ with $\| e \|_2 \leq \eta$ and suppose that $\hat{x}$ is a minimizer of the problem
\bes{
\min_{z \in \bbC^N} \| z \|_{1}\  \mbox{subject to $\| A z - y \|_2 \leq \eta$.}
}
Then the estimate
\be{
\label{l1_error_est}
\| \hat{x} - x \|_2 \leq  C_1 \| x - P_{\Delta} x \|_{1} + C_2 \left ( 1+ \sigma\sqrt{| \Delta |}\right ) \eta  ,
}
holds for constants $C_1$ and $C_2$ depending on $\alpha$, $\beta$, $\gamma$ and $\theta$ only.
}
We refer to such a $\rho$ as a dual certificate.

\subsection{Technical lemmas}
For the construction of an appropriate dual certificate, we require a series of technical lemmas:

\lem{
\label{l:tech1}
Let $F$ and $A \in \bbC^{m \times N}$ be as in \S \ref{sss:general_setup}, where $m = pD$, and suppose that $\Delta \subseteq \{1,\ldots,N\}$.  Then for $0 < \epsilon < 1$ and $\delta > 0$ we have
\bes{
\| P_{\Delta} A^* A P_{\Delta} - P_{\Delta} \|_2 < \delta,
}
with probability at least $1-\epsilon$, provided
\bes{
m  \geq D \cdot \Gamma_1(F,\Delta) \cdot (2 \delta^{-2} + 2 \delta^{-1} / 3 ) \cdot \log(2 | \Delta | / \epsilon ),
}
where $\Gamma_1(F,\Delta)$ is as in Definition \ref{d:coh_rel}
}
This result is in fact true under the somewhat weaker condition where $\Gamma_{1}(F,\Delta)$ is replaced by $\Gamma'_1(F,\Delta)$ being the smallest constant such that
\bes{
\| B^* P_{\Delta} \|^2_2  \leq \Gamma'_1(F,\Delta),\qquad B \sim F.
}
Since $\| B^* P_{\Delta} \|^2_2 = \| P_{\Delta} B B^* P_{\Delta} \|_{2} \leq \| P_{\Delta} B B^* P_{\Delta} \|_{\infty} \leq \| B B^* P_{\Delta} \|_{\infty}$ we see that $\Gamma'_1(F,\Delta) \leq \Gamma_1(F,\Delta)$.

\prf{
Observe that
\bes{
P_{\Delta} A^* A P_{\Delta} - P_{\Delta} = \frac1p \sum^{p}_{i=1} P_{\Delta} (B_i B^*_i - I ) P_{\Delta} = \frac1p \sum^{p}_{i=1} X_i.
}
The matrices $X_i$ are independent, and by \R{F_iso}, satisfy $\bbE X_i = 0$. Hence by the matrix Bernstein inequality (see, for example, \cite[Cor.\ 8.15]{Foucart&Rauhut:book})
\be{
\label{mat_Bern_A}
\bbP \left ( \| P_{\Delta} A^* A P_{\Delta} - P_{\Delta} \|_2 \geq \delta \right ) \leq 2 | \Delta | \exp \left ( - \frac{p^2 \delta^2/2}{\sigma^2+K p \delta/3} \right ),
}
provided $\| X_i \|_2 \leq K$ and $\nm{\sum^{p}_{i=1} \bbE(X^2_i)}_2 \leq \sigma^2$.  Observe that
\eas{
\| X_i \|_2 = \sup_{\substack{x \in \bbC^N \\ \| x \|_2 = 1}} | \ip{X_i x}{x} | \leq  \| B^*_i P_{\Delta}\|^2_2 \leq \Gamma'_1(F,\Delta) \leq \Gamma_1(F,\Delta).
}
Hence we may take $K = \Gamma_1(F,\Delta)$.  Also, since $\bbE(X^2_i )= \bbE(P_{\Delta} B_i B^*_i P_{\Delta} B_i B^*_i P_{\Delta}) - P_{\Delta}$ we have
\eas{
| \ip{\bbE(X^2_i) x}{x} | \leq \bbE \| P_{\Delta} B_i B^*_i x \|^2_2 \leq \| B^*_i P_{\Delta} \|^2_2 \bbE \| B^*_i x \|^2_2 =  \| B^*_i P_{\Delta} \|^2_2 \leq \Gamma_1(F,\Delta),
}
and therefore $\sigma^2 \leq p \Gamma_1(F,\Delta)$.  Since $p = m/D$, the result follows immediately from \R{mat_Bern_A}.
}

\lem{
\label{l:tech2}
Let $F$ and $A \in \bbC^{m \times N}$ be as in \S \ref{sss:general_setup}, where $m = pD$, and suppose that $\Delta \subseteq \{1,\ldots,N\}$.  Then for $0 < \epsilon < 1$, $\delta > 0$ and $z \in \bbC^N$ we have
\bes{
\| (P_{\Delta} A^* A P_{\Delta} - P_{\Delta})z \|_{\infty} < \delta \| z \|_{\infty},
}
with probability at least $1-\epsilon$, provided
\bes{
m \geq D \cdot \left ( 8 \Gamma_1(F,\Delta) \delta^{-1} /3+ 4 \Gamma_2(F,\Delta) \delta^{-2} \right ) \cdot \log(4 |\Delta| / \epsilon),
}
where $\Gamma_{1}(F,\Delta)$ and $\Gamma_{2}(F,\Delta)$ are as in Definition \ref{d:coh_rel}. 
}
\prf{
Let $\|z \|_{\infty} = 1$ without loss of generality.  Then
\be{
\label{uniform_relation_submatrix}
\| (P_{\Delta} A^* A P_{\Delta} - P_{\Delta})z \|_{\infty} = \max_{j \in \Delta} | \ip{e_j}{(P_{\Delta} A^* A P_{\Delta} - P_{\Delta})z} |.
}
Fix $j \in \Delta$ and observe that
\bes{
\ip{e_j}{(P_{\Delta} A^* A P_{\Delta} - P_{\Delta})z} = \frac{1}{p} \sum^{p}_{i=1} e^*_j (B_i B^*_i - I ) P_{\Delta} z = \frac{1}{p} \sum^{p}_{i=1} X_i,
}
where the random variable $X_i = e^*_j (B_i B^*_i - I ) P_{\Delta} z$ satisfies $\bbE(X_i) = 0$.  
Suppose that
\bes{
| X_i | \leq K,\qquad \sum^{p}_{i=1} \bbE | X_i |^2 \leq \sigma^2,
}
and we now find suitable constants $K$ and $\sigma$.  We have
\bes{
|X_i| \leq \| P_{\Delta} (B_i B^*_i - I )P_{\Delta} z \|_{\infty} \leq \| P_{\Delta} B_i B^*_i P_{\Delta} \|_{\infty} + 1 \leq 2 \Gamma_1(F,\Delta).
}
Also
\eas{
\bbE |X_i |^2 & =  \bbE | \ip{e_j}{B_i B^*_i P_{\Delta} z } |^2 - | \ip{e_j}{P_{\Delta} z } |^2 \leq \Gamma_{2}(F,\Delta),
}
and therefore we may take $\sigma^2 = m \Gamma_{2}(F,\Delta)$.  
In the standard way, we now separate the $X_i$ into real and imaginary parts and use the fact that the real and imaginary parts satisfy the same bounds as $X_i$.
Hence Bernstein's inequality gives
\ea{
\label{Bernstein_fixed_submatrix}
\bbP \left ( \left | \ip{e_j}{(P_{\Delta}A^* A P_{\Delta} - P_{\Delta})z} \right | \geq \delta \right ) 
& \leq \bbP \left ( \left |  \frac{1}{p} \sum^{p}_{i=1} \Re X_i \right | \geq \delta/\sqrt{2} \right ) + \bbP \left ( \left |  \frac{1}{p} \sum^{p}_{i=1} \Im X_i \right | \geq \delta/\sqrt{2} \right )
\nn \\
& \leq 4 \exp\left (-\frac{p^2 \delta^2/4}{\sigma^2+p K \delta/3} \right )
}
We apply \R{Bernstein_fixed_submatrix} and the union bound over all $j \in \Delta$ to give
\bes{
\bbP \left ( \| (P_{\Delta} A^* A P_{\Delta} - P_{\Delta})z \|_{\infty} > \delta \right ) \leq 4 |\Delta| \exp\left (-\frac{p^2\delta^2/4}{p \Gamma_2(F,\Delta)+2 p \Gamma_1(F,\Delta) \delta/3} \right ).
}
The result now follows immediately.
}

\lem{
\label{l:tech3}
Let $F$ and $A \in \bbC^{m \times N}$ be as in \S \ref{sss:general_setup} where $m=pD$, and suppose that $\Delta \subseteq \{1,\ldots,N\}$, $\Delta \neq \emptyset$.  Then for $0 < \epsilon < 1$ and $\delta > 0$ we have
\bes{
\max_{j \notin \Delta} \{ \| P_{\Delta} A^*A e_j \|_2 \} \leq \delta,
}
with probability at least $1-\epsilon$, provided
\be{
\label{tech3_bound}
m \geq D \cdot \Gamma_1(F,\Delta) \cdot \left (8 \delta^{-2} + 28 \delta^{-1} / 3 \right ) \cdot \log(2N / \epsilon)
}
where $\Gamma_{1}(F,\Delta)$ is as in Definition \ref{d:coh_rel}.
}
\prf{
Fix $j \notin \Delta$.  Then
\bes{
\| P_{\Delta} A^*A e_j \|_2 = \frac{1}{p} \nm{ \sum^{p}_{i=1} P_{\Delta} B_i B^*_i e_j }_2 = \frac{1}{p} \nm{\sum^{p}_{i=1} X_i }_2 = \frac1p Z,
}
where $X_i =P_{\Delta} B_i B^*_i e_j  $.  Note that the $X_i$'s are independent copies of the random vector $X = P_{\Delta} B B^* e_j$.  Also, since $j \notin \Delta$, we have $\bbE X = 0$.  Observe that
\bes{
\| X \|_2 = \| P_{\Delta} B B^* e_j \|_2 \leq \| P_{\Delta} B B^* P_{\Delta} \|_2 \leq \| B B^* P_{\Delta} \|_{\infty} \leq \Gamma_1(F,\Delta),
}
and
\bes{
\bbE \| X \|^2_2 \leq \| P_{\Delta} B \|^2_2 \bbE \| B^* e_j \|^2_2 = \| P_{\Delta} B \|^2_2 \leq \Gamma_1(F,\Delta).
}
Note also that $\bbE Z^2 = p \bbE \| X \|^2_2 \leq p \Gamma_1(F,\Delta)$.  Suppose now that
\be{
\label{bound_Z_var}
p \geq 4 \Gamma_1(F,\Delta) \cdot \delta^{-2}.
}
Then it follows from \cite[Cor.\ 8.45]{Foucart&Rauhut:book}\footnote{We restate this result here for convenience.  Let $X_1,\ldots,X_p$ be independent copies of a random vector $X$ on $\bbC^{n}$ that satisfies $\bbE(X) = 0$.  Suppose that $\nm{X}_{2} \leq K$ for some $K >0$.  Let $Z = \nm{\sum^{p}_{i=1} X_i }_{2}$ and $\sigma^2 = \sup_{\nm{x}_{2} \leq 1} \bbE | \ip{x}{X} |^2$.  Then, for $\delta > 0$, one has $\bbP(Z \geq \sqrt{\bbE Z^2} + t ) \leq \exp \left ( - \frac{\delta^2/2}{p \sigma^2 + 2 K \sqrt{\bbE Z^2} + \delta K/3} \right )$.} that
\eas{
\bbP \left ( \| P_{\Delta} A^*A e_j \|_2 \geq \delta \right ) &= \bbP \left ( Z \geq  p \delta \right ) 
\\
&\leq \bbP \left ( Z \geq \sqrt{\bbE Z^2} + p \delta / 2 \right ) 
\\
&\leq \exp \left ( - \frac{p^2 \delta^2/8}{p \Gamma_1(F,\Delta) + 2 \Gamma_1(F,\Delta) \sqrt{p \Gamma_1} + p \delta \Gamma_1(F,\Delta)/6 } \right )
\\
& \leq \exp \left ( - \frac{p}{\Gamma_1(F,\Delta)} \frac{\delta^2/8}{1+7 \delta/6} \right ).
}
Hence, after an application of the union bound, we deduce that
\bes{
\bbP \left ( \max_{j \notin \Delta} \{ \| P_{\Delta} A^*A e_j \|_2 \} \geq \delta \right ) \leq N \exp \left ( - \frac{p}{\Gamma_1(F,\Delta)} \frac{\delta^2/8}{1+7 \delta/6} \right ),
}
provided \R{bound_Z_var} holds.  Therefore $\bbP \left ( \max_{j \notin \Delta} \{ \| P_{\Delta} A^*A e_j \|_2 \} \geq \delta \right )  \leq \epsilon$ provided \R{bound_Z_var} holds and
\bes{
p \geq \Gamma_1(F,\Delta) \cdot \left (8 \delta^{-2} + 28 \delta^{-1} / 3 \right ) \cdot \log(N / \epsilon).
}
Clearly both this bound and \R{bound_Z_var} are implied by \R{tech3_bound}, hence we deduce the result.
}

\lem{
\label{l:tech4}
Let $F$ and $A \in \bbC^{m \times N}$ be as in \S \ref{sss:general_setup}, where $m = pD$, and suppose that $\Delta \subseteq \{1,\ldots,N\}$.  Then for $0 <  \epsilon < 1$, $\delta > 0$ and $z \in \bbC^N$ we have
\bes{
\| P^{\perp}_{\Delta} A^* A P_{\Delta} z \|_{\infty} \leq \delta \| z \|_{\infty},
}
with probability at least $1-\epsilon$, provided
\bes{
m \geq  D \cdot \left ( 4 \Gamma_1(F,\Delta) \delta^{-1} / 3 + 4 \Gamma_2(F,\Delta) \delta^{-2} \right ) \cdot \log(2N/\epsilon).
}
}
\prf{
Let $\|z\|_{\infty}=1$ without loss of generality. Then
\be{
\label{uniform_relation}
\| P^{\perp}_{\Delta} \tilde{A}^* \tilde{A} P_{\Delta} z \|_{\infty} = \max_{j \notin \Delta} | \ip{e_j}{\tilde{A}^*\tilde{A} P_{\Delta} z} |.
}
Fix $j \notin \Delta$ and observe that
\bes{
\ip{e_j}{A^* A P_{\Delta} z} = \ip{e_j}{(A^* A-I) P_{\Delta} z}  = \frac{1}{p} \sum^{p}_{i=1} e^*_j (B_i B^*_i-I) P_{\Delta} z = \frac{1}{p} \sum^{p}_{i=1} X_i,
}
where $X_i = e^*_j (B_i B^*_i-I) P_{\Delta} z$.  As in Lemma \ref{l:tech2}, note that
\bes{
|X_i| \leq 2 \Gamma_{1}(F,\Delta),
}
and also that
\bes{
\bbE |X_i|^2 \leq \Gamma_{2}(F,\Delta).
}
Hence, separating into real and imaginary parts and applying the union bound, this holds true once
\be{
\label{Bernstein_fixed}
\bbP \left ( \left | \ip{e_j}{\tilde{A}^*\tilde{A} P_{\Delta} z} \right | \geq \delta \right  ) \leq 4 \exp \left ( - \frac{p \delta^2/4}{ \Gamma_{2}(F,\Delta) + \Gamma_{1}(F,\Delta) \delta / 3 } \right ).
}
After another application of the union bound, we obtain
\bes{
\bbP \left ( \| P^{\perp}_{\Delta} A^*A P_{\Delta} z \|_{\infty} \geq \delta \right ) \leq 4 N \exp \left ( - \frac{p \delta^2/4}{ \Gamma_{2}(F,\Delta) + \Gamma_{1}(F,\Delta) \delta / 3 } \right ).
}
The result now follows.
}

\subsection{Dual certificate construction}
Our construction is based on the golfing scheme \cite{GrossGolfing:11TIT}, with a number of key modifications following ideas from\cite{Adcock&etal:16FMS}.
Recall that we assume $s \geq 2$ throughout.  In particular, $\log(s) > 0$.

\subsubsection{Setup}
For $L \in \bbN$, let $p_1,\ldots,p_L \in \bbN$ be such that $p = p_1 + \ldots + p_L$ and define the matrices
\bes{
A_{l} = \frac{1}{\sqrt{p_l}} \sum^{p_l}_{i=1} e_i \otimes B^*_i,\qquad l=1,\ldots,L,
}
so that
\bes{
A = \left [ \begin{array}{c} \sqrt{p_1/p} A_1 \\ \vdots \\ \sqrt{p_L/p}  A_L \end{array} \right ].
}
We construct the dual certificate iteratively as follows.  Let $\rho^{(0)} = 0$ and
\bes{
\rho^{(l)} =  (A_{l})^* A_l P_{\Delta} \left ( \sgn(P_{\Delta}(x)) - P_{\Delta} \rho^{(l-1)} \right ) + \rho^{(l-1)},\qquad l=1,\ldots,L,
}
for some $L \geq 1$ that will be defined later.  The dual certificate is then defined as $\rho = \rho^{(L)}$.  For ease of notation, we also set
\bes{
v^{(l)} =  \sgn(P_{\Delta}x) - P_{\Delta} \rho^{(l)} ,\qquad l=0,\ldots,L.
}
We now introduce the following events:
\eas{
A_l &:\quad \| (P_{\Delta} -  P_{\Delta} (A_{l})^* A_l P_{\Delta}) v^{(l-1)} \|_{\infty} \leq a_l \| v^{(l-1)} \|_{\infty},\qquad l=1,\ldots,L,
\\
 B_l &: \quad \|  P^{\perp}_{\Delta}  (A_{l})^* A_l P_{\Delta} v^{(l-1)} \|_{\infty} \leq b_l \| v^{(l-1)} \|_{\infty},\qquad l=1,\ldots,L.
\\
 C &: \quad \| P_{\Delta} A^* A P_{\Delta} - P_{\Delta} \|_2 \leq 1/4,
\\
 D &:\quad \max_{i \notin \Delta} \{ \| P_{\Delta} A^* A e_i \|_2 \} \leq 1,
\\
 E &: \quad A_1 \cap \cdots \cap A_L \cap B_1 \cap \cdots \cap B_L \cap C \cap D.
}
Note that the events $A_l$ and $B_l$ are different to those used in the original golfing scheme \cite{GrossGolfing:11TIT} (see also \cite{Candes&Plan:11IT}), and are based on a setup introduced in \cite{Adcock&etal:16FMS}.  A consequence of this is the slightly worse log factor \R{L_def} than that of \cite{Candes&Plan:11IT}.  But this approach allows us to deal successfully with the more complicated measurement model considered in this paper.  Unlike \cite{Adcock&etal:16FMS}, however, our iterative updates of $v^{(l)}$ are simpler and follow the first approach used in \cite{Candes&Plan:11IT}.  In the setup of this paper -- in particular, the slightly different model used for drawing the samples than that of \cite{Adcock&etal:16FMS}; see Remarks \ref{r:drawingmodel} and \ref{r:BreakingCoherenceRelation} -- we have found the more sophisticated construction employed in \cite{Adcock&etal:16FMS} does not lead to a better recovery guarantee.

Our aim is to choose the quantities $L$, $a_1,\ldots,a_L$, $b_1,\ldots,b_L$ and $p_1,\ldots,p_L$ so that conditions (i)--(v) of Lemma \ref{l:dual_certificate} are fulfilled for the parameter choices
\bes{
\alpha = 1/4,\quad \beta = 1,\quad \gamma = 1/4,\quad \theta = 1/2.
}
We choose these quantities as follows.  Set
\be{
\label{L_choice}
L = 2+ \lceil \log_2(\sqrt{s}) \rceil \geq 3.
}
\be{
\label{a_choice}
a_1 = a_2 = \frac{1}{2 \sqrt{\log_2(\sqrt{s})} },\qquad a_l = 1/2,\quad l =3,\ldots,L,
}
\be{
\label{b_choice}
b_1 = b_2 = \frac{1}{4},\qquad b_l = \frac{\log_2(\sqrt{s})}{4} ,\quad l =3,\ldots,L,
}
where $s = | \Delta |$, and
\bes{
p_1 = p_2 = \frac14 p,\qquad p_l = \frac{1}{2(L-2)} p,\quad l=3,\ldots,L,
}

\subsubsection{Event $E$ implies conditions (i)--(v)}
Suppose that event $E$ occurs.  Immediately, events $C$ and $D$ give that conditions (i) and (ii) hold with $\alpha = 1/4$ and $\beta = 1$.  Now consider condition (iii).  Note that
\ea{
 \label{v_relation}
v^{(l)} & = \sgn(P_{\Delta} x) - P_{\Delta} (A_l)^* A_l P_{\Delta} v^{(l-1)} - P_{\Delta} \rho^{(l-1)} = \left ( P_{\Delta} - P_{\Delta} (A_l)^* A_l P_{\Delta} \right ) v^{(l-1)}.
}
Hence
\be{
\label{v_bound}
\| v^{(l)} \|_{2} \leq \sqrt{s} \| v^{(l)} \|_{\infty} \leq \sqrt{s} \| v^{(0)} \|_{\infty} \prod^{l}_{j=1} a_j = \sqrt{s} \prod^{l}_{j=1} a_j.
}
Observe that
\be{
\label{a_prod}
\prod^{l}_{j=1} a_{j} = \frac{1}{2^l \log_2(\sqrt{s})} \leq \frac{1}{2^l} .
}
Hence setting $l = L$ in \R{v_bound} and noticing that
\bes{
\sgn(P_{\Delta}x)-P_{\Delta} \rho  = \sgn(P_{\Delta}x) -P_{\Delta} \rho^{(L)} = v^{(L)},
}
gives
\be{
\label{Propcond3}
\| P_{\Delta} \rho - \sgn(P_{\Delta}x) \|_2 \leq \sqrt{s} \prod^{L}_{j=1} a_{j} = \frac{\sqrt{s}}{2^L} \leq \frac{1}{4}.
}
Thus condition (iii) holds with $\gamma = 1/4$ as required.

Now consider condition (iv).  Observe that
\bes{
P^{\perp}_{\Delta} \rho^{(l)} = P^{\perp}_{\Delta} (A_l)^* A_l P_{\Delta} v^{(l-1)} + P^{\perp}_{\Delta}  \rho^{(l-1)}.
}
Therefore by definition of $ v^{(l)}$ and \R{v_bound},
\bes{
\| P^{\perp}_{\Delta} \rho^{(l)} \|_{\infty} \leq b_l \| v^{(l-1)} \|_{\infty} + \| P^{\perp}_{\Delta} \rho^{(l-1)} \|_{\infty} \leq b_l \prod^{l-1}_{j=1} a_j + \| P^{\perp}_{\Delta} \rho^{(l-1)} \|_{\infty}.
}
where we use the convention that $\prod^{l-1}_{j=1} a_{j} = 1$ when $l=1$.  Hence
\be{
\label{Propcond4}
\| P^{\perp}_{\Delta} \rho \|_{\infty} \leq \sum^{L}_{l=1} b_{l} \prod^{l-1}_{j=1} a_{j}.
}
Substituting the values of $a_l$ and $b_l$ into the right-hand side of \R{Propcond4} and using \R{a_prod} gives
\bes{
\| P^{\perp}_{\Delta} \rho \|_{\infty} \leq \frac{1}{4} \left ( 1 + \frac{1}{2 \sqrt{\log_2(\sqrt{s})}} + \frac{1}{4 } +  \frac{1}{8} + \ldots + \frac{1}{2^{L-1}}  \right ) \leq \frac{1}{2}.
}
Hence condition (iv) holds with $\theta = 1/2$, as required.

Finally consider condition (v).  Write $\rho^{(l)} =  A^* \xi^{(l)}$, where (with slight abuse of notation)
\bes{
\xi^{(l)} = \sqrt{\frac{p}{p_{l}}} A_{l} v^{(l-1)} + \xi^{(l-1)}.
}
It follows that
\be{
\label{wk_bound}
\| \xi^{(l)} \|_2 \leq \sqrt{\frac{p}{p_{l}}} \| A_{l} v^{(l-1)} \|_2 + \| \xi^{(l-1)} \|_2.
}
Consider the first term on the right-hand side.  We have
\eas{
\| A_{l} v^{(l-1)} \|^2_2 &= \ip{P_{\Delta} (A_{l})^* A_{l} P_{\Delta} v^{(l-1)}}{v^{(l-1)}}
 = \| v^{(l-1)} \|^2_2 - \ip{v^{(l)}}{v^{(l-1)}}
\leq \| v^{(l-1)} \|^2_2 + \| v^{(l-1)} \|_2 \| v^{(l)} \|_2,
}
where in the middle step we use \R{v_relation}.  By \R{v_bound}, it now follows that
\bes{
\| A_{l}  v^{(l-1)} \|^2_2 \leq s \left ( a_{l} + 1 \right ) \left ( \prod^{l-1}_{j=1} a_{j} \right )^2,
}
and therefore, returning to \R{wk_bound} and summing over $l=1,\ldots,L$, we get
\be{
\label{Propcond5}
\| \xi\|_2 \leq  \sqrt{s} \sqrt{p} \sum^{L}_{l=1} \sqrt{\frac{a_{l} + 1}{p_{l}} } \prod^{l-1}_{j=1} a_{j},
}
where $\xi = \xi^{(L)}$ is such that $\rho = A^* \xi$.  Now notice that
\bes{
\frac{a_{l} + 1}{p_{l}} = \frac{4}{p} \left ( 1 + \frac{1}{2 \sqrt{\log_2(\sqrt{s})}} \right ) \leq \frac{6}{p},\qquad l=1,2,
}
and
\bes{
\frac{a_{l} + 1}{p_{l}} \leq \frac{3(L-2)}{p},\qquad k=3,\ldots,L.
}
Hence it follows from \R{Propcond5} and \R{a_prod} that
\bes{
\| \xi \|_2  \leq \sqrt{s} \sqrt{p} \left ( \sqrt{\frac{6}{p}} (1+1/2) + \sqrt{\frac{3(L-2)}{p}} \sum^{L}_{l=3} \frac{1}{2^{l-1} \log_2(\sqrt{s})} \right ) \leq 4 \sqrt{s}  \left ( 1 + \frac{\sqrt{L-2}}{\log_2(\sqrt{s})} \right ) \leq 8 \sqrt{s}.
}
Hence condition (v) holds with $\sigma \leq 8$.

\subsubsection{Event $E$ holds with high probability} \label{sss:eventE}
In view of the arguments above, to complete the proof of Theorem \ref{t:abs_recov} it suffices to show that event $E$ holds with high probability.  By the union bound
\bes{
\bbP(E^c) \leq \sum^{L}_{l=1} \left ( \bbP(A^c_l) + \bbP(B^c_l) \right ) + \bbP(C^c) + \bbP(D^c).
}
To ensure that $\bbP(E^c) \leq \epsilon$, we shall derive conditions such that
\eas{
\bbP(A^c_l) , \bbP(B^c_l) &\leq \epsilon / 16,\qquad l=1,2,
\\
\bbP(A^c_l), \bbP(B^c_l) &\leq \epsilon / (8(L-2)),\qquad l=3,\ldots,L,
\\
\bbP(C^c) , \bbP(D^c) & \leq \epsilon / 4.
}

\noindent
\textit{Events $A_l$.}  We apply Lemma \ref{l:tech2} to the matrix $A_{l}$ with $\delta = 1/(2\sqrt{\log_2(\sqrt{s})})$ and $\epsilon / 16$ to deduce that $A_{1}$ and $A_{2}$ hold with probability at least $1-\epsilon/16$ provided
\bes{
p_l \gtrsim  \Gamma(F,\Delta) \cdot \log(s)  \cdot \log(64 s /\epsilon),\qquad l=1,2.
}
For $A_l$ with $l=3,\ldots,L$ we apply Lemma \ref{l:tech2} with $\delta = 1/2$ and $\epsilon / (8(L-2))$ to find that event $A_l$ holds with probability $1-\epsilon/(8(L-2))$ provided
\bes{
p_l \gtrsim \Gamma(F,\Delta) \cdot \log(32 s (L-2) / \epsilon),\qquad l=3,\ldots,L.
}
Applying the values for $p_l$ and using the fact that $s \geq 2$, $0 < \epsilon < 1$ and $L - 2 \lesssim \log(s)$ we thus deduce the following condition on $p$:
\be{
\label{A_m}
p \gtrsim \Gamma(F,\Delta) \cdot \log(s) \cdot \log(s \log(s) / \epsilon) 
}
\textit{Events $B_l$.}  For $l=1,2$, we apply Lemma \ref{l:tech4} to the matrix $A_{l}$ with $\delta = 1/4$ and $\epsilon / 16$ to see that $B_1$ and $B_2$ hold with probability at least $1-\epsilon/16$ provided
\bes{
p_l \gtrsim  \Gamma(F,\Delta) \cdot \log(32N/\epsilon),\qquad l=1,2.
}
Similarly, for $l=3,\ldots,L$ we apply the same lemma with $\delta = \log_2(\sqrt{s})/4$ and $\epsilon /(8(L-2))$ to get that $B_l$ holds with probability $1-\epsilon/(8(L-2))$ provided
\bes{
p_l \gtrsim  \Gamma(F,\Delta) \cdot (\log_2(\sqrt{s}))^{-1} \cdot \log(16(L-2) N/\epsilon),\qquad l=3,\ldots,L.
}
Using the values for $p_l$ once more, we deduce that
\be{
\label{B_m}
p \gtrsim \Gamma(F,\Delta) \cdot \log(\log(s)N/\epsilon).
}
\textit{Events $C$ and $D$.}  Lemma \ref{l:tech1} implies that event $C$ holds with probability at least $1-\epsilon/4$ provided
\be{
\label{D_m}
p \gtrsim \Gamma(F,\Delta) \cdot \log( s / \epsilon),
}
and Lemma \ref{l:tech3} implies that event $D$ holds with probability at least $1-\epsilon/4$ provided
\be{
\label{E_m}
p \gtrsim \Gamma(F,\Delta) \cdot \log(N/ \epsilon).
}
Combining \R{A_m}, \R{B_m}, \R{D_m} and \R{E_m}, we deduce the following condition on $m = p D$:
\bes{
m \gtrsim D \cdot \Gamma(F,\Delta) \cdot \left ( \log \left ( \log(s) N / \epsilon \right ) + \log(s) \log \left(s \log(s) / \epsilon \right ) \right ).
}
Since $\log(s) \leq s \leq s / \epsilon \leq N /\epsilon$, we see that this condition is implied by
\bes{
m \gtrsim D \cdot \Gamma(F,\Delta) \cdot \left ( \log(N/\epsilon) + \log(s) \log(s/\epsilon) \right ).
}
This completes the proof of Theorem \ref{t:abs_recov}.

\section{Conclusions and challenges}\label{s:conclusion}
In this paper we have presented a framework for parallel acquisition with CS. Multi-sensor systems arise in a variety of applications for a number of different reasons, including cost, scan time or power consumption reduction, or resolution enhancement.  Our main theoretical results quantify this improvement by giving nonuniform recovery guarantees for which the number of measurements required per sensor decreases linearly with the total number of sensors $C$, or equivalently, the total number of measurements $m$ is independent of $C$.  See Corollaries \ref{c:distinct_sparsity} and \ref{c:distinct_sparsity_in_levels} for distinct sampling, and Theorem \ref{t:identical_pcwse_const} for identical sampling.  For the specific case of diagonal or circulant sensor profile matrices, our results give sufficient conditions for such optimal guarantees, both in the case of distinct or identical sampling. Such results are in agreement with the numerical experiments performed in \S \ref{s:NumExp}.
In general, arguing which of the two sampling scenarios is better is not straightforward.  On the one hand, this is often dictated by the application; pMRI dictates identical sampling, for example, whereas distinct sampling may be possible in multi-view imaging.  Overall, our optimal recovery guarantees hold under weaker conditions for distinct sampling than for identical sampling.  Hence, given the choice, we generally recommend that over identical sampling.  However, our numerical experiments in \S \ref{s:NumExp} and more recent computable bounds in \cite{Chun&Adcock:16ITW, Chun&Li&Adcock:16MMSPARSE} suggest that identical sampling may succeed under weaker conditions than those of our current results, especially for sparse and distributed vectors.

There are a number of avenues for future work.  First, in this paper we have not considered the use of sparsifying transforms, such as wavelets, discrete cosine transforms or total variation.  Such transforms arise frequently in applications and often require more sophisticated CS analysis, due to issues such as varying coherence \cite{Adcock&etal:16FMS} and the lack of invariance of coherence to unitary transforms (see Remark \ref{r:sparsifying_transforms}). 
For some work in this direction on subgaussian sensing matrices and arbitrary unitary sparsifying transforms, see \cite{Chun&Adcock:16arXiv-CS&PA&RIP} (see also \cite{Eftekhari&etal:15ACHA} for the particular case of block-diagonal subgaussian sensing matrices). 

There are also number of extensions to the current theory that should also be explored.  First, in a manner typical of nonuniform guarantees, our error estimates are worse by a factor of $\sqrt{s}$ than those stemming from uniform guarantees; see, for example, \cite{Foucart&Rauhut:book}. In \cite{Candes&Plan:11IT} this is avoided by using the so-called weak RIP.  It is work in progress to extend the weak RIP to the multi-sensor setting and to the sparsity in levels model.  Second, we have only considered a discrete setup, where the signal to recover is a vector in $\bbC^N$.  Yet many physical sensing systems are continuous in nature, and thus require infinite-dimensional CS techniques.  For work in this direction in the single-sensor case, see \cite{Adcock&Hansen:15FCM, Adcock&etal:16FMS, Rauhut&Ward:12JAT, Rauhut&Ward:16ACHA}.  Third, our requirement that the sensor profiles satisfy $C^{-1} \sum^{C}_{c=1} H^*_c H_c = I$ (for distinct sampling) or $\sum^{C}_{c=1} H^*_c H_c = I$ (for independent sampling) is quite stringent.  Future work will explore the weakening of this condition, similar to \cite{Chun&Adcock:16arXiv-CS&PA&RIP}.  Fourth, in some applications it is possible to construct sensor profile matrices which are in some sense random.  We expect these will lead to better recovery guarantees, especially for the identical sampling case.  This is a question for future investigations.

On the algorithmic side, it is not always the case that the sensor profile matrices $H_c$ are exactly known in advance.  Joint estimation of the signal $x$ and the  $H_c$ is a topic for future work.  See \cite{Ling&Strohmer:15IP, Knoll&Clason&Bredies&Uecker&Stollberger:12MRM, She&Chen&Liang&DiBella&Ying:14MRM}  for related work in this direction, as well as \cite{Chun&Adcock&Talavage:15TMI, She&Chen&Liang&DiBella&Ying:14MRM} for the case of pMRI.  This aside, we expect further improvements can be made in the recovery algorithm.  In the context of pMRI in \cite{Chun&Adcock&Talavage:15TMI} the standard $\ell_1$ functional was replaced by a joint-sparsity promoting functional, reminiscent of ideas in distributed CS.  We expect a similar approach to be possible in the more general setting of this paper.

\section*{Acknowledgements}
The authors would like to thank Claire Boyer, Anders Hansen, Felix Krahmer, Holger Rauhut and Pierre Weiss for useful comments and suggestions.  BA wishes to acknowledge the support of Alfred P. Sloan Research Foundation and the Natural Sciences and Engineering Research Council of Canada through grant 611675.  BA and IYC acknowledge the support of the National Science Foundation through DMS grant 1318894.

\small
\bibliographystyle{abbrv}
\bibliography{referenceBibs_Bobby_abbrv}

\end{document}